\documentclass[traditabstract]{aa}
\usepackage{graphicx}
\usepackage{txfonts}
\usepackage{natbib}\usepackage{enumitem}
\usepackage[usenames]{color}
\bibpunct{(}{)}{;}{a}{}{,}

 \newcommand{\mic}{$\mu$m}
 \newcommand{\mics}{$\mu$m~}
 \newcommand{\lum}{$\nu L _\nu (60\mu $m$)$}
 \newcommand{\lums}{$\nu L _\nu (60\mu $m$)$~}
 \newcommand{\ergs}{erg s$^{-1}$}

\begin{document}

\title{Enhanced star formation rates in AGN hosts with respect to inactive galaxies from PEP-Herschel\thanks{Herschel is an ESA space observatory with science instruments provided
by European-led Principal Investigator consortia and with important
participation from NASA.} observations}

\author{
P. Santini\inst{1,2}
\and
D. J. Rosario\inst{1}
\and
L. Shao\inst{1}
\and
D. Lutz\inst{1}
\and
R. Maiolino\inst{2}
\and
D. M. Alexander\inst{3}
\and
B. Altieri\inst{4}
\and
P. Andreani\inst{5,6}
\and
H. Aussel\inst{7}
\and
F. E. Bauer\inst{8}
\and
S. Berta\inst{1}
\and
A. Bongiovanni\inst{9,10}
\and
W. N. Brandt\inst{11}
\and
M. Brusa\inst{1}
\and
J. Cepa\inst{9,10}
\and
A. Cimatti\inst{12}
\and
E. Daddi\inst{7}
\and
D. Elbaz\inst{7}
\and
A. Fontana\inst{2}
\and
N. M. F{\"o}rster Schreiber\inst{1}
\and
R. Genzel\inst{1}
\and
A. Grazian\inst{2}
\and
E. Le Floc'h\inst{7}
\and
B. Magnelli\inst{1}
\and
V. Mainieri\inst{5}
\and
R. Nordon\inst{1}
\and
A. M. P{\'e}rez Garcia\inst{9,10}
\and
A. Poglitsch\inst{1}
\and
P. Popesso\inst{1}
\and
F. Pozzi\inst{12}
\and
L. Riguccini\inst{7}
\and
G. Rodighiero\inst{13}
\and
M. Salvato\inst{14,1}
\and
M. Sanchez-Portal\inst{4}
\and
E. Sturm\inst{1}
\and
L. J. Tacconi\inst{1}
\and
I. Valtchanov\inst{4}
\and
S. Wuyts\inst{1}
}

   \offprints{P. Santini, \email{paola.santini@oa-roma.inaf.it}}

\institute{ Max-Planck-Institut f\"{u}r Extraterrestrische Physik (MPE), Postfach 1312, 85741 Garching, Germany.
\and INAF - Osservatorio Astronomico di Roma, via di Frascati 33, 00040 Monte Porzio Catone, Italy.
\and Department of Physics, Durham University, South Road, Durham, DH1 3LE, UK.
\and European Space Astronomy Centre, Villafranca del Castillo, Spain.
\and ESO, Karl-Schwarzschild-Str. 2, D-85748 Garching, Germany.
\and INAF-Osservatorio Astronomico di Trieste, via Tiepolo 11, 34131 Trieste, Italy.
\and Laboratoire AIM, CEA/DSM-CNRS-Universit{\'e} Paris Diderot, IRFU/Service d'Astrophysique, B\^at.709, CEA-Saclay, 91191 Gif-sur-Yvette Cedex, France.
\and Pontificia Universidad Cat\'olica de Chile, Departamento de Astronom\'ia y Astrof\'isica, Casilla 306, Santiago 22, Chile.
\and Instituto de Astrof{\'i}sica de Canarias, 38205 La Laguna, Spain.
\and Departamento de Astrof{\'i}sica, Universidad de La Laguna, Spain.
\and Department of Astronomy and Astrophysics, 525 Davey Lab, Pennsylvania State University, University Park, PA 16802, USA.
\and Dipartimento di Astronomia, Universit{\`a} di Bologna, Via Ranzani 1, 40127 Bologna, Italy.
\and Dipartimento di Astronomia, Universit{\`a} di Padova, Vicolo dell'Osservatorio 3, 35122 Padova, Italy.
\and IPP-Max-Planck-Institut f\"{u}r Plasmaphysik, Boltzmannstrasse 2, D-85748, Garching, Germany.
}

   \date{Received .... ; accepted ....}
   \titlerunning{Enhanced SFR in AGN hosts from PEP-Herschel observations}

 \abstract{  
We compare the average star formation (SF) activity in X-ray selected AGN hosts with a mass-matched control sample of inactive\thanks{Throughout the paper, the wording `inactive'/`active'  refers to galaxies lacking/showing nuclear activity (non-AGNs/AGNs), regardless of their star formation rate.}   
galaxies, including both star forming and quiescent sources, in the $0.5<z<2.5$ redshift range. 
Recent observations carried out by PACS, the $60-210$ \mics photometric camera on board the Herschel Space Observatory, in GOODS-S, GOODS-N and COSMOS allow us to obtain an unbiased estimate of the far-IR luminosity, and hence of the SF properties, of the two samples. Accurate AGN host stellar mass estimates are obtained by decomposing their total emission into the stellar and the nuclear components. 
We report evidence of a higher average SF activity in AGN hosts with respect to the control sample of inactive galaxies. 
The level of SF enhancement is modest ($\sim 0.26$ dex at $\sim 3 \sigma$ confidence level) at low X-ray luminosities ($L_X \lesssim 10^{43.5}$ \ergs) and more pronounced ($0.56$ dex at  $> 10 \sigma$ confidence level) in the hosts of luminous AGNs. 
However, when comparing to star forming galaxies only, AGN hosts are found broadly consistent with the locus of their `main sequence'. 
We  investigate the relative far-IR luminosity distributions of active and inactive galaxies, and find a higher fraction of PACS detected,  hence normal and highly star forming systems among AGN hosts. 
Although different interpretations are possible, we explain our findings as a  consequence of a twofold AGN growth path: 
faint AGNs  evolve through secular processes, with instantaneous AGN accretion not tightly linked to the current total SF in the host galaxy,   
while the luminous AGNs  co-evolve with their hosts through periods of enhanced AGN activity and star formation, possibly  through major mergers. 
While an increased SF activity  with respect to inactive galaxies of similar mass is expected in the latter, we interpret the modest SF offsets measured in low-$L_X$ AGN hosts as either 
$a)$ generated by non-synchronous accretion and SF histories in a merger scenario 
or $b)$ due to possible connections between instantaneous SF and accretion that can be induced by smaller scale (non-major merger) mechanisms. Far-IR luminosity distributions favour the latter scenario. }


\maketitle

\section{Introduction} \label{sec:intro}

Star formation (SF) and the phenomenon of Active Galactic Nuclei (AGNs) are two well-known and well studied aspects of galaxies. The former is related to the growth of stellar mass while the latter is tied to the growth of nuclear super-massive black holes
(SMBHs). Both processes play primary roles in the formation and evolution of galaxies. 

Most star forming galaxies over most of the age of the Universe follow a strong correlation between their star formation rate (SFR) and stellar mass, defining the so-called main sequence of star formation. This tight relationship appears to hold up to $z\sim 6$ \citep[e.g.][]{brinchmann04,noeske07,elbaz07,daddi07a,santini09,stark09}. 
While its normalization is known to increase with redshift up to $z\sim 2-3$, an epoch which corresponds to the peak of SF \citep[e.g.][]{hopkins06}, and to flatten at higher redshifts, its slope is still debated \citep[see discussion in][]{dunne09,pannella09,santini09,rodighiero10,karim11}. 
Outliers from the main sequence are also  observed, such as Ultraluminous Infrared Galaxies \citep[ULIRGs,][]{sanders96} and Submillimeter Galaxies (SMGs) \citep{chapman03,pope08,tacconi08}. These are characterized by very large SFRs and specific SFRs (SFR per unit stellar mass). Studies based on CO emission interpret  outliers as gas rich systems likely  experiencing a different SF regime with respect to normal star forming galaxies  on the main sequence: while the latter population is thought to form by smooth, secular accretion with long duty cycles \citep[e.g.][]{dekel09}, outliers  are believed to undergo short and intense starbursts occurring typically during mergers\footnote{Hereafter, unless otherwise stated, we refer to gas-rich major merging when talking about `mergers'.} or in dense nuclear star forming regions, which increase their SF efficiency \citep{daddi10,tacconi10,genzel10,wuyts11}. 

Observational evidence indicates that all massive galaxies experience one or more phases of nuclear activity \citep[e.g.][]{richstone98} during which their SMBHs grow. 
Several observations support a scenario of close connection between AGNs and their hosts. 
Tight correlations are known to exist between SMBH mass and galaxy bulge properties (stellar mass, velocity dispersion, etc.)  in the local Universe \citep[e.g.][]{gebhardt00,ferrarese00,marconi03,kormendy09,gueltekin09,graham11} and are also traced to higher redshift \citep[e.g.][]{merloni10,bennert11}. 
In addition, galaxies and SMBHs demonstrate similarities in their luminosity-dependent evolution with redshift: 
the space density of luminous AGNs peak at $z\sim 2$, while that of lower luminosity AGNs has its maximum at $z\sim 1$ \citep[e.g.][]{cowie03,fiore03,lafranca05,hasinger05,brandt05,bongiorno07}. This so-called \textit{anti-hierarchical} evolution is similar to 
the $downsizing$ behaviour of galaxy SF activity (see  
 \citealt{fontanot09} and references therein). 
Finally, AGN growth seems to be mostly due to matter accretion from the host galaxies during the
active phases of the AGN \citep[although mergers of black holes may be responsible for part of the growth,][]{marconi04}. Bright quasi-stellar objects (QSOs) are sometimes found to be associated with intense starburst events  \citep[e.g.][]{rowanrobinson95,barvainis02,omont03,priddey03,page04,stevens05}. 
Post-starburst spectral signatures have been reported in some QSOs, as well as in lower luminosity local AGNs \citep{canalizo00,kauffmann03,davies07}. 
In \cite{shao10}, we found a substantial level of star formation in both low and high luminosity AGNs, increasing with redshift at similar rate as the star formation in inactive galaxies (see also \citealt{lutz10,mullaney12} and \citealt{mullaney10}). 
Finally, a number of studies reported evidence for a correlation between AGN and far-infrared (FIR) luminosity, as well as between AGN luminosity and PAH feature strength \citep{rowanrobinson95,schweitzer06,netzer07,lutz08,bonfield11}.

Although there is general agreement that relationships exist between SMBHs and their host properties, the mechanism responsible for triggering the active phase is still debated. 
At high AGN luminosities, a process is at play which correlates AGN activity with star formation in the host galaxy. Major mergers are a good possibility according to early observational studies \citep[e.g.][]{stockton82,heckman84,stockton91,hutchings92} and to more recent theoretical predictions \citep[e.g.][]{springel05b}; these transport large amounts of gas to the centre of the merging galaxies, feeding SMBHs and triggering intense SF episodes. 
Such a co-evolutionary scenario is supported by studies of local ULIRGs \citep[e.g.][]{sanders88a,sanders88b,sanders96,canalizo00,canalizo01}, which are all in major mergers, and
whose exceptionally high SFR is often accompanied by powerful AGN activity. 
Moreover, there is  good evidence for a close connection between AGN activity and SF in SMGs \citep[e.g.][]{chapman05,alexander08,coppin08}. 
However, these studies may be subject to selection effects, since such bright IR galaxies require, by construction, large gas amounts, which are supplied by major mergers. 
Alternative secular mechanisms of gas inflow, `non-merger' mechanisms hereafter, have also been suggested as means to drive SF and trigger AGNs simultaneously. These mechanisms include  minor mergers, disk instabilities and bars, supernova explosions or the infall of recycled gas returned to the interstellar medium \citep[e.g.][]{wada04,martini04,jogee06,genzel08,dekel09,johansson09,ciotti10}. 
While major mergers are usually invoked to produce bright quasars \citep[e.g.][]{hopkins09,veilleux09a,fiore12}, secular processes are asserted to be sufficient to trigger the active nucleus in low luminosity AGNs \citep[e.g.][]{hopkins09,mullaney12}. However, recent studies \citep[e.g.][]{cisternas11,allevato11} support evidence for major mergers not being the leading triggering mechanism even in moderately luminous AGNs ($L_X \sim 10^{43.5}$ erg s$^{-1}$).

AGNs are believed to play an important role in regulating their host galaxy's SF activity, and  are therefore considered to be fundamental ingredients of theoretical models of galaxy formation and evolution. 
Nuclear emission is believed to be responsible for gas heating and consequently SF suppression. 
This `negative AGN feedback' is necessary for theoretical models to reproduce many observations, such as the strong suppression
of SF in the most massive galaxies \citep[e.g.][]{dimatteo05,springel05b,bower06,croton06,menci08,hopkins10}. 
Moreover, `positive feedback'  may also occur, where AGN-driven winds induce SF in the host galaxy \citep[e.g.][]{begelman89,silk05,feain07,silk09,elbaz09}.

For all the reasons described above, a detailed knowledge of the interplay between 
AGN accretion and star formation processes of the host galaxy is needed to reach a full understanding of galaxy formation and evolution. 
To shed light on their connection, we study the SF activity in X-ray selected AGN hosts from $z\sim0.5$ to $z\sim2.5$, i.e. in 
the range of redshifts over which most of the stellar and SMBH content of the Universe has been put into place. 
In \cite{shao10} we made the first effort in this direction, and  we used Herschel SDP data in GOODS-N to study the dependence of host star formation on redshift and AGN luminosity. We found that the level of SF in AGN host galaxies increases with redshifts at a similar rate as in inactive galaxies. Moreover, at each given redshift, we found no dependency between the host FIR luminosity (used as a proxy of the SFR) and the AGN luminosity up to $L_X\sim 10^{44}$ \ergs, while a correlation is observed in  brighter AGNs  \citep[Fig. 6 of][]{shao10}. We interpreted these results as reflecting the interplay of two paths of AGN-host co-evolution:   
very luminous AGNs are closely coupled to their host galaxy's growth by an evolutionary mechanism, for which one possible explanation is major merging; on the other hand, no close coupling is observed between AGN accretion and total host SFR in low luminosity AGNs, reflecting a secular evolution. The level of this secular star formation increasingly dominates over the correlation at increasing redshift.

This study expands upon our previous work. 
We take the next step and systematically compare the SF properties of AGN hosts to inactive galaxies of similar stellar mass in order to account for the important covariances of
stellar mass with SF. 
For this purpose, accurate estimates of the host stellar masses were obtained  
by means of a specific decomposition technique developed to separate stellar and nuclear emission.  
Measuring the SFR in AGN hosts is  not trivial, since nuclear emission can frequently outshine both the ultra-violet (UV) and mid-infrared (MIR) emission from young stars. However, the FIR continuum is shown to be dominated by the host galaxy emission for all but the most extreme AGN-dominated systems \citep{netzer07, lutz08,mullaney11} and to be a proxy of its SF activity \citep{schweitzer06,lutz08} (see also Sect. \ref{sec:l60}). 
The Herschel Space Observatory \citep{pilbratt10} allows us to probe the FIR nature of galaxies with greater depth and angular resolution than possible with previous FIR and sub-mm facilities. 
The Photodetector Array Camera and Spectrometer \citep[PACS,][]{poglitsch10}, with its 60-210 \mics window, is able to detect the FIR emission of  dust heated by UV photons from young stars. 
Herschel allows a clean measurement of the SF  activity in a wavelength range that is relatively unaffected by  AGN contamination or uncertain UV dust corrections. In addition, PACS studies are not restricted to the modest sample sizes and 
severe flux-limited selection of MIR spectroscopic campaigns.

The structure of the paper is the following: 
after introducing the  sample and the selections applied to it in Sect \ref{sec:dataset}, as well as the methodology used to infer stellar masses and FIR luminosities (and hence SFRs) in Sect. \ref{sec:method}, we present our results in Sect. \ref{sec:res}.  
In Sect. \ref{sec:disc} we discuss the implications of this analysis and suggest possible interpretations. 
Finally, Sect. \ref{sec:summ} summarizes the major steps and results of this work.

We adopt a $H_0=70$ km s$^{-1}$ Mpc$^{-1}$, $\Omega_M=0.3$ and $\Omega_\Lambda=0.7$ cosmology and assume a Salpeter IMF.

\section{Data set} \label{sec:dataset}

This work exploits the excellent multi-wavelength coverage available in three deep extragalactic fields: GOODS-South, GOODS-North and COSMOS. The depth of the two GOODS fields in the X-ray, FIR and several other bands is essential to probe faint and high redshift galaxies and AGNs. The shallower, wide-area COSMOS field provides us with good statistics among massive and bright sources, which are rare in the GOODS fields. 

\subsection{Far-infrared (FIR) data}
 
The far-infrared data used in this work were collected by the PACS instrument \citep{poglitsch10} on board the Herschel Space Observatory \citep{pilbratt10}, as part of the PACS Evolutionary Probe \citep[PEP,][]{lutz11} survey. 
Observations of GOODS-S were carried out in all three PACS bands (70, 100 and 160 \mic), while the other two fields were only  observed in the two long wavelength bands. 
We made use of PACS catalogues extracted based on the prior knowledge of the positions and fluxes of sources detected in deep archival 
MIPS 24 \mics imaging in these fields \citep{magnelli09,lefloch09}. This allows us to accurately deblend PACS sources in images characterized by a large PSF, especially in crowded fields, and permits us to 
improve the completeness and reduce the number of spurious sources at faint levels 
compared to a blind extraction. 
Fluxes in GOODS-S reach 1.1, 1.2 and 2.4 mJy at 3$\sigma$ in the 70, 100 and 160 \mics bands respectively. In the other two fields the 3$\sigma$ flux limit at 100/160 \mics is 3.0/5.7 mJy in GOODS-N and 5.0/10.2 mJy in COSMOS.  
We consider sources below these limits as undetected by PACS. 
Detailed information on the PEP survey, observed fields, data processing and source extraction may be found in \cite{lutz11}.

The technique used to  associate PACS fluxes  with optical counterparts (i.e. PACS-to-24 \mics and 24 \mic -to-IRAC and  optical association based on IRAC positions) could in principle introduce some biases when applied to the specific topic of this work, namely the comparison of FIR properties of AGN hosts and inactive galaxies. 
The first reason is that, given their strong MIR emission \citep[e.g.][]{daddi07b,fiore08}, AGN hosts are more represented in the 24 \mics prior list compared to inactive galaxies (for a given FIR host's luminosity, they have a larger probability of being detected at 24 \mic). 
We might therefore 
associate FIR flux to an AGN host rather than an inactive galaxy if the latter is not represented in the prior list. 
However, the number of PACS sources with no 24 \mics counterpart is very small \citep[of the order of $<2\%$,][]{magdis11,lutz11}. 
The second source of possible bias is that, in case of severe blending in the PACS images, the  flux is associated with the brighter MIR counterpart, which, again, is more likely to be an AGN (in case of one inactive and one active galaxy of similar host's luminosity falling in the same beam).  However, such blending issues are quite rare (only $\sim 4$\% of PACS sources would be affected in the reddest PACS band in GOODS while the fraction would be negligible in COSMOS).
We performed two tests to understand the degree of these possible biases on our results. First, we repeated our analysis only using PACS catalogues obtained by blind extraction \citep{lutz11}. Then, we performed a prior extraction without using a MIR flux weighting scheme and re-ran our analysis. 
In both cases we found no significant difference in our results, given the very similar FIR flux distributions. 
These tests guarantee that the results presented in this work are not caused by artefacts occurring during source extraction and/or optical associations.

\subsection{Ancillary data} \label{sec:ancil}
 
In order to infer redshifts and stellar masses needed for this study, we complement PACS observations with multiwavelength photometric catalogues. 

In the two GOODS fields we used  catalogues which provide photometry in 14 bands, from the $U$-band to 8 \mic.  
For GOODS-S we used the updated GOODS-MUSIC catalogue \citep{santini09,grazian06}. 
Since the latter covers an area smaller than the entire GOODS-S field, for AGNs lying outside the GOODS-MUSIC footprint  we used the photometric information compiled by \cite{luo10} from other public photometric datasets in the GOODS-S field. We verified that systematic differences in the photometric catalogues were not generating serious inconsistencies in the stellar mass estimates (Sect. \ref{sec:mass}) by comparing masses of sources in common to GOODS-MUSIC and \cite{luo10} derived from both sets of multi-wavelength photometry.  
For GOODS-N we  used the catalogue compiled by the PEP Team and described in \cite{berta10,berta11}.

For the COSMOS field we used the multiwavelength public catalog available at http://irsa.ipac.caltech.edu/data/COSMOS/tables/photometry/. The data reduction is described in \cite{capak07}. However the new catalog uses better algorithms for source detection and photometry measurements. 
Still, by comparing the COSMOS colour--mass diagram to that of GOODS, we found a suspiciously large overabundance of faint red galaxies in the COSMOS catalogue, especially at high redshifts. 
The Subaru $I$-band photometry, which is the selection basis for this catalogue, shows an extended tail towards fainter values for faint objects when compared to the ACS $I$-band. Extremely low Subaru $I$ fluxes can then give rise to fake breaks in the observed spectral shape which may be confused with red and massive galaxy features. The requirement of at least a detection in the ACS $I$-band solved this problem, so we restricted both  the AGN and the non-AGN sample to the area covered by ACS in order to ensure good quality photometry for all object in the COSMOS field. 
The COSMOS optical catalog is  supplemented with $K$ photometry from \cite{mccracken10}, IRAC photometry from \cite{sanders07} and \cite{ilbert09}, and 24 \mics photometry from \cite{lefloch09}. 
As far as the control sample of inactive galaxies is concerned (see Sect. \ref{sec:selections}), instead of using the entire (more than two million sources) dataset, we randomly extracted from the central and most covered area a catalogue listing $\sim 65000$ ACS $I$--selected sources.

\begin{figure*}[!t]
 \resizebox{\hsize}{!}{\includegraphics[angle=270]{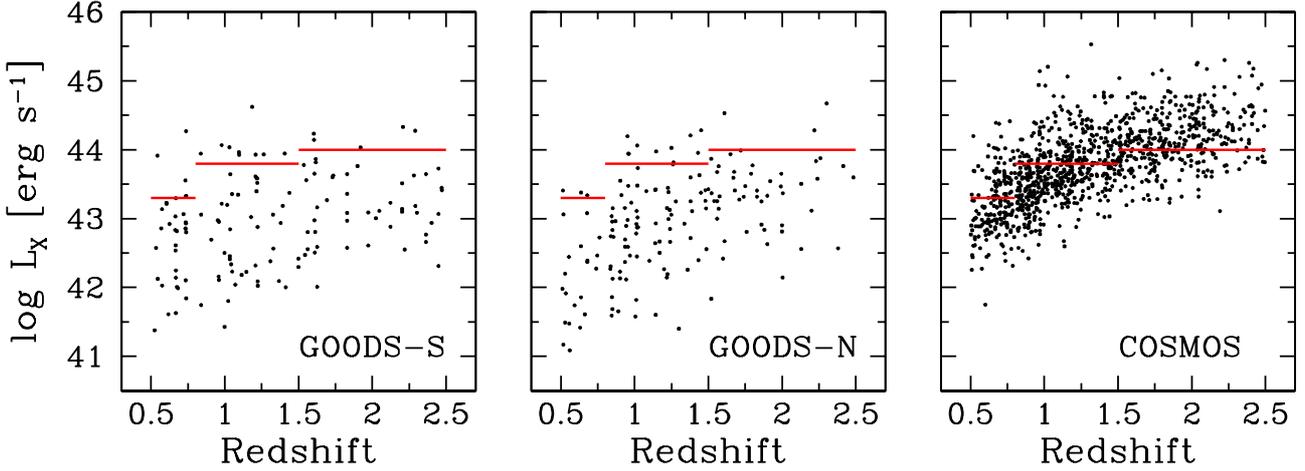}}
 \caption{Intrinsic X-ray luminosities at 2-10 keV ($L_X$) as a function of redshift in the three  fields for the final AGN sample used in the analysis (Sect. \ref{sec:selections}). Horizontal red lines show the thresholds used to split the sample into high- and low-$L_X$ subsamples (Sect. \ref{sec:dichotomy}). 
  }
 \label{fig:xray}
 \end{figure*}

All the catalogues are supplemented with either spectroscopic redshifts, or photometric redshifts when spectroscopic ones are unavailable. 
Photometric redshifts in GOODS-S were computed by fitting the multiwavelength photometry to the PEGASE 2.0 templates (\citealt{fioc97}, see details in \citealt{grazian06} updated as in \citealt{santini09}) and by adopting the EAZY code \citep{brammer08} in GOODS-N and COSMOS \citep[details in][]{berta11}. The fraction of  total outliers, defined as objects having $|\Delta(z)|/(1 + z_{spec}) > 0.2$, is 5.3\%, 5.8\% and 0.7\% in GOODS-S, GOODS-N and COSMOS, respectively. Once these outliers have been removed, the average  $|\Delta(z)|/(1 + z_{spec})$ is equal to $0.04\pm0.04$, 
$0.03\pm0.03$ and $0.01\pm0.02$, respectively in the three fields. We refer to the papers cited above for more detailed information about the photometric redshift estimate and their quality.

\subsection{X-ray information}\label{sec:xray}

Detailed X-ray point source catalogues were compiled for all three fields. These provide rest-frame X-ray luminosities and obscuring column densities $N_H$ derived from X-ray spectral analysis, source classification, optical associations as well as, where needed, photometric redshifts specifically suited for AGN hosts, based on optimised galaxy, AGN and galaxy/AGN hybrid templates.

For GOODS-S we used the \cite{luo08} catalogue from the 2 Msec Chandra Deep Field-South imaging program. Cross-matches, photometric redshifts and X-ray properties  are provided by \cite{luo10}.

For GOODS-N we used the 2 Msec Chandra Deep Field-North catalogue of \cite{alexander03} and an updated version of the classification into AGNs and other types of X-ray sources compiled by \cite{bauer04} \citep[see also][]{shao10}.  We refer to \cite{bauer04} for details on spectroscopic and photometric redshifts and X-ray properties.

Finally, for COSMOS, we used the XMM-COSMOS catalogue compiled by \cite{cappelluti09} and the X-ray optical associations and derived properties presented in \cite{brusa10} in its updated version, which includes new redshifts from ongoing (mostly DEIMOS/KECK) spectroscopic campaigns and few changes in the published redshifts and/or spectroscopic classifications from a re-analysis of the data. X-ray luminosities and obscuring column densities  are derived by \cite{mainieri07,mainieri11}.   
Photometric redshifts are  provided by \cite{salvato09}. 
As mentioned in Sect. \ref{sec:ancil},  
we restricted ourselves to X-ray selected AGNs that lie in the area covered by ACS $I$-band, which are roughly $82\%$ of the full XMM-COSMOS catalogue.

A sense of the differing depths of the X-ray data in the three fields may be gained from Fig. \ref{fig:xray}, which shows the distribution of absorption-corrected rest-frame 2-10 keV luminosities ($L_X$ hereafter) as a function of redshift for the final AGN sample used in this analysis (see Sect. \ref{sec:selections} for the selections applied to the total sample). GOODS (South+North) and COSMOS can be regarded as  complementary in terms of their X-ray properties: the deep GOODS fields allow a good sampling of faint AGNs, while in COSMOS we probe the  rarest and most powerful sources. 

\subsection{Sample selection} \label{sec:selections}

As discussed in Sect. \ref{sec:mass}, we use the classifications of X-ray sources into optical Type 1 and Type 2 as a prior in the
stellar mass estimates of AGN hosts. As a first step, we selected AGN host galaxies based on the classification provided by the X-ray catalogues. 
This classification is limited to optical spectroscopic criteria for most of the sources in the GOODS fields and for the entire COSMOS sample.  
In order to increase the statistics and to avoid spectroscopic biases, we expanded the AGN sample 
by including all X-ray sources with $L_X>10^{42}$ \ergs\ even if not classified as AGN. We considered them as Type 2 in GOODS, being highly unlikely that Type 1 AGNs were missed by the dedicated spectroscopic campaigns that targeted AGNs in these deep fields. In COSMOS we instead used the classification  provided by the photometric redshift procedure  of \cite{salvato09}, which has been demonstrated to be highly trustworthy \citep[e.g. ][]{cappelluti09,lusso10}. 
Following \cite{lusso10}, we considered  Type 1 all the objects best-fitted by a QSO or hybrid Type 1 - galaxy template (model SED $\geq 18$ in \citealt{salvato09}), and  Type 2 all the remaining objects. 
 
In order to derive reliable estimates of the FIR luminosity (Sect. \ref{sec:l60}), we only used sources in regions of our PACS images where the 100 \mics coverage (i.e. integration time) is at least half of its value at the centre of the image. 
In the redshift range 0.5 -- 2.5 studied in this work, this coverage restriction 
removes $\sim 21\%$ of AGNs in GOODS-S and $\sim 12\%$ in GOODS-N. 
We divided our sample into three mass bins ($\log M[M\odot] = 9-10, 10-11$ and $11-12$, see Sect. \ref{sec:mass} for the stellar mass estimate)   and  three redshift intervals ($z=0.5-0.8, 0.8-1.5$ and $1.5-2.5$) which are narrow enough to keep the K-correction in the FIR at reasonable values within any interval. 
As explained in Sect. \ref{sec:l60},  up to $z\sim 2.5$ PACS is able to probe a rest-frame window which traces the SFR in both active and inactive galaxies.

A key aim of this work is to compare AGN hosts with inactive galaxies of similar mass. For that purpose, we selected 
a reference sample of galaxies in the same PACS fields and in the same redshift and mass intervals as the AGN hosts' sample. 
We excluded X-ray selected AGNs and Galactic stars, as well as all sources with 2-10 keV X-ray luminosity larger than $10^{42}$ \ergs, even if they were not explicitly classified as AGNs. 
Although this degree of X-ray emission mostly traces AGN activity, it may also arise from intense star formation at the level of a few $10^2 M_{\odot}/yr$ \citep{ranalli03,alexander05,laird05,laird10}. Removal of these sources may therefore bias the control sample towards a lower average SFR by removing the most strongly SF systems. Such a bias would at the same time push the average SFR high for the AGN sample, if these intense starbursts were mistaken for AGNs.  We tested the importance of strong starbursts among the low-$L_X$ X-ray sources in two ways. Since such strongly star forming galaxies are certain to be detected with PACS and lie in the exponential tail of the population of SF galaxies at any given redshift, we first searched for strong variations in the PACS detection rates with $L_X$ among low luminosity X-ray sources in GOODS-S. None were found. We also repeated the entire analysis of this paper after including X-ray sources with no AGN classification in the control sample instead. Our results were unchanged. We conclude that including or excluding these very strong starbursts has a negligible effect on our reference catalogue and does not affect our results to any degree.

A sizeable fraction of very MIR bright galaxies, usually selected from their 24 \mics excess, are likely to be highly obscured AGNs, which are too dust enshrouded to be detected by X-ray surveys \citep[e.g.][]{daddi07b,georgantopoulos08,fiore08,fiore09}. 
Removing these objects as AGN candidates would bias our FIR luminosity estimate, since some of these galaxies are also actively forming stars \citep[e.g.][]{lutz05,sajina08,dey08,donley10, georgakakis10}. However, no more than 0.2\% of the sources in our inactive sample satisfy the selection criterion suggested by \cite{fiore08}.  
We kept these sources in our control sample, and we checked that the results presented below remain unchanged when they are removed. 

For consistency with the COSMOS sample, which is restricted to $I$-detected sources in order to ensure good photometric quality (Sect. \ref{sec:ancil}), we applied the same optical selection to all the samples considered, namely all AGNs and all non-AGNs in all the fields. 
The $I$-band filters used in GOODS and COSMOS are slightly different, with the latter having 20\% more power on the red side. However, given the good spectral sampling in the optical range, we do not expect significant bias effects from this small difference.

The $I$-band selection may in principle introduce a bias in the analysis, because, for $z \gtrsim 1$ galaxies, the observed $I$-band moves to UV rest-frame wavelengths. Dust obscured star forming and quiescent galaxies might therefore not  enter the $I$-band cut and be missed in the analysis, and it is not clear whether and at what level this affects our comparison. However, we checked that, given the depth of the GOODS fields and the correlation between the $I$-band emission and the stellar mass, the samples extracted from these fields are insignificantly affected by the $I$-band cut in the mass range studied. The same does not hold for the shallower COSMOS field. However, the similarities observed between GOODS fields and COSMOS (see Sect. \ref{sec:dichotomy}) make us confident that our analysis is unbiased by this effect.

As we show in the next section, AGNs are typically hosted by high mass galaxies \citep[e.g.][]{bundy08,alonsoherrero08,brusa09,silverman09b,xue10,mainieri11}. The mass bins used in this analysis are large enough that a difference in the mass distributions of AGNs and control galaxies, even within a given bin, could add some bias to our estimates of mean FIR properties. 
We therefore restricted ourselves to a mass-matched reference sample of control galaxies, to take out any possible covariance between SFR and stellar mass due to differing mass distributions between active and inactive galaxies. 
We matched each AGN with 6 non-AGNs with a mass within $\pm 0.2$ dex of the AGN host mass. 
Since AGNs typically reside in massive galaxies, we might run out of comparison galaxies at the high mass end, especially at high redshift. Under this circumstance, we allowed control galaxies to be picked more than once and associated to these sources the appropriate weight. This way, we only lose one AGN in GOODS-S (due to the lack of inactive galaxies of similar mass). 
The mass tolerance and the number of matches were chosen to maximize the final control sample and at the same time not have too many repeated galaxies among the control sample ($4\%$ in the worst case, typically $<2\%$).

We also checked whether the mass-matched samples are affected by any redshift effect within each redshift bin. 
We considered the difference between the average values of the redshift distributions of AGNs and mass-matched non-AGN in the $10^9-10^{12} M_\odot$ mass interval and in each redshift bin. 
The difference $\Delta \langle z\rangle = \langle z\rangle_{nonAGN} - \langle z\rangle_{AGN}$ is of the order of $\sim 10^{-3}-10^{-2}$ and  consistent with zero (errors are computed by assuming normal distributions) in most of the bins. 
The only redshift intervals where $\Delta \langle z\rangle$ is  significant are the high-$z$ bin in COSMOS and the low-$z$ and high-$z$ bins in GOODS-N. The latter bin shows the largest disagreement between non-AGN and AGN redshift distributions ($0.18 \pm 0.04$). 
However, as we will show in Sect. \ref{sec:res}, this mild redshift inconsistency, with AGNs lying at an average redshift lower than non-AGNs, even improves the robustness of our results.

To summarize, both AGN and mass-matched non-AGN samples are characterized by: 
\begin{itemize}
 \item $I$-band detection;
 \item good PACS coverage;
 \item 0.5 $\leq z <$ 2.5;
 \item $10^9 \leq M[M_\odot]< 10^{12}$.
\end{itemize}
Within these requirements, the two samples are selected according to the following criteria:
\begin{itemize}
 \renewcommand{\labelitemi}{$\bullet$}
 \item AGNs:
 \begin{itemize}
  \renewcommand{\labelitemi}{$-$}
  \item detected in X-ray;
  \item classified as AGNs from optical spectroscopy; 
  \item additionally includes unclassified X-ray sources if $L_X > 10^{42}$ \ergs, considering them as Type 2 in GOODS and using the classification from the  photometric redshift fitting with hybrid templates in COSMOS; 
 \end{itemize}
 \item non-AGNs:
 \begin{itemize}
  \renewcommand{\labelitemi}{$-$}
  \item excluded all AGN-classified sources;
  \item excluded Galactic stars. 
 \end{itemize} 
\end{itemize}

After applying all the selections above, 
we ended up with 136, 159 and 1052 AGNs in GOODS-S, GOODS-N and COSMOS respectively in the stellar mass range of $10^9 - 10^{12} M_\odot$. 
Of these, 19, 11 and 439 are Type 1 and 117, 148 and 613 are Type 2. 
The non-AGN reference samples are, by definition, $\sim 6$ times larger than the active ones.   
The fraction of spectroscopic AGNs  is 74\%, 31\% and 57\%, and the fraction of spectroscopic control inactive galaxies is 47\%, 43\% and 5\%, respectively in the three fields.  
By performing, as a control check, the analysis described below on purely spectroscopic sources, we verified that the difference in the spectroscopic fractions among the various fields and between AGN and non-AGN samples  does not bias our results.

The statistics in the different redshift and mass bins for all three fields is summarized in Tab. \ref{tab:sample}. 

\begin{table*}
\centering
\caption{Statistics of AGNs and inactive sample.
}
\begin{tabular} {c|ccc|ccc}
\hline \hline \noalign{\smallskip} 
\multicolumn{7}{c}{GOODS-South + GOODS-North}\\
\noalign{\smallskip} \hline \noalign{\smallskip} 
& \multicolumn{3}{|c|}{AGN hosts}& \multicolumn{3}{|c}{inactive galaxies}\\
\noalign{\smallskip}  \hline \noalign{\smallskip} 
 & $10^9-10^{10} M_\odot$ & $10^{10}-10^{11} M_\odot$ & $10^{11}-10^{12} M_\odot$ &  $10^9-10^{10} M_\odot$ & $10^{10}-10^{11} M_\odot$ & $10^{11}-10^{12} M_\odot$ \\
\noalign{\smallskip} \hline \noalign{\smallskip}
$z=0.5 - 0.8$ & 9 (33.3\%)  & 34 (20.6\%) & 14 (35.7\%) & 49 (4.1\%)  & 228 (18.9\%) & 59 (15.3\%)  \\
$z=0.8 - 1.5$ & 19 (15.8\%) & 70 (18.6\%) & 49 (30.6\%) & 114 (0.0\%) & 458 (11.6\%) & 230 (12.6\%)\\
$z=1.5 - 2.5$ & 5 (0.0\%)   & 59 (6.8\%) &  36 (33.3\%) & 34 (0.0\%)  & 377 (4.2\%) & 168 (17.3\%)  \\
\noalign{\smallskip}  \hline \hline  \noalign{\smallskip} 
\multicolumn{7}{c}{COSMOS}\\
\noalign{\smallskip}  \hline \noalign{\smallskip} 
& \multicolumn{3}{|c|}{AGN hosts}& \multicolumn{3}{|c}{inactive galaxies}\\
\noalign{\smallskip}  \hline \noalign{\smallskip} 
 & $10^9-10^{10} M_\odot$ & $10^{10}-10^{11} M_\odot$ & $10^{11}-10^{12} M_\odot$ &  $10^9-10^{10} M_\odot$ & $10^{10}-10^{11} M_\odot$ & $10^{11}-10^{12} M_\odot$ \\
\noalign{\smallskip} \hline \noalign{\smallskip}
$z=0.5 - 0.8$ & 19 (0.0\%) & 75 (8.0\%)  & 89 (14.6\%)   & 114 (0.0\%)  & 474 (4.0\%)  & 507 (5.7\%)  \\
$z=0.8 - 1.5$ & 15 (6.7\%) & 249 (4.0\%) & 245 (13.9\%)  & 117 (0.0\%) & 1689 (1.2\%) & 1071 (4.4\%)  \\
$z=1.5 - 2.5$ & 3 (33.3\%) & 170 (4.1\%) &  187 (9.1\%) & 24 (0.0\%)  & 1126 (0.3\%)  & 592 (1.9\%)  \\
\noalign{\smallskip} \hline \noalign{\smallskip}
\end{tabular}
\tablefoot{Number of AGNs and inactive galaxies in each redshift (rows) and mass (columns) bin in the different fields. Numbers in brackets show the fraction of sources fully detected by PACS (i.e. detected in the two bands used to infer \lums as explained in Sect. \ref{sec:l60}).
}\label{tab:sample}
\end{table*}

\section{Method} \label{sec:method}

In this section we describe how we derived the fundamental ingredients of this analysis, namely stellar masses and far-infrared luminosities. 
Results presented in Sect. \ref{sec:mass} apply to the entire sample of AGNs and non-AGNs in the redshift range 0.5 -- 2.5. The selection based on PACS coverage (Sect. \ref{sec:selections}) was neglected since it does not affect stellar mass estimates. 

\subsection{Stellar masses} \label{sec:mass}

Stellar masses are most robustly estimated by fitting observed photometry with a library of stellar synthetic templates \citep[e.g.][]{fontana06}. For our set of inactive galaxies, masses were derived using the same procedure as in \cite{santini09}, performing a $\chi^2$ minimization of \cite{bc03} synthetic models, assuming a Salpeter IMF and parameterizing the star formation histories (SFH) as exponentially declining laws. 
Each band is weighted with the inverse of the photometric uncertainty. 
1$\sigma$ errors determined by photometric uncertainties were computed by accounting for all the solutions within $\chi^2_{reduced\ min}+1$. 
Since \cite{bc03} models do not include emission from dust reprocessing, we fitted the observed fluxes up to 5.5 \mics rest-frame, fixing the redshift to the photometric or spectroscopic one, where available.

\begin{figure}[!t]
 \resizebox{\hsize}{!}{\includegraphics{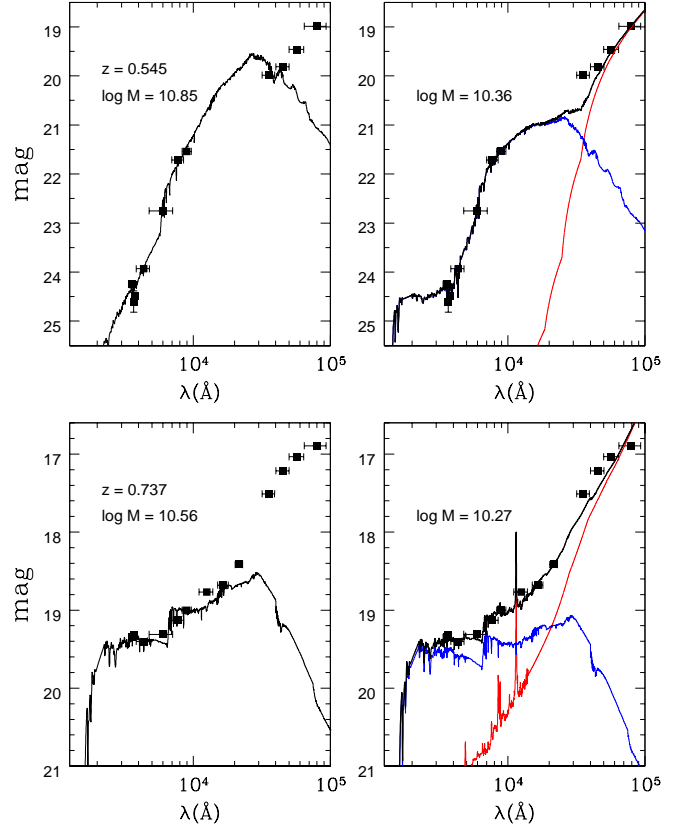}}
 \caption{Observed and best-fit SEDs of a Sy1 galaxy ($bottom$) and a Compton Thin Sy2 ($top$), both lying in the GOODS-S field.  
Left panels show the fit with pure stellar templates, while the right ones show the decomposition between the stellar (blue) and nuclear (red) components. In the latter, the black lines represent the total emission (stars + AGN). The inferred stellar masses are reported in each panel.
 }
 \label{fig:sed}
 \end{figure}

Of course, this procedure is only reliable under the assumption that the bulk of the light emitted by the galaxy comes from its stellar component. This assumption is no longer trustworthy in the case of AGN host galaxies. In Type 2 AGNs, near- and mid-infrared bands are boosted by the dust torus emission, typically biasing the mass estimate towards higher values 
(although at some level  star formation can also affect the 3.6 \mics band, \citealt{joseph84}, \citealt{glass85}).   
The situation is even more critical for Type 1 AGNs, whose Spectral Energy Distribution (SED) is often completely dominated by the nuclear emission at even shorter wavelengths. 
To account for the nuclear emission and to achieve a reliable estimate of the stellar mass in AGN host galaxies, we created a special template library by combining the stellar emission from \cite{bc03} model with the nuclear emission described by the \cite{silva04} templates. 
Once the total observed SED is decomposed into a stellar and a nuclear component, the stellar mass, as well as the other stellar parameters, are inferred from the stellar component only.

The \cite{silva04} templates consist of four (one unobscured, three affected by different levels of obscuration) nuclear SEDs of Seyfert galaxies, for which the stellar contribution was removed, normalized to the hard X-ray (2-10 keV) intrinsic luminosity and averaged within bins of absorbing $N_H$. We averaged the SEDs with $10^{22} < N_H < 10^{23}$ cm$^{-2}$ and $10^{23} < N_H < 10^{24}$ cm$^{-2}$ to yield one Sy1 and two Sy2 (Sy2$_{CThin}$ and Sy2$_{CThick}$ with respectively  $N_H$ $<$ and  $>10^{24}$ cm$^{-2}$) templates. Before combining each of these SEDs with the \cite{bc03} library,  to mimic real galaxy emission we reddened the Sy1 template using a \cite{calzetti00} law with E(B-V) values between 0 and 0.3 \citep{hopkinsp04,merloni10} and step of 0.1. 
With the aim of reducing the degeneracies given by the large number of free parameters (stellar parameters, type of nuclear template, normalization of the nuclear contribution vs the stellar one, reddening), we created three different libraries (one for each AGN template) and used prior knowledge of X-ray properties (classification and $N_H$) to fit each AGN host galaxy with the appropriate two-components library: 
objects classified as Type 1 were fitted with the stellar+Sy1 library, 
Type 2 AGNs with $N_H<10^{24}cm^{-2}$ were fitted with the stellar+Sy2$_{CThin}$ library while for heavily obscured AGNs ($N_H>10^{24}cm^{-2}$) we adopted the stellar+Sy2$_{CThick}$ library.

We used redshifts from the X-ray catalogues, which provide updated/new spectroscopy  as well as photometric redshifts obtained using specifically suited AGN templates (see references in Sect. \ref{sec:xray}).

 In order to make the fitting procedure computationally feasible, we resorted to a library for the stellar component with a reduced number of parameters. We kept the same range of stellar parameters as in \cite{santini09} \citep[see also Table 1 in][]{fontana04}, but adopted a coarser grid, and  we assumed a \cite{calzetti00} extinction law. 
We checked that the use of this coarse grid did not introduce significant differences with respect to the finer grid used in previous works on the GOODS-S field. 
The distribution of $(M_{coarse~grid} - M_{fine~grid})/M_{coarse~grid}$  for GOODS-S non-AGNs  in the redshift range used in this work is sharply peaked around 0, with a very low median/mean value of $\sim 7 \cdot 10^{-3} /  1.2 \cdot 10^{-2}$ and a semi-interquartile range/standard deviation of $\sim 0.09 / 0.28$; % $\sigma  < 0.3$; 
the ratio of the two stellar mass estimates is higher than a factor of 2 in 3.6\% of objects. 
For consistency with the AGN sample we apply the same coarse parameter grid to mass estimates for the non-AGN reference sample as well.

 We are aware that the exponentially declining models (i.e. $\tau$-models) adopted for building the stellar component are likely an oversimplified description for SFH \citep[e.g.,][]{maraston10}. However, \cite{lee10} showed that the resulting stellar masses can still be considered robust because of a combination of effects in the estimate of the galaxy star formation rates and ages. Moreover, they are widely used in the literature and allow an easy comparison with previous works. Nonetheless, their application to active galaxies can be even more problematic, since we expect that, at least in a fraction of them, a burst of star formation is triggered at some point during their evolution, making the declining tail of the SFH an even less accurate description.  We therefore repeated the mass estimate by using exponentially increasing star formation histories (i.e. the so-called inverted $\tau$-models, \citealt{maraston10}), which represent the opposite approximation and may resemble better the SFH in AGN hosts galaxies. We found very low differences in the best-fit stellar masses in GOODS-S: the median/mean value of $(M_{\tau} - M_{inverted~\tau})/M_{\tau}$ is equal to 0.07/0.03 with a semi-interquartile range of 0.14 and a standard deviation of 0.31 (these numbers refer to the $0.5-2.5$ redshift range).  
 The effect of the SFH is slightly larger in COSMOS, where many more bright AGNs are detected: the median/mean value of $(M_{\tau} - M_{inverted~\tau})/M_{\tau}$ is equal to 0.11/0.11 with a semi-interquartile range of 0.18 and a standard deviation of 0.35 (again for the $0.5-2.5$ redshift range). However, we can consider the star formation history parameterization to have an effect on the stellar mass estimate which is lower than its intrinsic  $1\sigma$ uncertainty in both fields (see below).  Moreover, $\tau$-models are preferred over inverted $\tau$-models from an analysis of the $\chi^2$, in agreement with the results of \cite{rosario11}.

The adoption of the widely used \cite{bc03} library instead of its more updated version including TP-AGB stars contribution (\citealt{bruzual07}, see also \citealt{maraston05}) allows an easier comparison with previous works. The improved TP-AGB stars treatment predicts a larger rest-frame $K$-band flux, and hence slightly lower stellar mass estimates. By comparing the stellar masses obtained with the 2003 and the 2007 libraries, we found that the latter underpredicts stellar masses  by 0.12 dex on average, although the scatter between the two estimates is as large as 0.17 dex and there is no clear trend with redshift or stellar mass itself \citep{santini12}. However, AGNs and inactive galaxies should be affected by the same possible systematics, and since we aim at performing an unbiased comparison between the two populations, the analysis is internally consistent.

 \begin{figure}[!t]
 \resizebox{\hsize}{!}{\includegraphics[angle=270]{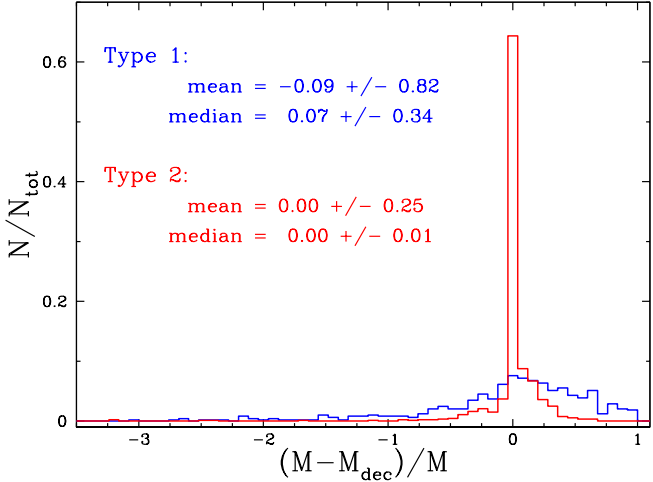}}
 \caption{Normalized distributions of the relative difference between the stellar mass estimates when using pure stellar templates ($M$) and the decomposition method ($M_{dec}$). Blue and red histograms represent the total sample of Type 1 and Type 2, respectively. 
 We report the mean/median values and standard deviation/semi-interquartile range of each sample. 
 }
 \label{fig:mrap}
 \end{figure}

Figure \ref{fig:sed} shows an example of the SED decomposition method described above. It presents a Sy1 galaxy (bottom) and a Compton Thin Sy2 (top). Left panels show the fit with pure stellar templates, while the right ones show the decomposition between the stellar (blue) and nuclear (red) components (the black curves showing the total best-fit SED). From these plots it is evident that in these objects pure stellar templates can not properly reproduce  
the emission at long wavelength (and even at short wavelengths  in the most powerful AGN hosts).

The typical Type 1 SED is usually highly dominated by the nuclear component, and sometimes almost completely outshone by it. Disentangling the stellar and the nuclear components in Type 1 AGNs is therefore very complex, and the derived stellar masses are affected by large uncertainties. 
The average $1\sigma$ relative uncertainty ($\pm$ standard deviation) in the stellar mass estimate $\Delta M /M$, where $\Delta M = (M_{max} - M_{min})/2$ for inactive galaxies and for Type 2 is $0.16 \pm 0.10 $ and $0.18 \pm 0.11$ respectively in GOODS-S, $0.19 \pm 0.10 $ and $0.22 \pm 0.09$ respectively in GOODS-N,  and  $0.30 \pm 0.13 $ and $0.30 \pm 0.11$ respectively in COSMOS. The average $1\sigma$ relative uncertainty in the stellar mass estimate for Type 1 is significantly larger, being equal to  $0.49 \pm 0.22 $,  $0.52 \pm 0.40 $,  and $0.64 \pm 0.35$, respectively, in GOODS-S, GOODS-N, and COSMOS. (These numbers refer to AGNs and mass-matched non-AGNs in the $0.5-2.5$ redshift range). 
Given  these caveats, we checked our results against possible biases introduced by uncertain estimates of the stellar mass in Type 1 AGNs. By repeating the analysis described in Section 4, but only on Type 2 AGNs, we verified that the principal conclusions of this 
work are unaffected by these large uncertainties. 
We also compared our results with the sample of Type 1 AGNs from \cite{merloni10}, and we found good agreement when the same $\chi^2$ minimization method is used, despite the different AGN template used (see also Bongiorno et al. in prep.).

In order to characterize the improvements in the stellar mass estimates achieved through the decomposition technique, we plotted in Fig. \ref{fig:mrap} the distribution of the relative difference $(M-M_{dec})/M$, where $M$ is the stellar mass computed from pure stellar templates and $M_{dec}$ is the output of the decomposition. 
The red histogram refers to Type 2 AGN hosts, the blue one to Type 1. 
When only Type 2  hosts are considered, the mean and median values for $(M-M_{dec})/M$ are consistent with zero, and only in $1.3\%$ of objects the difference in the stellar masses is larger than a factor of 2. 
However, the shift becomes more severe for Type 1 AGNs, modulo the  uncertainties  discussed above. 
Although the  distribution goes in the direction of recovering lower stellar masses through decomposition (the median assumes a positive value, although still consistent with zero), the opposite is also found: in some cases the pure stellar fits might favour bluer, younger and lower mass objects in the attempt of reproducing the total (stellar + nuclear) light. 
The distribution of $(M-M_{dec})/M$ in Type 1 AGN hosts is broad, and 
the difference in the stellar masses is larger than a factor of 2 in $\sim 29\%$ of the objects: the decomposition procedure seems to be essential to recover a reliable stellar mass in these sources.

 \begin{figure*}[!t]
 \resizebox{\hsize}{!}{\includegraphics[angle=270]{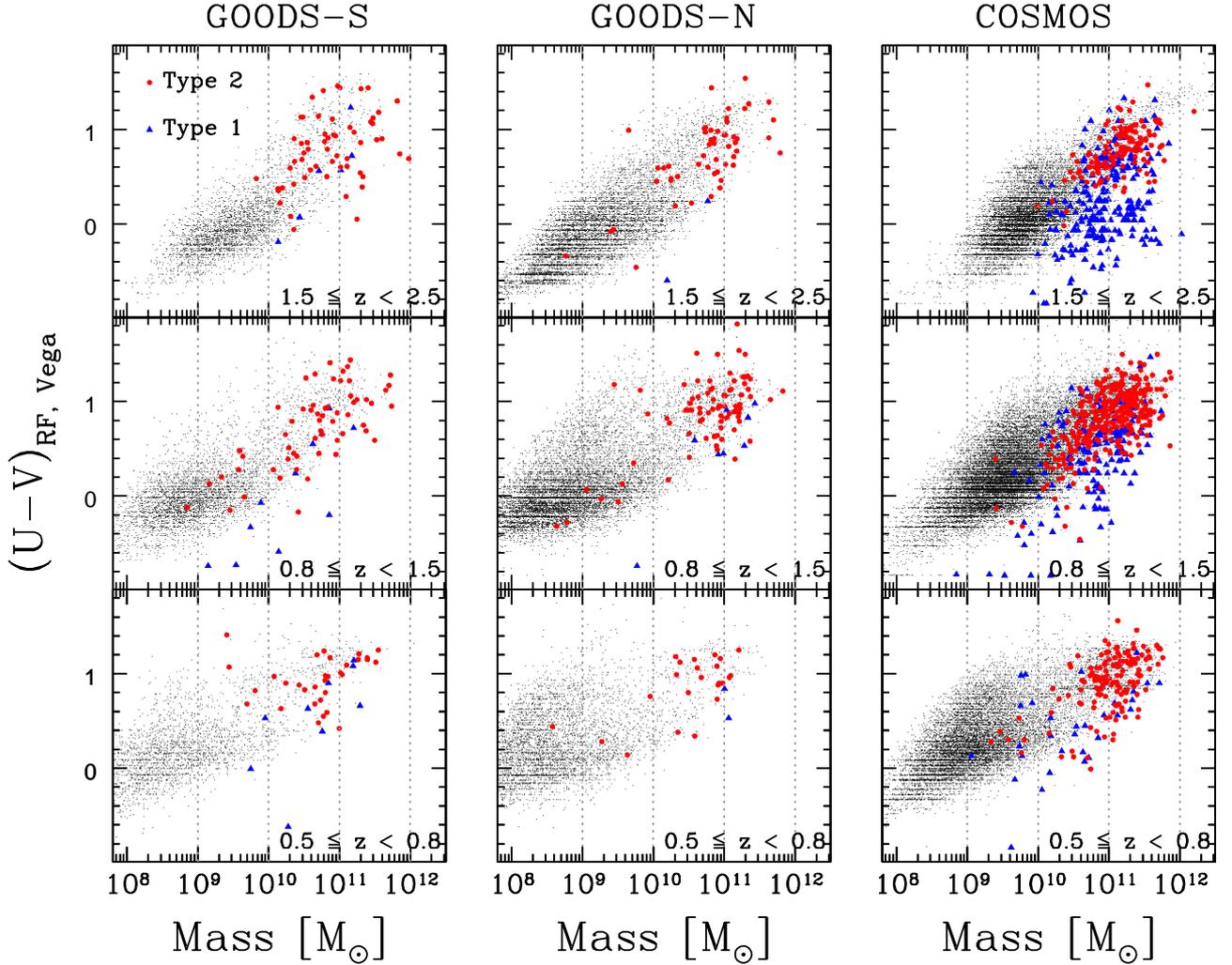}}
 \caption{$(U-V)_{rest-frame}$ versus stellar mass in GOODS-S, GOODS-N and COSMOS (from left to right) in different redshift intervals (increasing from bottom to top). 
Red circles and blue triangles  show Type 2 and Type 1 AGN hosts, respectively, and the y-axis represents host-only colours. 
Black small dots identify the inactive population. Vertical dashed lines indicate the mass bins used in this analysis.
 }
 \label{fig:colmass}
 \end{figure*}

We also explored different ways to estimate the stellar mass in AGN hosts. The first one consists in fitting observed fluxes with pure stellar templates only up to 2 \mics rest-frame ($M_{2\mu m}$), ignoring the near- and mid-infrared bands typically boosted by the AGN emission. As a second method  ($M_{AGNsubtraction}$), 
we used the X-ray 2-10 keV luminosity and the obscuring column $N_H$ to estimate the nuclear contribution to the total light in each observed band according to the \cite{silva04} templates, we subtracted it from the observed photometry, and fitted the residual fluxes with the pure stellar  library. 
Stellar masses inferred with these approaches agree satisfactorily with those obtained through the decomposition method when only Type 2 AGNs are taken into account.  
The median/mean value of $(M_{dec}-M_{2\mu m})/M_{dec}$ equals to -0.004/-0.046, with a dispersion (i.e. semi-interquartile range) of 0.045, while the median/mean of $(M_{dec}-M_{AGNsubtraction})/M_{dec}$ is equal to 0.040/0.023, with a dispersion of 0.190 (these numbers refer to the $z=0.5-2.5$ redshift range in GOODS-S). 
However, both these approaches ($M_{2\mu m}$ and $M_{AGNsubtraction}$) fail in reproducing the Type 1 SEDs. Indeed, nuclear emission in Sy1 galaxies does affect all bands, not only the near- and mid-infrared ones, and it often outshines the stellar light, making the subtraction method unsuccessful. The decomposition technique that we adopt is instead the only way to properly reproduce Type 1 SEDs (see Fig. \ref{fig:sed}).

Finally, we applied the decomposition technique to non-AGN galaxies to check that this procedure is not introducing any systematics in the stellar mass estimates. The adoption of the stellar+Sy2$_{CThin(CThick)}$ library on non-AGNs provides basically the same stellar masses, the ratio of those estimated with and without decomposition being $0.99 \pm 0.10$ $(1.00 \pm 0.08)$. A larger scatter is obtained when using the stellar+Sy1 library, the previous ratio being equal to $0.98 \pm 0.32$, with only a very mild tendency for stellar masses to be smaller when the decomposition is applied. However, a 2\% error in the stellar mass estimate is perfectly consistent with its uncertainty, and represents a further confirmation of the reliability of our mass estimate procedure. However, in order not to increase the uncertainties, we do not include the nuclear component when fitting non-AGN galaxies.

\subsubsection{Colour--mass diagram} \label{sec:cmd}

In Fig. \ref{fig:colmass} we show the colour--stellar mass relation for AGN hosts 
(large coloured symbols) and inactive galaxies (small black dots) in the three fields, divided into redshift bins.
The figure plots the rest-frame $(U-V)$ colour of the best-fit stellar template for each object.

As far as the control sample of non-AGNs is considered, thanks to the restriction to good photometric quality, the colour--mass diagrams in the three fields are well consistent. 

Figure \ref{fig:colmass} shows that AGNs are preferentially hosted in massive galaxies, as expected \citep[e.g.][]{bundy08,alonsoherrero08,brusa09,silverman09b,xue10,mainieri11}, although they can have a tail towards lower masses. 
We carefully inspected fits to AGNs with suspiciously low stellar mass hosts. We found 8 Type 1 AGNs in COSMOS with $M<10^{10}M_\odot$ and no 8 \mics photometry. In powerful AGNs, MIR photometry is  crucial to constrain the nuclear component and recover a reliable stellar mass estimate. We therefore removed these sources from our sample 
(they are not plotted in Fig. \ref{fig:colmass} and not accounted when computing the statistics given in the previous section). 

The host-only colours of Type 1 AGNs appear very blue compared to the host stellar mass in about half of the population. 
Blue colours with respect to the underlying galaxy population in the host galaxies of bright AGNs were first reported by \cite{kauffmann03} and \cite{jahnke04}, and more recently by \cite{cardamone10}. 
They may hint to some physical properties, e.g. the existence of young stellar populations with low reddening. 
However, we also note that large uncertainties are  involved in the SED decomposition of these sources and it is difficult to interpret the best-fit stellar colours with any confidence. The key aspect is that the stellar mass distributions of Type 1 and 2 are quite similar and, therefore, the inclusion of Type 1 in our final AGN sample does not bias our subsequent results.

\subsection{Far-infrared luminosities and star formation rates} \label{sec:l60}

As a direct tracer of the far-infrared emission we concentrate on the mean luminosity \lums\ at a rest-frame wavelength of 60 \mic. This choice is a compromise between a wavelength long enough to avoid most of the AGN contamination (see discussion below) and at the same time short enough to be sampled by PACS 160 \mics observations even at the highest redshifts considered in this work. 

To preclude any assumptions about the SED shape, we computed \lums though a log-linear interpolation of PACS fluxes, after converting them to luminosities following the approach of \cite{shao10}. For the GOODS-S dataset, which includes 70 \mics observations as well, we interpolated between the two PACS bands bracketing rest-frame 60 \mic. 
Although the 70 \mics band is only  used in low-$z$ objects ($z\lesssim 0.67$), 
we checked that the addition of these observations in GOODS-S does not introduce any bias in the analysis: when interpolating rest-frame 60 \mics luminosity only from 100 and 160 \mics PACS photometry, our results remain unchanged, and  \lums in each mass and redshift bin do not differ by more than 50\%. 

We also explored fitting PACS fluxes with a typical IR template (e.g. from \citealt{ce01} or \citealt{dh02} libraries) to derive \lum. 
We found differences by a few tens per cent (only in 15\% of the bins is the difference larger than $50\%$), 
but the global picture is unaffected. Therefore, we adopt the simple interpolated \lums estimate throughout, making no assumptions on the detailed SED shape. 

For sources fully detected (i.e. detected in both PACS bands used to interpolate 60 \mics luminosity) we used the corresponding fluxes and computed individual $\nu L^i_\nu (60\mu $m$)$. 
For PACS undetected sources we computed average  fluxes by stacking in a given mass and redshift interval. We stacked at the X-ray positions on PACS residual maps using the \cite{bethermin10} libraries\footnote{http://www.ias.u-psud.fr/irgalaxies/downloads.php}. The use of residual maps, from which all detected objects were removed, avoids contamination by nearby brighter sources. PSF photometry was performed on the final stacked images. Uncertainties in the stacked fluxes were computed by means of a bootstrap procedure.  
For each PACS band $j$, we then averaged, by weighting with the number of sources, stacked fluxes with individual fluxes of partially detected objects (i.e. detected only in band $j$ but not in the other one). 
These stacked and averaged fluxes in each band are used to get $\nu L^{STACK}_\nu (60\mu $m$)$, in the same fashion as for the fully detected sources.

The final 60 \mics luminosity in each mass and redshift interval is computed by averaging over the linear luminosities of detections and non-detections, weighted by the number of sources. Only bins with more than 3 sources are used in our analysis.

Errors in the infrared luminosity were obtained by bootstrapping. A set of N sources, where N is equal to the number of sources per mass and redshift bin, is randomly chosen 100 times among detections and non-detections (allowing thus repetitions), and a \lums is computed per each iteration. The standard deviation of the obtained \lums values gives the error in the average 60 \mics luminosity in each bin. The error bars thus account for both measurement errors (uncertainty in the fluxes) and the error in the population distribution (abundances of various galaxy types in each bin). They do not, however,  account for cosmic variance. 
The stellar mass uncertainties are not accounted for either, since they are small compared to the mass bin widths.

This work is based on the assumption that the rest-frame far-infrared emission is dominated by the host galaxy and is therefore a proxy of the SF activity.  
A number of previous studies support our assumption based on different grounds.  
Many authors \citep[e.g.][]{schweitzer06,netzer07,lutz08}  showed a strong correlation between FIR luminosity and SFR tracers, such as PAH emission features, both in local and high redshift bright ($L_{AGN}/L_{FIR}$ up to $\sim 10$) QSOs. \cite{hatziminaoglou10}  used a SED decomposition technique based on the \cite{fritz06} dusty torus model, and showed that a starburst component is always needed to reproduce both Type 1 and Type 2 AGN FIR emission. 
For a detailed discussion see also the introductions of \cite{lutz10} and \cite{serjeant09}. 
As a further validation, we compared PACS colours (F160/F100) of AGNs with those of non-AGNs, and  found no significant difference between the two samples. 
Similar results were also obtained by 
\cite{hatziminaoglou10} using SPIRE (250, 350 and 500 \mic) colours. These checks, as well as the results of the studies cited above, make us confident that the AGN contamination in FIR emission is not dominant over the emission of dust heated by young stars, at least in the majority of sources. 
However,  
\cite{mullaney11} suggest that the total emission at 60 \mics may be dominated by nuclear light in at least three of their 11 AGN dominated galaxies. 
In Rosario et al. (in prep.), we demonstrate that, if adopting the templates of \cite{mullaney11}, the tail of the torus emission can extend to rest-frame 60 \mics in a fraction of the most luminous ($L_x>10^{44} erg/s$) AGNs. 
We will discuss in Sect. \ref{sec:distrib} and \ref{sec:dichotomy} if and how this contamination can affect our conclusions.

\section{Results} \label{sec:res}

 \begin{figure*}[!t]
 \resizebox{\hsize}{!}{\includegraphics[angle=90]{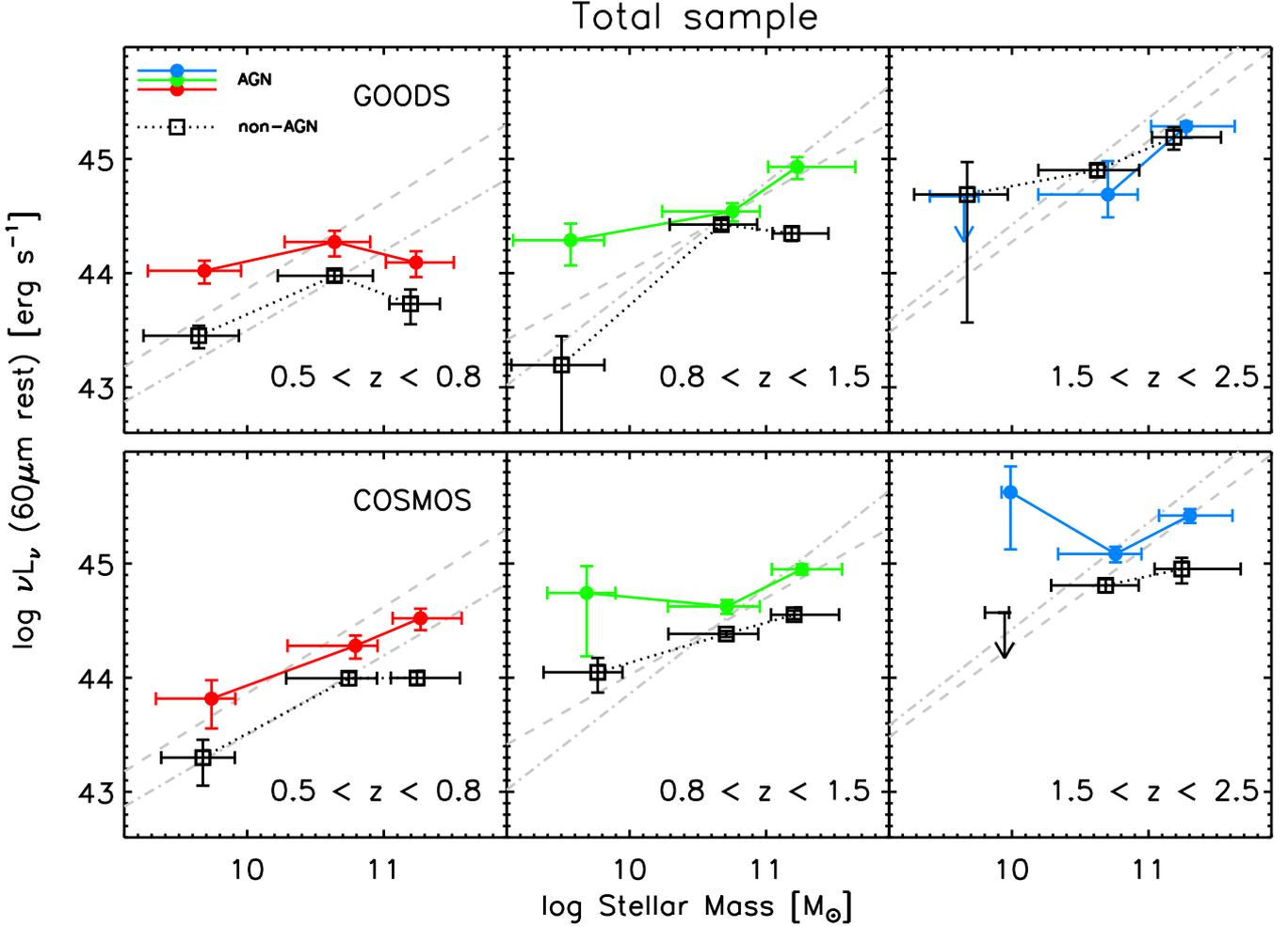}}
 \caption{Average \lum, proportional to the SFR, versus stellar mass in the different redshift bins, increasing from left to right, for the GOODS fields ($upper$ panels) and COSMOS ($lower$ panels). Coloured solid lines/solid circles and black dotted lines/open boxes show  AGN hosts and inactive galaxies, respectively. Stellar mass values reflect the median in each mass and redshift interval, and their error bars show the  80\% confidence intervals. Error bars on \lums were computed by bootstrapping (see text). Grey dashed lines represent the main sequences as inferred by \cite{santini09} in redshift bins similar to the ones used in this work, while dot-dashed ones are the results of \cite{noeske07}, \cite{elbaz07} and \cite{daddi07a} respectively at $z\sim 0.5$, $z\sim 1$ and $z\sim 2$. At all masses and redshifts, with the exception of the highest redshift interval in the GOODS fields, AGN hosts show an enhanced FIR emission with respect to inactive galaxies of similar stellar mass. In contrast, the locus of AGN hosts is broadly consistent with the main sequence of (only) star forming galaxies.
 }
 \label{fig:l60mass}
 \end{figure*}

In this section we address the main topic of this work, i.e. the comparison of average star formation properties of AGN host galaxies with inactive galaxies of similar mass.  
In Fig. \ref{fig:l60mass} we plot average \lums versus stellar mass for AGN hosts (coloured solid lines and circles) and mass-matched non-AGNs  (black dotted lines and open boxes) in the different redshift bins for the GOODS sample (South+North, upper panels) and the COSMOS one (lower panels).
Masses represent the median values in each mass and redshift interval. 
Error bars on the stellar masses show the 80\% confidence intervals.

Given the different depth and sky areas, the GOODS and COSMOS fields can be regarded as somewhat complementary samples (see Fig. \ref{fig:xray}). 
However, given the large difference in the IR, X-ray and optical depth between COSMOS and the GOODS fields, combining the catalogues would introduce biases which can not be easily taken into account. 
For these reasons we decided to keep the COSMOS analysis separate and to combine GOODS-S and GOODS-N, which have similar properties and show similar results if considered separately, to increase the statistics in the deep fields.  
In order not to stack on different PACS residual images with different noise levels,   
\lums was independently computed in each redshift and mass bin for each of the GOODS fields. We then computed the mean GOODS \lums for each bin by averaging the values obtained separately in the two fields, weighting them by the corresponding errors. 
Similarly, the typical mass associated to a given \lums was computed by averaging the individual median masses weighted with the errors in the individual \lum. 
The inter-percentile range of the stellar mass distribution at 80\% was computed on the full combined GOODS sample in each bin as an estimate of mass uncertainty. 

Each field  is affected by incompleteness at low masses in a different way. 
In the mass range studied ($10^9 - 10^{12} M_\odot$), our control sample lacks sources in the GOODS-S field  at the highest redshift bin, which is only complete above $ \sim 2.5\cdot 10^9 M_\odot$, and in COSMOS, where the limit is set around $\sim 4\cdot 10^9 M_\odot$ at $0.8<z<1.5$ and $\sim 5.5\cdot 10^9 M_\odot$ at $1.5<z<2.5$. 
However, mass incompleteness is not a serious issue for this kind of work for two reasons. Firstly,  we only deal with mass-matched samples, which, given the AGN mass distribution, only include extremely few sources (only 6 AGNs in COSMOS) below the mass incompleteness limits. 
Secondly, the aim of this study is not measuring the absolute SF  activity, but rather comparing SF properties between X-ray selected AGN hosts and  inactive galaxies of similar mass, and mass incompleteness can be ignored as long as it affects the two samples in a similar way: since in AGN hosts the rest-frame $K$-band, which is broadly proportional to the stellar mass, is generally dominated by stellar emission, there is no reason to believe that this is not the case.

Using the 60 \mics rest-frame luminosity as a proxy of the star formation rate, 
a \lum--stellar mass diagram is  analogous to one plotting SFR--stellar mass. 
From Fig. \ref{fig:l60mass} it is possible to observe the global increase of the far-infrared luminosity, and hence SFR, with redshift as well as with stellar mass, in both active and inactive galaxies \citep[e.g.][and references therein]{santini09,damen09,rodighiero10}. 
We  also plot the main sequence relations (grey curves) inferred by previous studies in similar redshift intervals as the ones used in this work: \cite{santini09} at $z=0.6-1.0$, $z=1.0-1.5$ and $z=1.5-2.5$, \cite{noeske07} at $z=0.2-0.7$, \cite{elbaz07} at $z=0.8-1.2$ and \cite{daddi07a} at $z=1.4-2.5$. For this purpose, we first converted the SFR into an estimate of the total infrared luminosity $L_{IR(8-1000 \mu m)}$ by dividing by $1.8\ 10^{-10}$ \citep[following][]{santini09} and then converted the latter into \lums by linearly fitting the values of  $L_{IR(8-1000 \mu m)}$ and \lums predicted by the \cite{ce01} library. 

Our control sample of inactive galaxies shows general agreement with the locus of the main sequence  with the exception of the highest masses. 
The slopes for our  sample indeed appear flatter than  the main sequence. 
This is explained by the fact that the main sequence relations are computed for purely star forming galaxies, while our mass-matched reference sample is made by a combination of star
 forming and quiescent galaxies. The latter, more abundant at high masses and low redshifts, are responsible for the bending of the relation. 

One may also notice that the locus of inactive galaxies (black curves) in both the average GOODS field and the COSMOS field are quite similar, especially
at intermediate masses (between $10^{10}$ -- 10$^{11} M_\odot$) where we have the best statistics. At higher masses, cosmic variance 
can lead to the small differences we observe. At low masses, incompleteness in our samples can play
a role. However, despite these effects, the trends for inactive galaxies between fields are generally consistent, despite their very different PACS depths. This may be taken as evidence for the validity of the procedure we employ to derive \lum.

The locus occupied by AGN hosts is also  roughly consistent with the main sequence, with the exception of the lower mass bins which show a larger SFR. They could be pushed up  by a higher fraction of strongly star forming galaxies at low masses. 
Broad consistency with the main sequence was also reported by the previous studies of \cite{mainieri11}, who measured comparable specific SFRs in obscured quasars and in an IRAC 3.6 \mics selected control sample, and \cite{mullaney12}, who found that the specific SFRs in AGN hosts with $L_X = 10^{42} - 10^{44} erg/s$ are only marginally lower (by 20\%) than those of main sequence galaxies at $0.5<z<3.0$.

The main target of  this study is however the difference between AGN hosts and a mass-matched control sample of inactive galaxies, which, being optically-selected, includes both the star forming and the quiescent populations. 
A comparison sample including both star forming and quiescent galaxies is motivated by, e.g., the wide spread of AGN loci in optical color-magnitude diagrams. 
The measured \lums in each redshift and mass bin and in each field are reported in Tab. \ref{tab:l60}. Both from the latter table and from Fig. \ref{fig:l60mass}, it is clear that at all redshifts and masses AGN hosts  sit above non-AGNs, with the exception of the GOODS high redshift interval, where the two samples show consistent FIR emission. 
The difference in the FIR emission between the two populations is larger 
in COSMOS, and more modest in GOODS. 
We also note that the \lums enhancement at low redshift seems to be stronger than at high redshift. 
This can be partly explained by the slightly larger ($0.1-0.2$) average redshift of the inactive control sample compared to AGNs at $z=1.5-2.5$ (especially in GOODS-N, see Sect. \ref{sec:selections}), and makes the global offset that we observe between the two populations even more robust.  
It is also in qualitative agreement with the  results of \cite{xue10}, who report a larger difference in the SFR between AGN hosts and mass-matched non-AGN galaxies in GOODS at $z\simeq 0-1$ ($\sim 0.3-0.4$ dex) than at $z\simeq 1-3$ ($\sim 0.1-0.2$ dex).  
We report in Tab. \ref{tab:offset} (second and fifth columns) the average 
\lums offsets\footnote{We computed linear averages of the differences between the logarithmic FIR luminosities in the individual bins; uncertainties in the average were computed by summing in quadrature the logarithmic error bars; upper limits were not considered in the average computation.} in each redshift bin and in the total redshift range. 
The FIR emission in AGN hosts is modestly enhanced in the GOODS fields. The average offset 
 is equal to $0.32 \pm 0.03$ dex (the average offsets in the two GOODS fields considered separately is equal to $0.30 \pm 0.03$ dex in GOODS-S and $0.17 \pm 0.03$ dex in GOODS-N\footnote{It is unsurprising that the two offsets considered separately are lower than the average offset. Indeed upper limits, more easily  in place when dealing with only one field and the number of sources is lower, are not considered in the average computation.}). 
In COSMOS the level of enhancement is  larger, with an average offset of $0.42\pm 0.04$ dex. 
We will discuss in Sect. \ref{sec:dichotomy} how these different results found for the GOODS vs COSMOS fields are driven by the different luminosities of the AGN samples (Fig. \ref{fig:xray}).

Enhanced star formation in high $z$ AGN hosts may at first glance be surprising given the  tendency of these hosts to preferentially reside in the so-called green valley in optical colour--magnitude diagrams \citep[e.g.][]{nandra07,coil09,hickox09}. It has been shown, however, that this tendency is strongly influenced by varying mass-to-light ratios and disappears when studying colours of mass-selected samples \citep{silverman09a,xue10} or even reverts to a preference for blue cloud/green valley hosts \citep{aird12}.

\begin{table*}
\centering
\caption{Average log(\lum[\ergs]) values in each redshift (rows) and mass (columns) bin in the different fields. %\lums is in units of [erg/s].
}\label{tab:l60}
\begin{tabular} {c|ccc|ccc}
\hline \hline \noalign{\smallskip} 
\multicolumn{7}{c}{GOODS-South + GOODS-North}\\
\noalign{\smallskip} \hline \noalign{\smallskip} 
& \multicolumn{3}{|c|}{AGN hosts}& \multicolumn{3}{|c}{inactive galaxies}\\
\noalign{\smallskip} \hline \noalign{\smallskip} 
 & $10^9-10^{10} M_\odot$ & $10^{10}-10^{11} M_\odot$ & $10^{11}-10^{12} M_\odot$ &  $10^9-10^{10} M_\odot$ & $10^{10}-10^{11} M_\odot$ & $10^{11}-10^{12} M_\odot$ \\
\noalign{\smallskip} \hline \noalign{\smallskip}
$z= 0.5 - 0.8$ & $44.02^{+0.09}_{-0.11}$ & $44.27^{+0.10}_{-0.13}$  & $44.09^{+0.10}_{-0.13}$& $43.45^{+0.09}_{-0.11}$ & $43.98^{+0.05}_{-0.05}$  & $43.73^{+0.13}_{-0.18}$  \\
$z= 0.8 - 1.5$ & $44.29^{+0.15}_{-0.22}$ & $44.54^{+0.07}_{-0.09}$  & $44.93^{+0.09}_{-0.11}$& $43.19^{+0.25}_{-0.68}$ & $44.43^{+0.04}_{-0.05}$  & $44.34^{+0.06}_{-0.07}$  \\
$z= 1.5 - 2.5$ & $44.67^u$ & $44.69^{+0.14}_{-0.20}$  & $45.29^{+0.08}_{-0.10}$& $44.69^{+0.28}_{-1.12}$ & $44.90^{+0.05}_{-0.06}$  & $45.19^{+0.09}_{-0.11}$  \\
\noalign{\smallskip} \hline \hline \noalign{\smallskip} 
\multicolumn{7}{c}{COSMOS}\\
\noalign{\smallskip} \hline \noalign{\smallskip} 
& \multicolumn{3}{|c|}{AGN hosts}& \multicolumn{3}{|c}{inactive galaxies}\\
\noalign{\smallskip} \hline \noalign{\smallskip} 
 & $10^9-10^{10} M_\odot$ & $10^{10}-10^{11} M_\odot$ & $10^{11}-10^{12} M_\odot$ &  $10^9-10^{10} M_\odot$ & $10^{10}-10^{11} M_\odot$ & $10^{11}-10^{12} M_\odot$ \\
\noalign{\smallskip} \hline \noalign{\smallskip}
$z= 0.5 - 0.8$ & $43.82^{+0.16}_{-0.26}$ & $44.28^{+0.09}_{-0.11}$  & $44.52^{+0.08}_{-0.10}$& $43.30^{+0.16}_{-0.25}$ & $44.00^{+0.05}_{-0.06}$  & $44.00^{+0.05}_{-0.06}$  \\
$z= 0.8 - 1.5$ & $44.74^{+0.24}_{-0.55}$ & $44.62^{+0.06}_{-0.07}$  & $44.95^{+0.04}_{-0.05}$& $44.05^{+0.13}_{-0.18}$ & $44.38^{+0.03}_{-0.03}$  & $ 44.55^{+0.03}_{-0.04}$  \\
$z= 1.5 - 2.5$ & $45.62^{+0.23}_{-0.50}$ & $45.08^{+0.06}_{-0.07}$  & $45.42^{+0.06}_{-0.07}$& $44.57^u$ & $44.81^{+0.06}_{-0.06}$  & $44.95^{+0.10}_{-0.13}$  \\
\noalign{\smallskip} \hline \noalign{\smallskip}
\multicolumn{7}{c}{$^u$: $1 \sigma$ upper limit }\\
\end{tabular}
\end{table*}

\begin{table*}
\centering
\caption{Average FIR emission offsets between AGNs and non-AGNs over all mass bins.  
}\begin{tabular} {c|ccc|ccc}
\hline \hline \noalign{\smallskip} 
\multicolumn{7}{c}{Average offsets}\\
\noalign{\smallskip} \hline \noalign{\smallskip} 
& \multicolumn{3}{|c|}{GOODS}& \multicolumn{3}{|c}{COSMOS}\\
\noalign{\smallskip} \hline \noalign{\smallskip} 
 $z$ & Total sample & Low-$L_X$ & High-$L_X$ &  Total sample  & Low-$L_X$ & High-$L_X$ \\
\noalign{\smallskip} \hline \noalign{\smallskip}
0.5 - 0.8 & $0.41\pm 0.04$  & $0.29\pm 0.06$  & ... & $0.44\pm 0.06$ & $0.21\pm 0.05$ & $0.81\pm 0.07$ \\
0.8 - 1.5 & $0.60\pm 0.06 $ & $0.56\pm 0.06$  & ... & $0.44\pm 0.10$ & $0.50\pm 0.09$ & $0.45\pm 0.04$  \\
 1.5 - 2.5 & $-0.06\pm 0.05$ & $-0.01\pm 0.06$ & ... & $0.37\pm 0.04$ & $0.02\pm 0.07$ & $0.42\pm 0.04$  \\
 0.5 - 2.5 & $0.32\pm 0.03$  & $0.28\pm 0.10$  & ... & $0.42\pm 0.04$ & $0.24\pm 0.04$ & $0.56\pm 0.03$ \\
\noalign{\smallskip} \hline \noalign{\smallskip}
\end{tabular}
\tablefoot{The different rows correspond to different redshift bins, while the columns correspond to the entire sample, the low- and the high-$L_X$ luminosity ones.   Average logarithmic offsets were computed as linear averages of the differences between the logarithmic FIR luminosities. Uncertainties  were computed by summing in quadrature the logarithmic error bars on \lum . Upper limits were not considered in the average computation. Bins with less than three sources were discarded.
}\label{tab:offset}
\end{table*}

\begin{table*}
\centering
\caption{Detailed comparison between AGNs and non-AGNs by separately considering  PACS detected and undetected sources.}.
\begin{tabular} {c|ccc|ccc}
\hline \hline \noalign{\smallskip} 
& \multicolumn{3}{|c|}{GOODS}& \multicolumn{3}{|c}{COSMOS}\\
\noalign{\smallskip} \hline \noalign{\smallskip} 
$z$ & (1) & (2) & (3)  & (1) & (2) & (3) \\
\noalign{\smallskip} \hline \noalign{\smallskip}
\multicolumn{7}{c}{Total sample}\\
\noalign{\smallskip} \hline \noalign{\smallskip} 
0.5 - 0.8 &  77\%  & $0.28 \pm 0.04$ &  $3.87 \pm 1.00$ & 64\%  & $0.43 \pm 0.06$ &  $2.27 \pm 1.00$ \\
0.8 - 1.5 &  1\%  & $0.38 \pm 0.06$ &  $2.02 \pm 0.99$ & 18\% & $0.22 \pm 0.06$ &  $3.28 \pm 1.02$ \\
1.5 - 2.5 &  82\%  & $-0.13 \pm 0.04$ &  $1.76 \pm 1.00$ & 36\% & $0.28 \pm 0.04$ &  $10.17 \pm 1.30$\\
\noalign{\smallskip} \hline \noalign{\smallskip}
\multicolumn{7}{c}{Low-$L_X$ subsample}\\
\noalign{\smallskip} \hline \noalign{\smallskip} 
0.5 - 0.8 &  76\% & $0.22 \pm 0.04$ &  $7.26 \pm 1.29$ &  99\%  & $0.23 \pm 0.05$ &  $1.90 \pm 1.03$ \\
0.8 - 1.5 &  1\% & $0.39 \pm 0.05$ &  $1.71 \pm 0.98$ &  83\%  & $0.17 \pm 0.08$ &  $3.78 \pm 1.06$ \\
1.5 - 2.5 &  59\% & $-0.14 \pm 0.05$ &  $1.44 \pm 1.39$ &  31\%  & $-0.08 \pm 0.08$ & $4.38 \pm 1.41$ \\
\noalign{\smallskip} \hline \noalign{\smallskip} 
\end{tabular}
\tablefoot{Columns: (1) probability, according to a Kolmogorov Smirnov test, that the $\log($\lum$)$ distributions of PACS detected AGNs and non-AGNs are consistent with describing the same population; (2) average offset  between the stacked $\log($\lum$)$ for PACS undetected AGNs and non-AGN; (3) ratio of PACS detection rates of AGNs and non-AGNs. The upper table refers to the low luminosity subsample (upper panel of Fig. \ref{fig:indiv}), whereas the lower table represents the total sample (lower panel of Fig. \ref{fig:indiv}).} 
\label{tab:distrib}
\end{table*}

\subsection{A closer look at SFR distributions} \label{sec:distrib}

\begin{figure*}[!t]
\centering
 \includegraphics[angle=90,width=15cm]{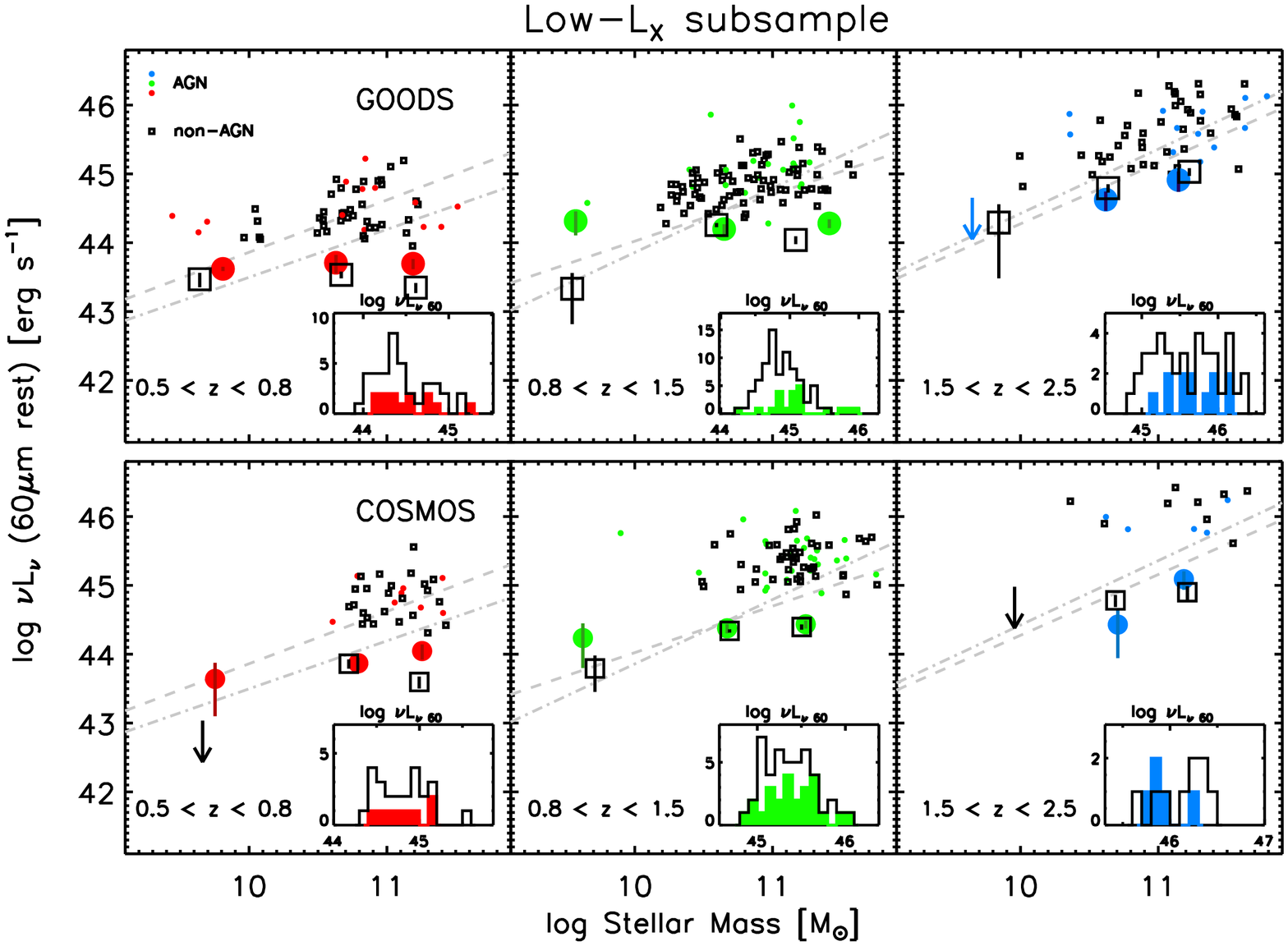}
 \includegraphics[angle=90,width=15cm]{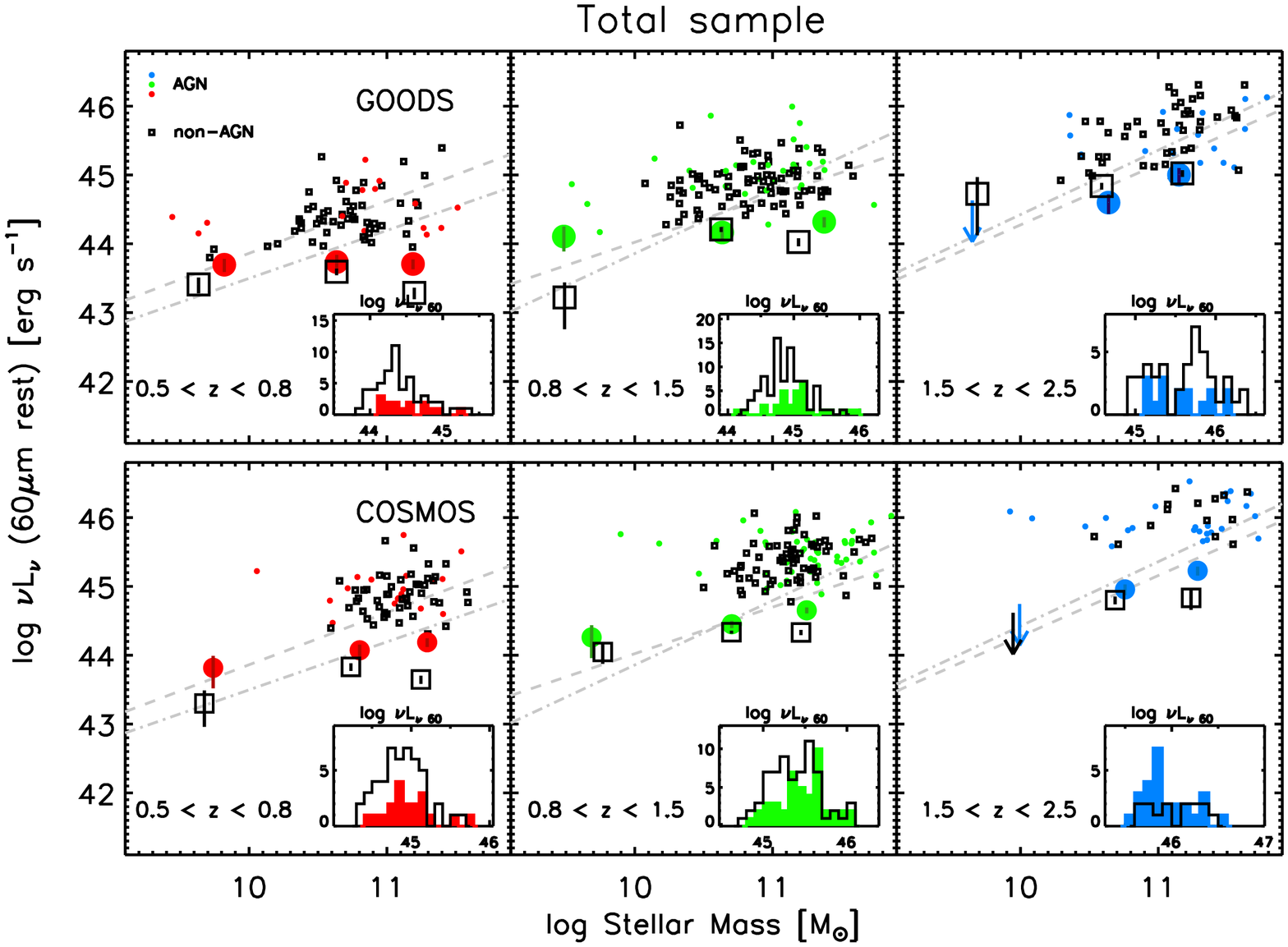}
 \caption{\lums versus stellar mass in the different redshift bins, for the low-$L_X$ subsample ($top$) and for the total one ($bottom$). The GOODS fields are represented in the $upper$ panels and COSMOS  in the $lower$ ones. Coloured solid circles and black open boxes show AGN hosts and inactive galaxies, respectively. Small symbols refer to individual PACS detections, while large ones indicate the stacked values (for sources undetected in at least one PACS band). Error bars on the stacked values are computed by bootstrapping. Stellar masses for stacked sources reflect the median masses of non detected sources in each mass and redshift interval. In the two GOODS fields, stacked luminosities and stellar masses were averaged and weighted with the number of sources. Main sequence relations (grey curves) are as in Fig. \ref{fig:l60mass}. The inset panels show the logarithmic \lums distribution for PACS detected AGN hosts (coloured solid histograms) and inactive galaxies (black plain ones).
 }
 \label{fig:indiv}
 \end{figure*}

So far we have discussed average SF properties of the AGN host  population compared to a control sample of inactive galaxies of similar mass. 
However, averages are insensitive to the detailed distribution of SFRs of the two samples, especially if they are not normally distributed, as is known to be the case for inactive galaxies \citep[e.g.][]{bell04,salimbeni08,santini09,ilbert10} and AGN hosts \citep[e.g.][]{nandra07,silverman08,cardamone10}. 
To gain a handle on the underlying SFR distributions of the two samples, we plot in Fig. \ref{fig:indiv} \lums versus stellar mass for individual PACS detections (small symbols) as well as the stacked values for non detections (large symbols), both for AGNs (coloured solid circles) and for the control sample (black open boxes).

We distinguish between the low-$L_X$ subsample (on the top), where the FIR luminosity can be safely regarded as a proxy of the SFR (see discussion in Sect. \ref{sec:l60}), and the total sample (on the bottom), where nuclear light may affect the \lums estimate in some of the most luminous AGNs.
In order to split the sample by X-ray luminosity, we chose redshift-dependent $L_X$ thresholds, equal to $\log L_X[$\ergs$] = $ 43.3, 43.8 and 44.0  respectively at $z \sim $ 0.65, 1.15 and 2, shown as red lines in Fig. \ref{fig:xray} (for a justification, please see Sect. \ref{sec:dichotomy}). 
As far as the low-$L_X$ subsample is considered, a new set of control galaxies was extracted from the parent inactive sample to match the mass distribution of low luminosity AGN hosts.

First of all, it must be noted that, in the deeper GOODS fields, individual PACS detections are able to probe over a good fraction of the main 
sequence of star forming galaxies, at least at higher masses. In shallower fields like COSMOS only the upper envelope is traced by individual detections, and we have to rely on stacks to investigate the lower envelope (as well as the quiescent population). 

Among PACS detected sources, no significant offset in FIR luminosity is observed between AGN hosts and inactive galaxies. 
This can be seen from the inset histograms in Fig. \ref{fig:indiv}, although the low number of detected galaxies in some bins does not allow a proper comparison. 
According to a Kolmogorov Smirnov test performed on both the total sample and the low-$L_X$ subsample, the  $\log($\lum$)$ distributions for PACS detected AGNs and non-AGNs are consistent  
with describing the same population of galaxies in most of the bins (the probability is reported in column (1) of Tab. \ref{tab:distrib}). In the GOODS intermediate redshift interval, these probabilities are biased by  three or four outliers: when removing such sources, the probabilities that the two distributions describe the same populations turn out to be 10\% and 3\%, respectively for the low-$L_X$ subsample and the total sample. 
In the low-$L_X$ subsample, unbiased by possible AGN contamination to the FIR, the distribution of PACS detected AGN hosts and non-AGNs are therefore consistent at all redshifts and in both fields.   
The slightly lower probabilities measured in the COSMOS total sample % small difference 
can be attributed to either an intrinsically larger SFR in high-$L_X$ PACS detected AGN hosts or nuclear contamination to \lums in the brightest AGNs.

An offset between AGN hosts and non-AGNs is instead observed  among the stacked non detections, including weak star forming as well as quiescent galaxies (see column (2) of Tab. \ref{tab:distrib}). 
The average\footnote{Average offsets were computed in the same way as  for the global (detections+non-detections) averages, and upper limits were not considered.}  offsets  at all masses and redshifts between the stacked \lums of AGNs and the control sample for low/all X-ray luminosity is $0.16\pm 0.03$/$0.18\pm 0.03$ dex in GOODS and $0.11\pm 0.04$/$0.31\pm 0.03$ dex in COSMOS. 
However, these offsets  are not large enough  to completely explain the SF enhancement observed in Fig. \ref{fig:l60mass}. 
This seems to exclude that a significant role may be played by a larger fraction of quiescent galaxies in the control sample, which would imply larger offsets among undetected sources. 

\begin{figure*}[!t]
\begin{tabular}{rl}
\resizebox{\hsize}{!}{
 \includegraphics[angle=90]{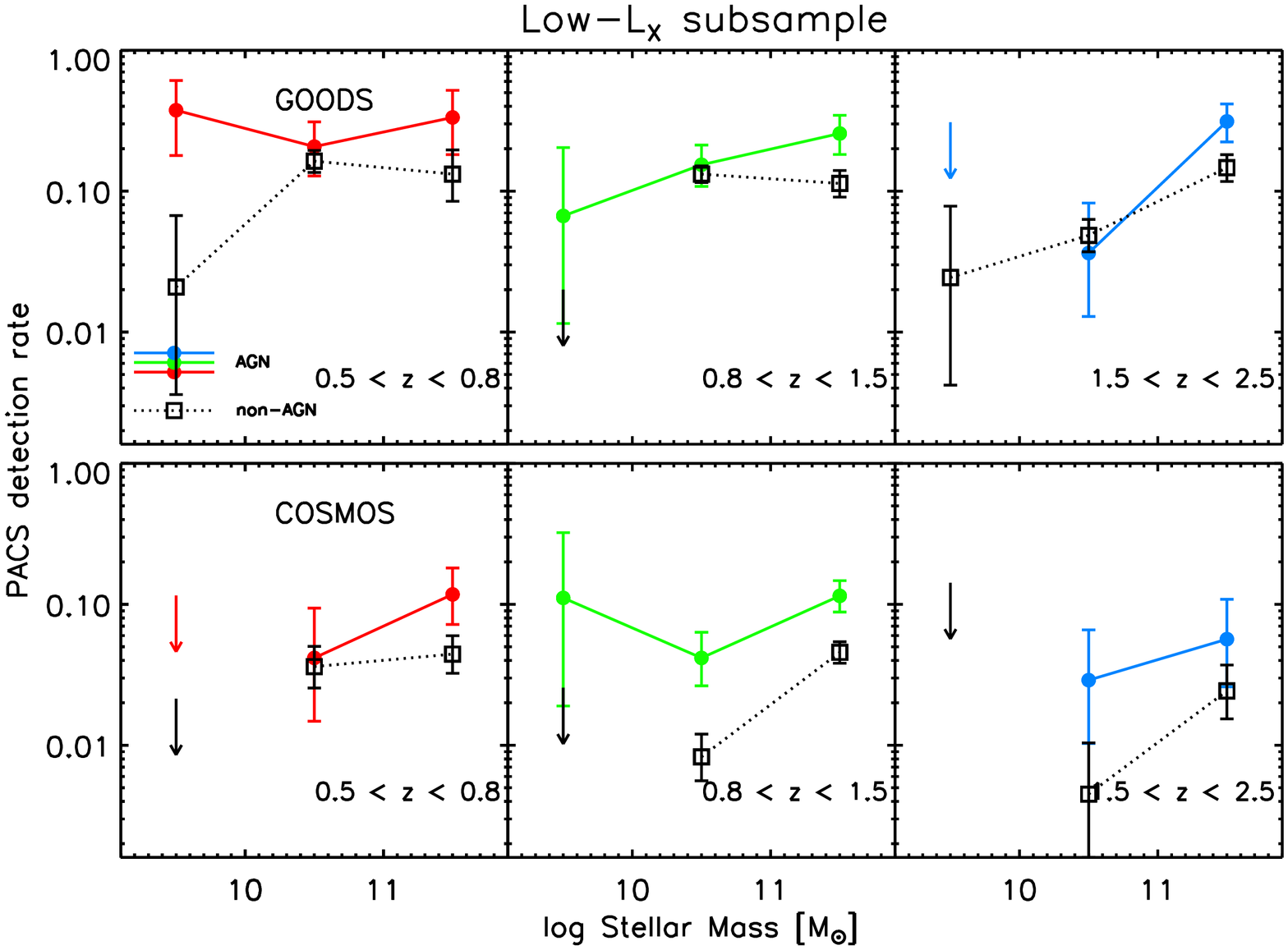}
 \includegraphics[angle=90]{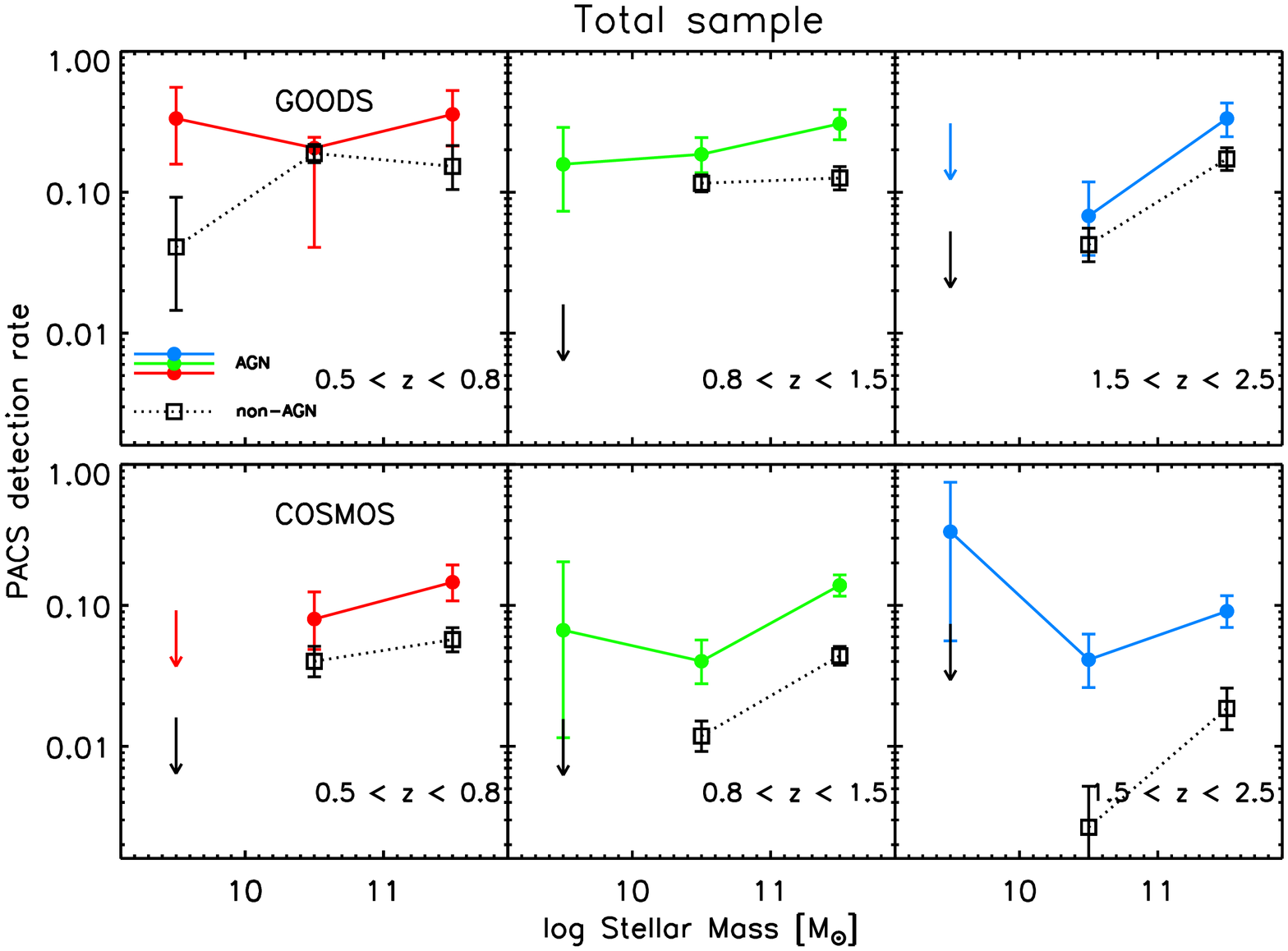}
}
\end{tabular}
 \caption{PACS detection rate in the different redshift bins, increasing from left to right, for the low-$L_X$ subsample ($left$) and for the total one ($right$). The GOODS fields are represented in the $upper$ panels and COSMOS  in the $lower$ ones. AGN hosts are showed as coloured solid lines and circles, inactive galaxies are represented by black dotted lines and open boxes. $1\sigma$ uncertainties are derived through binomial statistics. Masses indicate the centre of each mass bin. Upper limits indicate cases where none of the galaxies is detected by PACS. Detection in all PACS bands is required.
 }
 \label{fig:detfrac}
 \end{figure*}

Since averages are computed by weighting with the number of sources, it immediately follows that PACS detection rates must favour AGN hosts. To verify this statement, we plotted in Fig. \ref{fig:detfrac} the fraction of PACS detected sources among AGN hosts (coloured solid lines) and inactive galaxies (black dotted lines) in each of our redshift and mass bins. As before, we required PACS detections in both bands used for  \lums interpolation in order for an object to be `detected'. AGN hosts of low/all X-ray luminosities are on average\footnote{We computed linear averages from  the individual bins of the ratios between the detection rates of the two populations; uncertainties in the average were computed by summing in quadrature the binomial uncertainties in the individual detection rates; upper limits were not considered.} $3.5 \pm 0.7$/ $2.6 \pm 0.6$ times more likely in GOODS and $3.4 \pm 0.7$/$5.2 \pm 0.6$ times more likely in COSMOS to be detected by PACS than the control galaxies (the ratios in each redshift bin are reported in column (3) of  Tab. \ref{tab:distrib}). The PACS detection rates in each redshift and mass bin for the total sample (shown in the right panel of Fig. \ref{fig:detfrac}) can be found in Tab. \ref{tab:sample}. 
We  also checked this results against possible AGN contamination at the shortest PACS bands. 
The excess in the  detection rates of AGN hosts compared to non-AGNs is the same within the errors if detection is restricted only to the PACS160\mics or PACS100\mics bands, in agreement with the fact that AGN hosts and inactive galaxies have consistent F160/F100 colours (see Sect. \ref{sec:l60}). This further strengthens the assumption that AGN contamination does not strongly influence the \lums derivation, even for the total sample, and validates the interpretation of \lums as a SFR diagnostic. 

We can conclude that the SF enhancement that we observe in Fig. \ref{fig:l60mass} is on average not attributable to larger SFRs in rare very extreme starburst events in
AGN hosts than in inactive galaxies, which would shift the SFR distribution to higher values. 
It is instead due to a combination of modestly  brighter  stacked FIR emission in faint FIR sources  (i.e. undetected by PACS) and higher PACS detection rate (i.e. larger fraction of highly star forming galaxies) among AGN hosts. 
By means of a simple calculation described in Appendix A, the higher PACS detection rate turned out to be the dominant effect in the deep GOODS fields. Indeed, for a given detection rate, the linear average is highly dominated by the brightest sources, with PACS undetected sources only contributing in a minor fashion to the average offset.   
In other words, the distribution of SFRs among AGN hosts is not grossly different from normal star forming galaxies, consistent with the principal result from \cite{mullaney12}. Instead, on average, star forming galaxies are more likely to host AGNs at a given stellar mass. 
However, in the shallower COSMOS field, given the larger fraction of PACS undetected sources, the brighter FIR stacked emission and the higher PACS detection rate among AGNs can give comparable contributions to the average \lums offset between the two populations.

\section{Discussion} \label{sec:disc}

From the results presented in the previous section, it is clear that AGN hosts have  enhanced FIR emission with respect to a mass-matched sample of inactive galaxies including both star forming and quiescent populations. This is observed in all fields and at all masses and redshifts except the highest redshift bin in GOODS. 
The level of SF enhancement  between the two populations 
depends on the fields, with COSMOS showing a larger offset than GOODS. As shown in Fig. \ref{fig:xray}, COSMOS 
probes to higher X-ray luminosities 
compared to the GOODS fields, while missing most low-$L_X$ AGNs. In some ways, COSMOS and GOODS are complementary AGN samples. 
This hints to the possibility that the level of SF enhancement is related to the X-ray emission from the AGN. One interpretation, following the scenario suggested by our previous results inferred with SDP Herschel data  and presented in \cite{shao10}, is that the different levels of SF enhancement observed in the two fields reflect  different modes of AGN growth. 
However, before exploring this possibility, we mention that another viable interpretation of the larger SF activity observed in AGN hosts is in terms of positive feedback: AGN outflows,  such as winds and radio jets,  
compress and collapse interstellar clouds,  thus inducing SF, as suggested by a number of previous works \citep[e.g.][]{begelman89,silk05,feain07,silk09,elbaz09}. Radio-based studies of this population may help to uncover the role of positive feedback.

\subsection{Two modes of AGN evolution?} \label{sec:dichotomy}

\begin{figure*}[!t]
\begin{tabular}{rl}
\resizebox{\hsize}{!}{
 \includegraphics[angle=90]{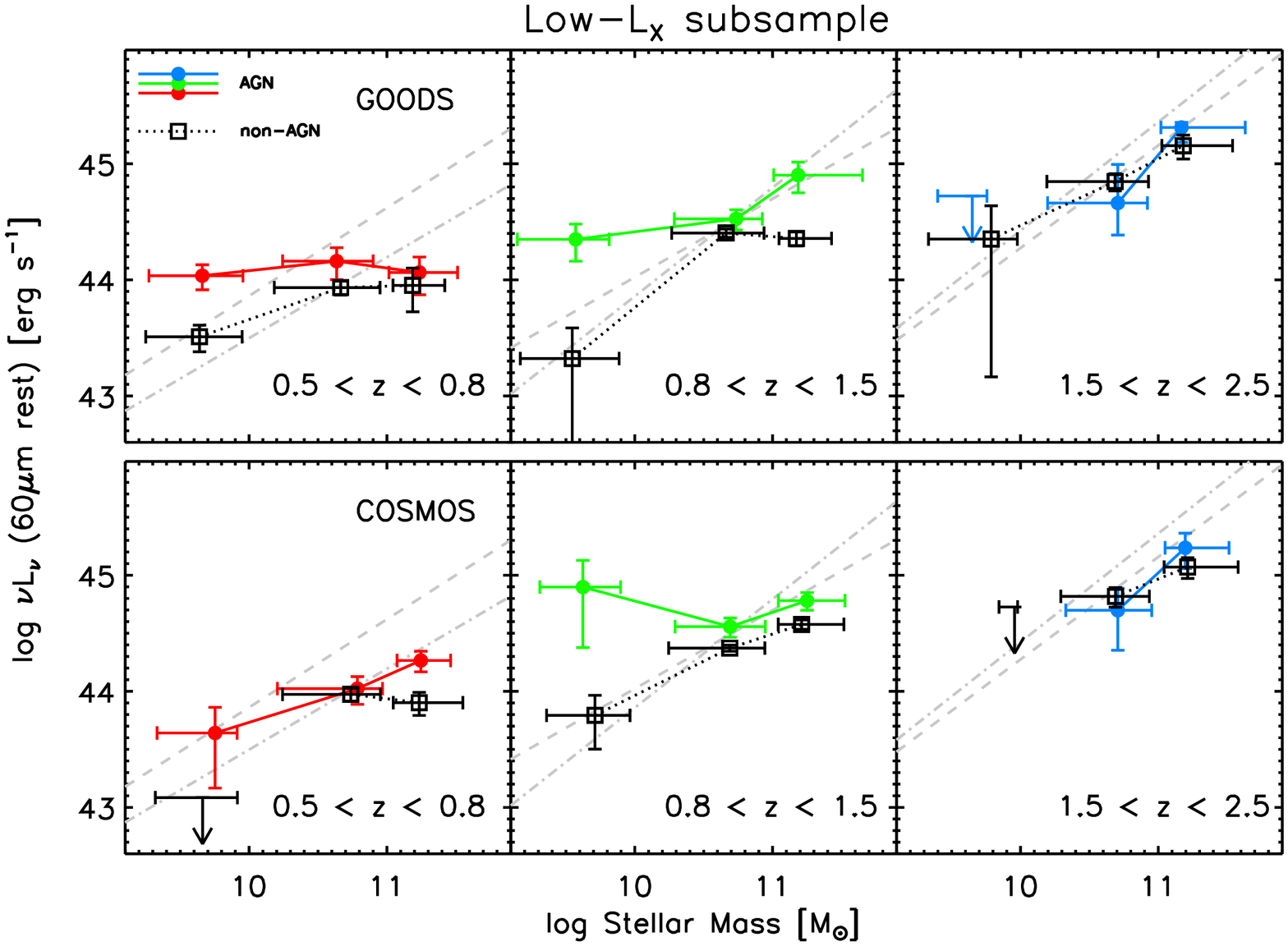}
 \includegraphics[angle=90]{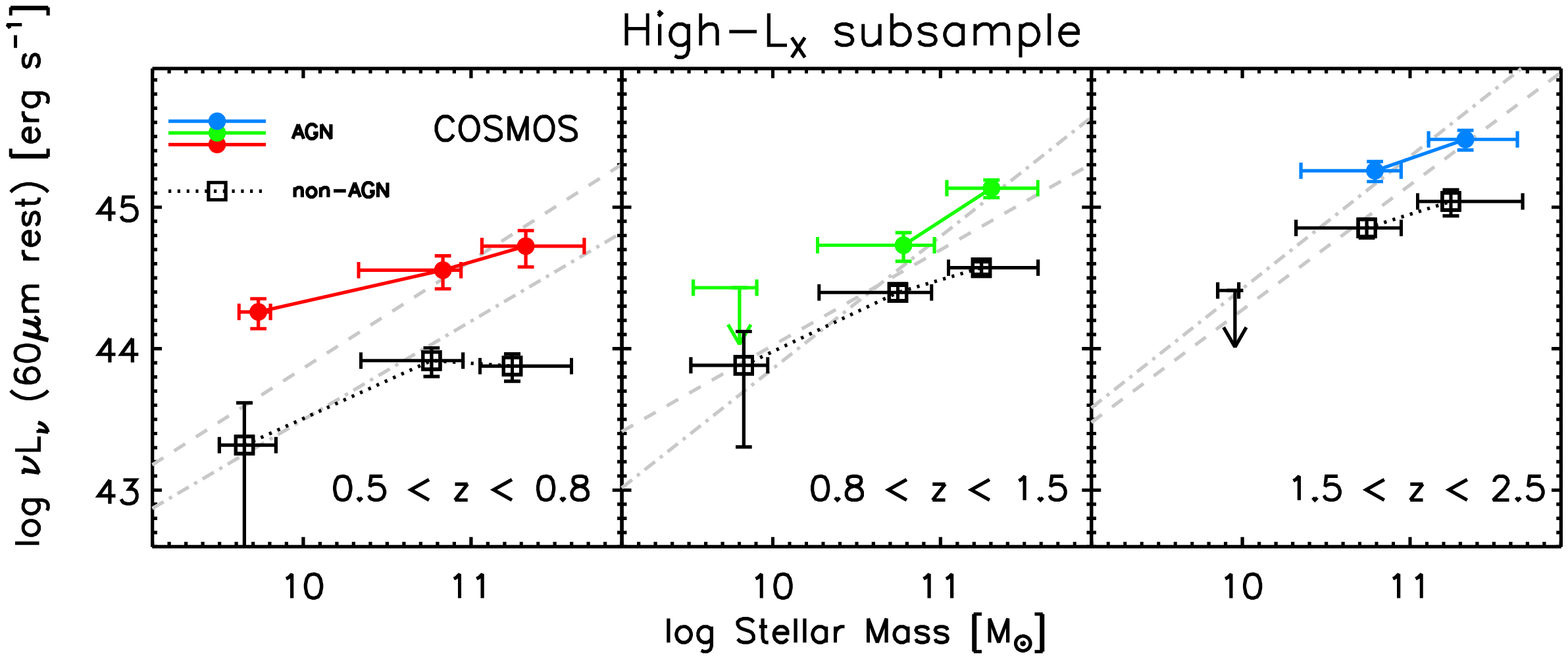}
}
\end{tabular}
 \caption{Same as Fig. \ref{fig:l60mass}, for low luminosity AGNs (see text) on the $left$ and high luminosity AGNs on the $right$. The inactive galaxies are  slightly changed, since they are randomly extracted from the total control sample to be mass-matched to the corresponding AGN subsample.
 }
 \label{fig:lumlx}
 \end{figure*}

To verify whether the different \lums offsets observed in GOODS and COSMOS are a consequence of the different AGN luminosity regimes mostly probed by each field, and hence possibly a consequence of different AGN evolutionary modes, 
we  repeated the analysis described in Sect. \ref{sec:res} by dividing the samples into high- and  low-$L_X$ subsamples 
by using the redshift-dependent $L_X$ thresholds reported in Sect. \ref{sec:distrib}.  
As done for the complete sample and as mentioned above, we extracted  mass-matched subsamples from the total control dataset to avoid any bias due to the different stellar mass distribution of AGN hosts and inactive galaxies. 

Another motivation for splitting the dataset based on the X-ray luminosity is to arrive at samples free of possible AGN contamination to the FIR emission: in the low-$L_X$ host galaxies, the \lums enhancement can be safely regarded as a SFR enhancement (see discussion in Sect. \ref{sec:l60}). 

The results are presented in Fig. \ref{fig:lumlx} and the average \lums offsets are summarized in Tab. \ref{tab:offset}. 
Once the same AGN luminosities are sampled, GOODS and COSMOS fields provide consistent results. 
The GOODS fields  are almost entirely ($\sim 88\%$) made of galaxies fainter than the luminosity thresholds adopted to split the sample. Consequently, the average \lums offset observed in the low-$L_X$ subsample ($0.28 \pm 0.03$ dex, top left panels) does not differ significantly from that observed in the total sample. The high-$L_X$ subsample does not offer enough statistics to extract meaningful numbers, owing to the paucity of luminous sources in the small GOODS fields. 
In COSMOS, the average FIR offset measured in the high-$L_X$ subsample (bottom right panels) increases to $0.56 \pm 0.03$ dex,  while at lower X-ray luminosities (bottom left panels) it is reduced to $0.24 \pm 0.04$ dex, consistent with the FIR enhancement observed in GOODS at similar AGN luminosities. 

The consistency between GOODS and COSMOS results, when the same AGN population is considered, is a further confirmation that uncertainties involving the host stellar mass estimate in Type 1 AGNs are not introducing serious biases in the analysis. 
Indeed, while the GOODS low-$L_x$ subsample is almost completely ($\sim 90\%$) made of Type 2 AGNs, the fraction of Type 1 AGNs in the COSMOS low-$L_X$ one is $\sim 30\%$. 
It also ensures that the analysis in the COSMOS field is insignificantly affected by the $I$-band selection applied to the sample (Sect. \ref{sec:selections}).

\cite{silverman09b} used [O II] ($\lambda = 3727 \AA$) as a star formation indicator to look for enhanced star formation in $z \lesssim 1$ AGN hosts. 
They found a weak but not significant enhancement for hosts of moderate luminosity ($10^{42}<L_X[$\ergs$]<10^{43.7}$) AGNs, while indication for enhanced SF if considering hosts of more luminous AGNs. Intricacies of [O II] as an AGN host star formation indicator \citep{kim06} likely reconcile this with our findings. 

%\cite{silverman09b} used [O II] ($\lambda = 3727 \AA$) as a star formation indicator to look for enhanced star formation in $z \lesssim 1$ AGN hosts. They did not find a significant enhancement for hosts of moderate luminosity ($10^{42}<L_X[$\ergs$]<10^{43.7}$) AGNs but indication for enhanced SF if including hosts of more luminous AGNs. Intricacies of [O II] as an AGN host star formation indicator \citep{kim06} likely reconcile this with our findings. 

In order to understand whether the different \lums enhancement that we observe in the two subsamples is indeed due to the different accretion regimes, a few effects must be taken into account. 
Although a morphological analysis is beyond the scope of the present study, we note that the morphological mix in the low-$L_X$ and high-$L_X$ AGNs subsamples may in general be different \citep[e.g.][]{hutchings02,hutchings09,veilleux09b,mainieri11,kocevski12}. The same can in principle be true also for the mass-matched control galaxies, even though the mass distribution of low-$L_X$ and high-$L_X$ AGN host galaxies are very similar, with the latter having an average stellar mass $\sim 0.14 \pm 0.03$ dex higher than their lower luminosity counterparts (the comparison has been done in the COSMOS field, where we have the better statistics). Since the combination of these effects is hard to account for, 
we compared the FIR emission among low-$L_X$ vs high-$L_X$ AGN host galaxies, and, similarly, among the inactive galaxies mass-matched to low-$L_X$ AGN hosts vs those mass-matched to high-$L_X$ AGN hosts. As can also be seen from Fig. \ref{fig:lumlx}, the two control samples are totally consistent, while an offset is observed among low-$L_X$ and high-$L_X$ AGN hosts, the latter showing the larger FIR emission. The different behaviour of the two subsamples can therefore be associated with differences in the SF activity among the host galaxies of faint and bright AGNs.

The $L_X$-dependent FIR enhancement can be interpreted in terms of the AGN evolutionary picture outlined by the previous step of the present work \citep{shao10} and described in the introduction. There, we used Herschel SDP data from GOODS-N to study the relation between host star formation and bolometric AGN luminosity at different redshifts and we showed that AGN growth can be explained by a combination of two paths of evolution. 
Very luminous AGNs are closely coupled to their host galaxy's growth, 
through a process that leads to a correlation between AGN luminosity and  SF activity in the host  determining a tight co-evolutionary phase. A good possibility are gas-rich major mergers: these  transport to the centre large amounts of gas which trigger both AGN fuelling and intense SF  bursts. 
On the other side, at lower luminosities AGN hosts experience  secular evolution, in the sense that instantaneous nuclear accretion and total secular, non-merger driven SF are not strongly correlated and expected to evolve independently. 
In Rosario et al. (in prep.), which represents our third step in the study of the interplay between AGN accretion and host's star formation, we extend this analysis by including more Herschel fields and make use of a larger dataset which allows sampling of both low and high luminosity AGNs. We  find similar results with respect to our previous study, although, as already suggested by \cite{shao10}, the transition between the two modes of AGN growth seems to be shallower at high redshift. This transition  is measured around $L_{AGN} \sim 10^{44.5 - 45.5}$ \ergs\, which corresponds\footnote{Bolometric AGN luminosities are converted into intrinsic 2-10 keV luminosities  by using Eq. (5) of \cite{maiolino07} and adopting a ratio of 7 between bolometric and 5100 \AA\ luminosity.} to $L_X \sim 10^{43.3 - 44}$ \ergs. These redshift-dependent luminosity thresholds were used to split the sample and are shown as red lines in Fig. \ref{fig:xray}.

The two-mode evolutionary scenario above explains the larger \lums enhancement (with respect to mass-matched inactive galaxies) observed in high-$L_X$ AGN hosts. However, we remind that  contamination from the AGN could affect the FIR luminosity estimate in a fraction of the brightest ($L_X>10^{44}$ \ergs) hosts (see Rosario et al. in prep. for further details). 
The aforementioned shallower transition between the two growth modes observed at high redshift 
also explains the larger SF enhancement measured at low $z$ than at high $z$. 
This behaviour is expected for two reasons: first, given the lower average SFR at a given mass, 
we are more sensitive to merger-induced SF bursts; second, larger errors at high $z$ and possible gas feeding  through secular processes (besides major mergers) can wash out the observed SF offset.

On the other hand, the \cite{shao10} picture does not predict any \lums enhancement, i.e. SF enhancement, in low-$L_X$ AGN hosts. In this view, the offset in low-$L_X$ is therefore rather surprising.
We discuss possible interpretations  in the next section.

There is another possibility to explain the difference in the  FIR excess observed with respect to inactive galaxies  between hosts of low-$L_X$ and high-$L_X$ AGNs, without assuming the existence of separate modes of evolution (i.e. merger co-evolution vs secular). AGNs may be more likely to exist in galaxies with higher gas fractions at a given stellar mass, only because cold gas in these galaxies has a higher probability of reaching the nucleus. Indeed, in order to feed the AGN, a substantial amount of gas is required to inflow to the centre, plausibly from large scales. Low-$L_X$ would  require smaller gas reservoir than brighter AGNs, and hence would be characterized by less star formation activity on large scales. Under this hypothesis, we would expect to see a correlation between AGN and FIR luminosity at all scales, while there is no evidence for such trend at $\log(L_X[$\ergs$])\lesssim 43.5$ (\citealt{shao10}, Rosario et al. in prep.). For this reason, we decided to interpret our data in the view of the picture presented by \cite{shao10}. However, deeper data and larger statistics is needed to definitely accept or exclude this hypothesis.

\subsection{SF enhancement in low-$L_X$ AGN hosts: non-synchronous accretion/SF histories  or break in the dichotomy?
}

While the FIR enhancement in bright AGN hosts can be explained by merger-induced SF bursts,  the twofold AGN evolutionary path  
that we suggested in \cite{shao10} cannot explain any SF enhancement in low-$L_X$ AGN hosts. However, a modest but significant offset of $\sim 0.26$ dex at  $\sim 3\sigma$ confidence level is measured with respect to the mass-matched inactive control sample. 
We suggest here some possible reasons for this enhancement. They may be broadly classified as enhancements in a subset of AGN hosts, due to  gas-rich  major 
galaxy mergers, or enhancements in most AGN hosts, due to secularly driven relationships between accretion and star formation.

The first group of possibilities implies that in objects following a `merger' co-evolution, the instantaneous AGN accretion and star formation rate are not well synchronized, even if causally linked. This will lead to complex evolutionary tracks in a $L_{FIR}/L_{AGN}$ diagram  \citep[such as Fig 6 of][]{shao10}, and to merger-enhanced SFRs being present in objects with low AGN luminosity. Specifically, it is plausible to discuss the case of delayed 
AGN feeding with respect to the onset of the SF episodes  induced by the gas inflow   during major interactions. This circumstance is suggested by several models \citep[e.g.][]{dimatteo05,hopkinsp06,netzer09}, which assume that AGNs are fuelled only when substantial amounts of gas can reach the central black holes.  
It is in line with the classical local ULIRG-to-QSO transition scenario \citep{sanders88b,sanders89,sanders96}, and would concern a small number of strongly SF galaxies. It is also supported by simulations such as those presented by \cite{hayward11}, which are based on the observational evidences provided by  \cite{alexander08} and show that accretion/SF histories in SMGs are not strictly  synchronized. 
\cite{netzer09} supports this scenario by arguing the existence of sources, such as ULIRGs and SMGs, characterized by huge SFRs and modest AGN contribution (see also Fig. 6 of \citealt{lutz10}). 
According to the evolutionary sequence for the starburst--AGN co-evolution suggested by \cite{netzer09}, the SF episode starts first, and only after a certain time lag ($\Delta t_{SF}$) does enough cold gas reach the centre and the black hole accretion begin. 
Simulations prefer a long $\Delta t_{SF}$ compared to the rise time of the AGN phase ($\Delta t_{AGN}$). Such a model is qualitatively compatible with the SF offset  we measure only if AGN duty cycles are long with respect to the SF episode (i.e., $\Delta t_{AGN} \lesssim \Delta t_{SF}$). This will lead to a number of merger-induced starbursts with low levels of accretion, which would raise the mean SFR of low luminosity AGN hosts compared to inactive galaxies. If, on the other hand, $\Delta t_{AGN} << \Delta t_{SF}$, a number of the strongest star forming galaxies would be missing from the AGN sample, leading to a higher average SFR for non-AGNs, at variance with what we find. 
Alternatively, one may relax the requirement that bright AGNs occur only after most of the star formation has progressed in mergers, and instead postulate that the general synchronization between the two phases is rather weak. While this can qualitatively explain the observed mean offset, it is not well-motivated by most models of AGN fuelling by tidal torques.

There is a further explanation for the low but significant SF enhancement observed at low X-ray luminosities. 
Alternative AGN feeding mechanisms, at work during secular evolution and not invoking major mergers, 
 may lead to a weaker physical connection, but still a certain level of correlation, between instantaneous accretion rate and total star formation. This will make the dichotomy between secular evolution and  co-evolution less sharp than outlined in Sect. \ref{sec:dichotomy}. 
The nature of the relationship is such that the level of accretion is not intimately correlated with the level of star formation; it may be causal or may be mediated through other processes. For example, one may imagine star formation in the circum-nuclear environment of SMBHs can disturb gas by supernovae or stellar winds and lead to inflow to the nuclear regions. In this picture, only star formation within the inner kpc or so may be related to AGN activity. 
Alternatively, processes that drive star formation in inactive galaxies, such as fresh gas infall, bar instabilities, stellar feedback,  galaxy harassment or satellite accretion, may be able to inspire some inflow to the SMBH, but to a degree  that is not closely related to the level of star formation at large scales. An example of these processes, which applies to  $z \gtrsim 2$ hosts, is gravitational instability in clumpy disks. 
The recent work of \cite{bournaud11}  (see also \citealt{dimatteo11} for bright QSOs at $z>5$) suggests that both AGN accretion and bulge building, often characterized by giant star forming clumps, can be associated with violent gravitational instability in high redshift disk galaxies. This can be generated by cold streams and high gas fractions, and it includes clumps migration \citep[e.g.][]{genzel08}. 
A prediction of their simulation, which is supported by observational evidences (see below), is that high-$z$ moderate luminosity ($L_X\sim 10^{42-43} erg/s$) AGNs are hosted by star forming clumpy disk galaxies with no signs of major mergers. Finally, dry mergers, with low gas fractions, could also in principle explain the reduced SF enhancement in low-$L_X$ AGN hosts. However, most low luminosity AGNs are in disks at these redshifts \citep{schawinski11,kocevski12}, so it is unlikely that they underwent a (dry) merger recently.

According to the first class of explanations, a subset of low-$L_x$ AGNs are associated with strong merger-induced starbursts, while the rest are in normal star forming galaxies. Therefore, we would expect low-$L_X$ AGNs to be over-represented at the upper envelope of the star forming main sequence, where most starbursting systems lie. Our analysis in Sect. \ref{sec:distrib} suggests, however, that bright FIR AGN hosts and inactive galaxies have rather similar distributions in the SFR-stellar mass plane. Thus, a merger-driven scenario for the offset we measure is not strongly supported by our observations. 
The lack of an excess of major mergers among low luminosity AGNs is consistent with several observational studies, according to which moderate luminosity AGNs have similar morphologies to inactive galaxies, not consistent with disturbed morphologies \citep{grogin05,gabor09,cisternas11}, and are frequently found in disks at high redshifts \citep{schawinski11,kocevski12}. The latter property is consistent with AGNs being on average hosted by main sequence galaxies \citep[e.g.][]{wuyts11}.  
The second class of scenarios applies to the majority of normal star forming galaxies and predicts that AGN hosts have similar star formation properties as inactive galaxies, but that AGNs are a bit more likely to be found in hosts where some degree of recent star formation is in effect. 
This is qualitatively consistent with our study of the SFR distributions, as well as the fairly constant SFR that we measured in low-$L_X$ AGNs across a range of $L_{X}$ \citep{shao10}.
However, further statistics, deeper FIR and multiwavelength observations and careful morphological studies are needed in order to definitely distinguish between the two hypotheses, understand the importance of mergers as AGN hosts, as well as explore the processes of secular fuelling in AGNs at these redshifts.

 \section{Summary} \label{sec:summ}

We  used Herschel PACS observations carried out in the two GOODS fields and in COSMOS, in combination with ancillary optical-to-MIR and X-ray data, to study the star forming properties of AGN hosts as a function of stellar mass and redshift with respect to inactive galaxies. A fitting procedure to estimate the stellar masses of galaxies was developed to disentangle  stellar and nuclear emission, in order to measure stellar masses of host galaxies unbiased by the AGN light. The FIR emission, which can be considered as a proxy of the SFR in all galaxies except in a fraction of the brightest AGNs, was derived by interpolating PACS fluxes to derive a rest-frame 60 \mics luminosity, without assuming a particular FIR SED shape. The FIR emission of AGN hosts was compared to that of non-AGN control galaxies with similar masses and at similar redshifts including both star forming and quiescent populations.

We found that the FIR emission, and hence the SF  activity, in AGN hosts is enhanced with respect to that of mass-matched inactive galaxies. 
We  checked our analysis against possible sources of bias (PACS source extraction techniques, different spectroscopic fractions among the various samples, large uncertainties in the host stellar mass estimates in Type 1 AGNs, different methods to measure the FIR luminosity) and verified it to be robust. 
We  investigated in detail the FIR emission of PACS detected and undetected sources. 
The average offset between the AGN hosts and inactive galaxies  is not attributable to larger SFRs in rare very extreme starburst events in AGN hosts than in inactive galaxies. 
It is instead due to   a combination of modestly brighter  stacked FIR emission in faint FIR sources (i.e. undetected by PACS) and higher PACS detection rate among AGN hosts compared to inactive sources, the latter feature being the main driver of the SF enhancement. In other words, AGNs are more likely to be hosted by star forming galaxies with a SFR above the PACS detection limit.

The two GOODS fields, mainly characterized by low-$L_X$ AGNs, and COSMOS, dominated by the brighter sources, show different levels of enhancement.  
We demonstrated that the SF offset is highly dependent on the X-ray luminosity: it is equal to $\sim 0.26$ dex at $\sim 3 \sigma$ confidence level at $\log(L_X[$\ergs$])\lesssim 43.5$, and $0.56$ dex at  $> 10 \sigma$ confidence level for more luminous AGN hosts.  
The different levels of measured FIR enhancement as a function of X-ray luminosity support the twofold AGN growth picture suggested by our previous results presented in \cite{shao10}, although different interpretations cannot be ruled out with the present data. 
According to \cite{shao10}, low-$L_X$ AGN hosts undergo secular evolution, with instantaneous AGN feeding not strongly correlated with the total instantaneous SF activity in the host galaxy, and no SF offset is expected compared to inactive galaxies; luminous AGNs, on the other hand, co-evolve with their hosts, possibly through (major) merger interactions, and their host's SF activity is correlated with the AGN accretion rate. 

We suggested two  hypotheses to explain the FIR enhancement observed at low X-ray luminosities.  
The first possibility ascribes the enhancement to a subset of FIR bright AGN hosts, and assumes that, in a major merger scenario, accretion and SF histories are not well synchronized, even if causally linked; 
a plausible example of this is  a delayed AGN fuelling with respect to the onset of SF  episodes after gas inflow, but with a long or uneven rise time for AGN accretion, compared to the starburst timescale.  
A second  explanation involves a larger number of AGN hosts, whose properties are similar to those of main sequence galaxies, and it implies that a certain level of correlation between total instantaneous SF and accretion is also induced by smaller scale (non-merger) mechanisms; for example, clump migration in massive turbulent disks drive instabilities which are responsible for both SF and AGN accretion. 
Although the present data do not allow us to distinguish among these two scenarios with certainty, similarities in the SFR distributions of bright FIR AGN hosts and non-AGNs seem to prefer the second option.

Future studies, especially careful morphological studies (e.g. in the rest-frame optical with WFC3) of luminous AGNs at these redshifts, will directly probe the relationship between AGN hosts and galaxy interactions and shed light on the process of AGN growth. Moreover, they will permit a discrimination to be made between the two scenarios suggested to explain the modest level of SF enhancement measured in low-$L_X$ AGN hosts.

\begin{acknowledgements}
We thank the referee, whose suggestions improved the quality of the analysis. 
We also thank Hagai Netzer and Fabrizio Fiore for interesting discussions.  
PACS has been developed by a consortium of institutes led by MPE (Germany) and including UVIE
(Austria); KU Leuven, CSL, IMEC (Belgium); CEA, LAM (France); MPIA (Germany); INAF-IFSI/
OAA/OAP/OAT, LENS, SISSA (Italy); IAC (Spain). This development has been supported by the
funding agencies BMVIT (Austria), ESA-PRODEX (Belgium), CEA/CNES (France), DLR (Germany),
ASI/INAF (Italy), and CICYT/MCYT (Spain). 
We acknowledge financial contribution from the agreement ASI/INAF I/005/011/0.
\end{acknowledgements}

%\bibliographystyle{aa}
%\bibliography{biblio}

\begin{thebibliography}{176}
\expandafter\ifx\csname natexlab\endcsname\relax\def\natexlab#1{#1}\fi

\bibitem[{{Aird} {et~al.}(2012){Aird}, {Coil}, {Moustakas}, {Blanton},
  {Burles}, {Cool}, {Eisenstein}, {Smith}, {Wong}, \& {Zhu}}]{aird12}
{Aird}, J., {Coil}, A.~L., {Moustakas}, J., {et~al.} 2012, \apj, 746, 90

\bibitem[{{Alexander} {et~al.}(2003){Alexander}, {Bauer}, {Brandt},
  {Hornschemeier}, {Vignali}, {Garmire}, {Schneider}, {Chartas}, \&
  {Gallagher}}]{alexander03}
{Alexander}, D.~M., {Bauer}, F.~E., {Brandt}, W.~N., {et~al.} 2003, \aj, 125,
  383

\bibitem[{{Alexander} {et~al.}(2005){Alexander}, {Bauer}, {Chapman}, {Smail},
  {Blain}, {Brandt}, \& {Ivison}}]{alexander05}
{Alexander}, D.~M., {Bauer}, F.~E., {Chapman}, S.~C., {et~al.} 2005, \apj, 632,
  736

\bibitem[{{Alexander} {et~al.}(2008){Alexander}, {Brandt}, {Smail}, {Swinbank},
  {Bauer}, {Blain}, {Chapman}, {Coppin}, {Ivison}, \&
  {Men{\'e}ndez-Delmestre}}]{alexander08}
{Alexander}, D.~M., {Brandt}, W.~N., {Smail}, I., {et~al.} 2008, \aj, 135, 1968

\bibitem[{{Allevato} {et~al.}(2011){Allevato}, {Finoguenov}, {Cappelluti},
  {Miyaji}, {Hasinger}, {Salvato}, {Brusa}, {Gilli}, {Zamorani}, {Shankar},
  {James}, {McCracken}, {Bongiorno}, {Merloni}, {Peacock}, {Silverman}, \&
  {Comastri}}]{allevato11}
{Allevato}, V., {Finoguenov}, A., {Cappelluti}, N., {et~al.} 2011, \apj, 736,
  99

\bibitem[{{Alonso-Herrero} {et~al.}(2008){Alonso-Herrero},
  {P{\'e}rez-Gonz{\'a}lez}, {Rieke}, {Alexander}, {Rigby}, {Papovich},
  {Donley}, \& {Rigopoulou}}]{alonsoherrero08}
{Alonso-Herrero}, A., {P{\'e}rez-Gonz{\'a}lez}, P.~G., {Rieke}, G.~H., {et~al.}
  2008, \apj, 677, 127

\bibitem[{{Barvainis} \& {Ivison}(2002)}]{barvainis02}
{Barvainis}, R. \& {Ivison}, R. 2002, \apj, 571, 712

\bibitem[{{Bauer} {et~al.}(2004){Bauer}, {Alexander}, {Brandt}, {Schneider},
  {Treister}, {Hornschemeier}, \& {Garmire}}]{bauer04}
{Bauer}, F.~E., {Alexander}, D.~M., {Brandt}, W.~N., {et~al.} 2004, \aj, 128,
  2048

\bibitem[{{Begelman} \& {Cioffi}(1989)}]{begelman89}
{Begelman}, M.~C. \& {Cioffi}, D.~F. 1989, \apjl, 345, L21

\bibitem[{{Bell} {et~al.}(2004){Bell}, {Wolf}, {Meisenheimer}, {Rix}, {Borch},
  {Dye}, {Kleinheinrich}, {Wisotzki}, \& {McIntosh}}]{bell04}
{Bell}, E.~F., {Wolf}, C., {Meisenheimer}, K., {et~al.} 2004, \apj, 608, 752

\bibitem[{{Bennert} {et~al.}(2011){Bennert}, {Auger}, {Treu}, {Woo}, \&
  {Malkan}}]{bennert11}
{Bennert}, V.~N., {Auger}, M.~W., {Treu}, T., {Woo}, J., \& {Malkan}, M.~A.
  2011, \apj, 742, 107

\bibitem[{{Berta} {et~al.}(2010){Berta}, {Magnelli}, {Lutz}, {Altieri},
  {Aussel}, {Andreani}, {Bauer}, {Bongiovanni}, {Cava}, {Cepa}, {Cimatti},
  {Daddi}, {Dominguez}, {Elbaz}, {Feuchtgruber}, {F{\"o}rster Schreiber},
  {Genzel}, {Gruppioni}, {Katterloher}, {Magdis}, {Maiolino}, {Nordon},
  {P{\'e}rez Garc{\'{\i}}a}, {Poglitsch}, {Popesso}, {Pozzi}, {Riguccini},
  {Rodighiero}, {Saintonge}, {Santini}, {Sanchez-Portal}, {Shao}, {Sturm},
  {Tacconi}, {Valtchanov}, {Wetzstein}, \& {Wieprecht}}]{berta10}
{Berta}, S., {Magnelli}, B., {Lutz}, D., {et~al.} 2010, \aap, 518, L30+

\bibitem[{{Berta} {et~al.}(2011){Berta}, {Magnelli}, {Nordon}, {Lutz}, {Wuyts},
  {Altieri}, {Andreani}, {Aussel}, {Casta{\~n}eda}, {Cepa}, {Cimatti}, {Daddi},
  {Elbaz}, {F{\"o}rster Schreiber}, {Genzel}, {Le Floc'h}, {Maiolino},
  {P{\'e}rez-Fournon}, {Poglitsch}, {Popesso}, {Pozzi}, {Riguccini},
  {Rodighiero}, {Sanchez-Portal}, {Sturm}, {Tacconi}, \&
  {Valtchanov}}]{berta11}
{Berta}, S., {Magnelli}, B., {Nordon}, R., {et~al.} 2011, \aap, 532, A49+

\bibitem[{{B{\'e}thermin} {et~al.}(2010){B{\'e}thermin}, {Dole}, {Beelen}, \&
  {Aussel}}]{bethermin10}
{B{\'e}thermin}, M., {Dole}, H., {Beelen}, A., \& {Aussel}, H. 2010, \aap, 512,
  A78+

\bibitem[{{Bonfield} {et~al.}(2011){Bonfield}, {Jarvis}, {Hardcastle},
  {Cooray}, {Hatziminaoglou}, {Ivison}, {Page}, {Stevens}, {de Zotti}, {Auld},
  {Baes}, {Buttiglione}, {Cava}, {Dariush}, {Dunlop}, {Dunne}, {Dye}, {Eales},
  {Fritz}, {Hopwood}, {Ibar}, {Maddox}, {Micha{\l}owski}, {Pascale}, {Pohlen},
  {Rigby}, {Rodighiero}, {Serjeant}, {Smith}, {Temi}, \& {van der
  Werf}}]{bonfield11}
{Bonfield}, D.~G., {Jarvis}, M.~J., {Hardcastle}, M.~J., {et~al.} 2011, \mnras,
  416, 13

\bibitem[{{Bongiorno} {et~al.}(2007){Bongiorno}, {Zamorani}, {Gavignaud},
  {Marano}, {Paltani}, {Mathez}, {M{\o}ller}, {Picat}, {Cirasuolo},
  {Lamareille}, {Bottini}, {Garilli}, {Le Brun}, {Le F{\`e}vre}, {Maccagni},
  {Scaramella}, {Scodeggio}, {Tresse}, {Vettolani}, {Zanichelli}, {Adami},
  {Arnouts}, {Bardelli}, {Bolzonella}, {Cappi}, {Charlot}, {Ciliegi},
  {Contini}, {Foucaud}, {Franzetti}, {Guzzo}, {Ilbert}, {Iovino}, {McCracken},
  {Marinoni}, {Mazure}, {Meneux}, {Merighi}, {Pell{\`o}}, {Pollo}, {Pozzetti},
  {Radovich}, {Zucca}, {Hatziminaoglou}, {Polletta}, {Bondi}, {Brinchmann},
  {Cucciati}, {de la Torre}, {Gregorini}, {Mellier}, {Merluzzi}, {Temporin},
  {Vergani}, \& {Walcher}}]{bongiorno07}
{Bongiorno}, A., {Zamorani}, G., {Gavignaud}, I., {et~al.} 2007, \aap, 472, 443

\bibitem[{{Bournaud} {et~al.}(2011){Bournaud}, {Dekel}, {Teyssier}, {Cacciato},
  {Daddi}, {Juneau}, \& {Shankar}}]{bournaud11}
{Bournaud}, F., {Dekel}, A., {Teyssier}, R., {et~al.} 2011, \apjl, 741, L33

\bibitem[{{Bower} {et~al.}(2006){Bower}, {Benson}, {Malbon}, {Helly}, {Frenk},
  {Baugh}, {Cole}, \& {Lacey}}]{bower06}
{Bower}, R.~G., {Benson}, A.~J., {Malbon}, R., {et~al.} 2006, \mnras, 370, 645

\bibitem[{{Brammer} {et~al.}(2008){Brammer}, {van Dokkum}, \&
  {Coppi}}]{brammer08}
{Brammer}, G.~B., {van Dokkum}, P.~G., \& {Coppi}, P. 2008, \apj, 686, 1503

\bibitem[{{Brandt} \& {Hasinger}(2005)}]{brandt05}
{Brandt}, W.~N. \& {Hasinger}, G. 2005, \araa, 43, 827

\bibitem[{{Brinchmann} {et~al.}(2004){Brinchmann}, {Charlot}, {White},
  {Tremonti}, {Kauffmann}, {Heckman}, \& {Brinkmann}}]{brinchmann04}
{Brinchmann}, J., {Charlot}, S., {White}, S.~D.~M., {et~al.} 2004, \mnras, 351,
  1151

\bibitem[{{Brusa} {et~al.}(2010){Brusa}, {Civano}, {Comastri}, {Miyaji},
  {Salvato}, {Zamorani}, {Cappelluti}, {Fiore}, {Hasinger}, {Mainieri},
  {Merloni}, {Bongiorno}, {Capak}, {Elvis}, {Gilli}, {Hao}, {Jahnke},
  {Koekemoer}, {Ilbert}, {Le Floc'h}, {Lusso}, {Mignoli}, {Schinnerer},
  {Silverman}, {Treister}, {Trump}, {Vignali}, {Zamojski}, {Aldcroft},
  {Aussel}, {Bardelli}, {Bolzonella}, {Cappi}, {Caputi}, {Contini},
  {Finoguenov}, {Fruscione}, {Garilli}, {Impey}, {Iovino}, {Iwasawa},
  {Kampczyk}, {Kartaltepe}, {Kneib}, {Knobel}, {Kovac}, {Lamareille},
  {Leborgne}, {Le Brun}, {Le Fevre}, {Lilly}, {Maier}, {McCracken}, {Pello},
  {Peng}, {Perez-Montero}, {de Ravel}, {Sanders}, {Scodeggio}, {Scoville},
  {Tanaka}, {Taniguchi}, {Tasca}, {de la Torre}, {Tresse}, {Vergani}, \&
  {Zucca}}]{brusa10}
{Brusa}, M., {Civano}, F., {Comastri}, A., {et~al.} 2010, \apj, 716, 348

\bibitem[{{Brusa} {et~al.}(2009){Brusa}, {Fiore}, {Santini}, {Grazian},
  {Comastri}, {Zamorani}, {Hasinger}, {Merloni}, {Civano}, {Fontana}, \&
  {Mainieri}}]{brusa09}
{Brusa}, M., {Fiore}, F., {Santini}, P., {et~al.} 2009, \aap, 507, 1277

\bibitem[{{Bruzual}(2007)}]{bruzual07}
{Bruzual}, G. 2007, in Astronomical Society of the Pacific Conference Series,
  Vol. 374, From Stars to Galaxies: Building the Pieces to Build Up the
  Universe, ed. A.~{Vallenari}, R.~{Tantalo}, L.~{Portinari}, \& A.~{Moretti},
  303--+

\bibitem[{{Bruzual} \& {Charlot}(2003)}]{bc03}
{Bruzual}, G. \& {Charlot}, S. 2003, Royal Astronomical Society, Monthly
  Notices, 344, 1000

\bibitem[{{Bundy} {et~al.}(2008){Bundy}, {Georgakakis}, {Nandra}, {Ellis},
  {Conselice}, {Laird}, {Coil}, {Cooper}, {Faber}, {Newman}, {Pierce},
  {Primack}, \& {Yan}}]{bundy08}
{Bundy}, K., {Georgakakis}, A., {Nandra}, K., {et~al.} 2008, \apj, 681, 931

\bibitem[{{Calzetti} {et~al.}(2000){Calzetti}, {Armus}, {Bohlin}, {Kinney},
  {Koornneef}, \& {Storchi-Bergmann}}]{calzetti00}
{Calzetti}, D., {Armus}, L., {Bohlin}, R.~C., {et~al.} 2000, \apj, 533, 682

\bibitem[{{Canalizo} \& {Stockton}(2000)}]{canalizo00}
{Canalizo}, G. \& {Stockton}, A. 2000, \aj, 120, 1750

\bibitem[{{Canalizo} \& {Stockton}(2001)}]{canalizo01}
{Canalizo}, G. \& {Stockton}, A. 2001, \apj, 555, 719

\bibitem[{{Capak} {et~al.}(2007){Capak}, {Aussel}, {Ajiki}, {McCracken},
  {Mobasher}, {Scoville}, {Shopbell}, {Taniguchi}, {Thompson}, {Tribiano},
  {Sasaki}, {Blain}, {Brusa}, {Carilli}, {Comastri}, {Carollo}, {Cassata},
  {Colbert}, {Ellis}, {Elvis}, {Giavalisco}, {Green}, {Guzzo}, {Hasinger},
  {Ilbert}, {Impey}, {Jahnke}, {Kartaltepe}, {Kneib}, {Koda}, {Koekemoer},
  {Komiyama}, {Leauthaud}, {Le Fevre}, {Lilly}, {Liu}, {Massey}, {Miyazaki},
  {Murayama}, {Nagao}, {Peacock}, {Pickles}, {Porciani}, {Renzini}, {Rhodes},
  {Rich}, {Salvato}, {Sanders}, {Scarlata}, {Schiminovich}, {Schinnerer},
  {Scodeggio}, {Sheth}, {Shioya}, {Tasca}, {Taylor}, {Yan}, \&
  {Zamorani}}]{capak07}
{Capak}, P., {Aussel}, H., {Ajiki}, M., {et~al.} 2007, \apjs, 172, 99

\bibitem[{{Cappelluti} {et~al.}(2009){Cappelluti}, {Brusa}, {Hasinger},
  {Comastri}, {Zamorani}, {Finoguenov}, {Gilli}, {Puccetti}, {Miyaji},
  {Salvato}, {Vignali}, {Aldcroft}, {B{\"o}hringer}, {Brunner}, {Civano},
  {Elvis}, {Fiore}, {Fruscione}, {Griffiths}, {Guzzo}, {Iovino}, {Koekemoer},
  {Mainieri}, {Scoville}, {Shopbell}, {Silverman}, \& {Urry}}]{cappelluti09}
{Cappelluti}, N., {Brusa}, M., {Hasinger}, G., {et~al.} 2009, \aap, 497, 635

\bibitem[{{Cardamone} {et~al.}(2010){Cardamone}, {Urry}, {Schawinski},
  {Treister}, {Brammer}, \& {Gawiser}}]{cardamone10}
{Cardamone}, C.~N., {Urry}, C.~M., {Schawinski}, K., {et~al.} 2010, \apjl, 721,
  L38

\bibitem[{{Chapman} {et~al.}(2005){Chapman}, {Blain}, {Smail}, \&
  {Ivison}}]{chapman05}
{Chapman}, S.~C., {Blain}, A.~W., {Smail}, I., \& {Ivison}, R.~J. 2005, \apj,
  622, 772

\bibitem[{{Chapman} {et~al.}(2003){Chapman}, {Windhorst}, {Odewahn}, {Yan}, \&
  {Conselice}}]{chapman03}
{Chapman}, S.~C., {Windhorst}, R., {Odewahn}, S., {Yan}, H., \& {Conselice}, C.
  2003, \apj, 599, 92

\bibitem[{{Chary} \& {Elbaz}(2001)}]{ce01}
{Chary}, R. \& {Elbaz}, D. 2001, The Astrophysical Journal, 556, 562

\bibitem[{{Ciotti} {et~al.}(2010){Ciotti}, {Ostriker}, \& {Proga}}]{ciotti10}
{Ciotti}, L., {Ostriker}, J.~P., \& {Proga}, D. 2010, \apj, 717, 708

\bibitem[{{Cisternas} {et~al.}(2011){Cisternas}, {Jahnke}, {Inskip},
  {Kartaltepe}, {Koekemoer}, {Lisker}, {Robaina}, {Scodeggio}, {Sheth},
  {Trump}, {Andrae}, {Miyaji}, {Lusso}, {Brusa}, {Capak}, {Cappelluti},
  {Civano}, {Ilbert}, {Impey}, {Leauthaud}, {Lilly}, {Salvato}, {Scoville}, \&
  {Taniguchi}}]{cisternas11}
{Cisternas}, M., {Jahnke}, K., {Inskip}, K.~J., {et~al.} 2011, \apj, 726, 57

\bibitem[{{Coil} {et~al.}(2009){Coil}, {Georgakakis}, {Newman}, {Cooper},
  {Croton}, {Davis}, {Koo}, {Laird}, {Nandra}, {Weiner}, {Willmer}, \&
  {Yan}}]{coil09}
{Coil}, A.~L., {Georgakakis}, A., {Newman}, J.~A., {et~al.} 2009, \apj, 701,
  1484

\bibitem[{{Coppin} {et~al.}(2008){Coppin}, {Swinbank}, {Neri}, {Cox},
  {Alexander}, {Smail}, {Page}, {Stevens}, {Knudsen}, {Ivison}, {Beelen},
  {Bertoldi}, \& {Omont}}]{coppin08}
{Coppin}, K.~E.~K., {Swinbank}, A.~M., {Neri}, R., {et~al.} 2008, \mnras, 389,
  45

\bibitem[{{Cowie} {et~al.}(2003){Cowie}, {Barger}, {Bautz}, {Brandt}, \&
  {Garmire}}]{cowie03}
{Cowie}, L.~L., {Barger}, A.~J., {Bautz}, M.~W., {Brandt}, W.~N., \& {Garmire},
  G.~P. 2003, \apjl, 584, L57

\bibitem[{{Croton} {et~al.}(2006){Croton}, {Springel}, {White}, {De Lucia},
  {Frenk}, {Gao}, {Jenkins}, {Kauffmann}, {Navarro}, \& {Yoshida}}]{croton06}
{Croton}, D.~J., {Springel}, V., {White}, S.~D.~M., {et~al.} 2006, \mnras, 367,
  864

\bibitem[{{Daddi} {et~al.}(2007{\natexlab{a}}){Daddi}, {Alexander},
  {Dickinson}, {Gilli}, {Renzini}, {Elbaz}, {Cimatti}, {Chary}, {Frayer},
  {Bauer}, {Brandt}, {Giavalisco}, {Grogin}, {Huynh}, {Kurk}, {Mignoli},
  {Morrison}, {Pope}, \& {Ravindranath}}]{daddi07b}
{Daddi}, E., {Alexander}, D.~M., {Dickinson}, M., {et~al.} 2007{\natexlab{a}},
  \apj, 670, 173

\bibitem[{{Daddi} {et~al.}(2007{\natexlab{b}}){Daddi}, {Dickinson}, {Morrison},
  {Chary}, {Cimatti}, {Elbaz}, {Frayer}, {Renzini}, {Pope}, {Alexander},
  {Bauer}, {Giavalisco}, {Huynh}, {Kurk}, \& {Mignoli}}]{daddi07a}
{Daddi}, E., {Dickinson}, M., {Morrison}, G., {et~al.} 2007{\natexlab{b}},
  \apj, 670, 156

\bibitem[{{Daddi} {et~al.}(2010){Daddi}, {Elbaz}, {Walter}, {Bournaud},
  {Salmi}, {Carilli}, {Dannerbauer}, {Dickinson}, {Monaco}, \&
  {Riechers}}]{daddi10}
{Daddi}, E., {Elbaz}, D., {Walter}, F., {et~al.} 2010, \apjl, 714, L118

\bibitem[{{Dale} \& {Helou}(2002)}]{dh02}
{Dale}, D.~A. \& {Helou}, G. 2002, The Astrophysical Journal, 576, 159

\bibitem[{{Damen} {et~al.}(2009){Damen}, {Labb{\'e}}, {Franx}, {van Dokkum},
  {Taylor}, \& {Gawiser}}]{damen09}
{Damen}, M., {Labb{\'e}}, I., {Franx}, M., {et~al.} 2009, \apj, 690, 937

\bibitem[{{Davies} {et~al.}(2007){Davies}, {M{\"u}ller S{\'a}nchez}, {Genzel},
  {Tacconi}, {Hicks}, {Friedrich}, \& {Sternberg}}]{davies07}
{Davies}, R.~I., {M{\"u}ller S{\'a}nchez}, F., {Genzel}, R., {et~al.} 2007,
  \apj, 671, 1388

\bibitem[{{Dekel} {et~al.}(2009){Dekel}, {Birnboim}, {Engel}, {Freundlich},
  {Goerdt}, {Mumcuoglu}, {Neistein}, {Pichon}, {Teyssier}, \&
  {Zinger}}]{dekel09}
{Dekel}, A., {Birnboim}, Y., {Engel}, G., {et~al.} 2009, \nat, 457, 451

\bibitem[{{Dey} {et~al.}(2008){Dey}, {Soifer}, {Desai}, {Brand}, {Le Floc'h},
  {Brown}, {Jannuzi}, {Armus}, {Bussmann}, {Brodwin}, {Bian}, {Eisenhardt},
  {Higdon}, {Weedman}, \& {Willner}}]{dey08}
{Dey}, A., {Soifer}, B.~T., {Desai}, V., {et~al.} 2008, \apj, 677, 943

\bibitem[{{Di Matteo} {et~al.}(2011){Di Matteo}, {Khandai}, {DeGraf}, {Feng},
  {Croft}, {Lopez}, \& {Springel}}]{dimatteo11}
{Di Matteo}, T., {Khandai}, N., {DeGraf}, C., {et~al.} 2011, arXiv: 1107.1253

\bibitem[{{Di Matteo} {et~al.}(2005){Di Matteo}, {Springel}, \&
  {Hernquist}}]{dimatteo05}
{Di Matteo}, T., {Springel}, V., \& {Hernquist}, L. 2005, \nat, 433, 604

\bibitem[{{Donley} {et~al.}(2010){Donley}, {Rieke}, {Alexander}, {Egami}, \&
  {P{\'e}rez-Gonz{\'a}lez}}]{donley10}
{Donley}, J.~L., {Rieke}, G.~H., {Alexander}, D.~M., {Egami}, E., \&
  {P{\'e}rez-Gonz{\'a}lez}, P.~G. 2010, \apj, 719, 1393

\bibitem[{{Dunne} {et~al.}(2009){Dunne}, {Ivison}, {Maddox}, {Cirasuolo},
  {Mortier}, {Foucaud}, {Ibar}, {Almaini}, {Simpson}, \& {McLure}}]{dunne09}
{Dunne}, L., {Ivison}, R.~J., {Maddox}, S., {et~al.} 2009, \mnras, 394, 3

\bibitem[{{Elbaz} {et~al.}(2007){Elbaz}, {Daddi}, {Le Borgne}, {Dickinson},
  {Alexander}, {Chary}, {Starck}, {Brandt}, {Kitzbichler}, {MacDonald},
  {Nonino}, {Popesso}, {Stern}, \& {Vanzella}}]{elbaz07}
{Elbaz}, D., {Daddi}, E., {Le Borgne}, D., {et~al.} 2007, \aap, 468, 33

\bibitem[{{Elbaz} {et~al.}(2009){Elbaz}, {Jahnke}, {Pantin}, {Le Borgne}, \&
  {Letawe}}]{elbaz09}
{Elbaz}, D., {Jahnke}, K., {Pantin}, E., {Le Borgne}, D., \& {Letawe}, G. 2009,
  \aap, 507, 1359

\bibitem[{{Feain} {et~al.}(2007){Feain}, {Papadopoulos}, {Ekers}, \&
  {Middelberg}}]{feain07}
{Feain}, I.~J., {Papadopoulos}, P.~P., {Ekers}, R.~D., \& {Middelberg}, E.
  2007, \apj, 662, 872

\bibitem[{{Ferrarese} \& {Merritt}(2000)}]{ferrarese00}
{Ferrarese}, L. \& {Merritt}, D. 2000, \apjl, 539, L9

\bibitem[{{Fioc} \& {Rocca-Volmerange}(1997)}]{fioc97}
{Fioc}, M. \& {Rocca-Volmerange}, B. 1997, \aap, 326, 950

\bibitem[{{Fiore} {et~al.}(2003){Fiore}, {Brusa}, {Cocchia}, {Baldi},
  {Carangelo}, {Ciliegi}, {Comastri}, {La Franca}, {Maiolino}, {Matt},
  {Molendi}, {Mignoli}, {Perola}, {Severgnini}, \& {Vignali}}]{fiore03}
{Fiore}, F., {Brusa}, M., {Cocchia}, F., {et~al.} 2003, \aap, 409, 79

\bibitem[{{Fiore} {et~al.}(2008){Fiore}, {Grazian}, {Santini}, {Puccetti},
  {Brusa}, {Feruglio}, {Fontana}, {Giallongo}, {Comastri}, {Gruppioni},
  {Pozzi}, {Zamorani}, \& {Vignali}}]{fiore08}
{Fiore}, F., {Grazian}, A., {Santini}, P., {et~al.} 2008, \apj, 672, 94

\bibitem[{{Fiore} {et~al.}(2009){Fiore}, {Puccetti}, {Brusa}, {Salvato},
  {Zamorani}, {Aldcroft}, {Aussel}, {Brunner}, {Capak}, {Cappelluti}, {Civano},
  {Comastri}, {Elvis}, {Feruglio}, {Finoguenov}, {Fruscione}, {Gilli},
  {Hasinger}, {Koekemoer}, {Kartaltepe}, {Ilbert}, {Impey}, {Le Floc'h},
  {Lilly}, {Mainieri}, {Martinez-Sansigre}, {McCracken}, {Menci}, {Merloni},
  {Miyaji}, {Sanders}, {Sargent}, {Schinnerer}, {Scoville}, {Silverman},
  {Smolcic}, {Steffen}, {Santini}, {Taniguchi}, {Thompson}, {Trump}, {Vignali},
  {Urry}, \& {Yan}}]{fiore09}
{Fiore}, F., {Puccetti}, S., {Brusa}, M., {et~al.} 2009, \apj, 693, 447

\bibitem[{{Fiore} {et~al.}(2012){Fiore}, {Puccetti}, {Grazian}, {Menci},
  {Shankar}, {Santini}, {Piconcelli}, {Koekemoer}, {Fontana}, {Boutsia},
  {Castellano}, {Lamastra}, {Malacaria}, {Feruglio}, {Mathur}, {Miller}, \&
  {Pannella}}]{fiore12}
{Fiore}, F., {Puccetti}, S., {Grazian}, A., {et~al.} 2012, \aap, 537, A16+


\bibitem[{{Fontana} {et~al.}(2004){Fontana}, {Pozzetti}, {Donnarumma},
  {Renzini}, {Cimatti}, {Zamorani}, {Menci}, {Daddi}, {Giallongo}, {Mignoli},
  {Perna}, {Salimbeni}, {Saracco}, {Broadhurst}, {Cristiani}, {D'Odorico}, \&
  {Gilmozzi}}]{fontana04}
{Fontana}, A., {Pozzetti}, L., {Donnarumma}, I., {et~al.} 2004, Astronomy \&
  Astrophysics, 424, 23

\bibitem[{{Fontana} {et~al.}(2006){Fontana}, {Salimbeni}, {Grazian},
  {Giallongo}, {Pentericci}, {Nonino}, {Fontanot}, {Menci}, {Monaco},
  {Cristiani}, {Vanzella}, {De Santis}, \& {Gallozzi}}]{fontana06}
{Fontana}, A., {Salimbeni}, S., {Grazian}, A., {et~al.} 2006, \aap, 459, 745

\bibitem[{{Fontanot} {et~al.}(2009){Fontanot}, {De Lucia}, {Monaco},
  {Somerville}, \& {Santini}}]{fontanot09}
{Fontanot}, F., {De Lucia}, G., {Monaco}, P., {Somerville}, R.~S., \&
  {Santini}, P. 2009, \mnras, 397, 1776

\bibitem[{{Fritz} {et~al.}(2006){Fritz}, {Franceschini}, \&
  {Hatziminaoglou}}]{fritz06}
{Fritz}, J., {Franceschini}, A., \& {Hatziminaoglou}, E. 2006, \mnras, 366, 767

\bibitem[{{Gabor} {et~al.}(2009){Gabor}, {Impey}, {Jahnke}, {Simmons}, {Trump},
  {Koekemoer}, {Brusa}, {Cappelluti}, {Schinnerer}, {Smol{\v c}i{\'c}},
  {Salvato}, {Rhodes}, {Mobasher}, {Capak}, {Massey}, {Leauthaud}, \&
  {Scoville}}]{gabor09}
{Gabor}, J.~M., {Impey}, C.~D., {Jahnke}, K., {et~al.} 2009, \apj, 691, 705

\bibitem[{{Gebhardt} {et~al.}(2000){Gebhardt}, {Kormendy}, {Ho}, {Bender},
  {Bower}, {Dressler}, {Faber}, {Filippenko}, {Green}, {Grillmair}, {Lauer},
  {Magorrian}, {Pinkney}, {Richstone}, \& {Tremaine}}]{gebhardt00}
{Gebhardt}, K., {Kormendy}, J., {Ho}, L.~C., {et~al.} 2000, \apjl, 543, L5

\bibitem[{{Genzel} {et~al.}(2008){Genzel}, {Burkert}, {Bouch{\'e}}, {Cresci},
  {F{\"o}rster Schreiber}, {Shapley}, {Shapiro}, {Tacconi}, {Buschkamp},
  {Cimatti}, {Daddi}, {Davies}, {Eisenhauer}, {Erb}, {Genel}, {Gerhard},
  {Hicks}, {Lutz}, {Naab}, {Ott}, {Rabien}, {Renzini}, {Steidel}, {Sternberg},
  \& {Lilly}}]{genzel08}
{Genzel}, R., {Burkert}, A., {Bouch{\'e}}, N., {et~al.} 2008, \apj, 687, 59

\bibitem[{{Genzel} {et~al.}(2010){Genzel}, {Tacconi}, {Gracia-Carpio},
  {Sternberg}, {Cooper}, {Shapiro}, {Bolatto}, {Bouch{\'e}}, {Bournaud},
  {Burkert}, {Combes}, {Comerford}, {Cox}, {Davis}, {Schreiber},
  {Garcia-Burillo}, {Lutz}, {Naab}, {Neri}, {Omont}, {Shapley}, \&
  {Weiner}}]{genzel10}
{Genzel}, R., {Tacconi}, L.~J., {Gracia-Carpio}, J., {et~al.} 2010, \mnras,
  407, 2091

\bibitem[{{Georgakakis} {et~al.}(2010){Georgakakis}, {Rowan-Robinson},
  {Nandra}, {Digby-North}, {P{\'e}rez-Gonz{\'a}lez}, \&
  {Barro}}]{georgakakis10}
{Georgakakis}, A., {Rowan-Robinson}, M., {Nandra}, K., {et~al.} 2010, \mnras,
  406, 420

\bibitem[{{Georgantopoulos} {et~al.}(2008){Georgantopoulos}, {Georgakakis},
  {Rowan-Robinson}, \& {Rovilos}}]{georgantopoulos08}
{Georgantopoulos}, I., {Georgakakis}, A., {Rowan-Robinson}, M., \& {Rovilos},
  E. 2008, \aap, 484, 671

\bibitem[{{Glass} \& {Moorwood}(1985)}]{glass85}
{Glass}, I.~S. \& {Moorwood}, A.~F.~M. 1985, \mnras, 214, 429

\bibitem[{{Graham} {et~al.}(2011){Graham}, {Onken}, {Athanassoula}, \&
  {Combes}}]{graham11}
{Graham}, A.~W., {Onken}, C.~A., {Athanassoula}, E., \& {Combes}, F. 2011,
  \mnras, 412, 2211

\bibitem[{{Grazian} {et~al.}(2006){Grazian}, {Fontana}, {De Santis}, {Nonino},
  {Salimbeni}, {Giallongo}, {Cristiani}, {Gallozzi}, \& {Vanzella}}]{grazian06}
{Grazian}, A., {Fontana}, A., {De Santis}, C., {et~al.} 2006, Astronomy \&
  Astrophysics, 449, 951

\bibitem[{{Grogin} {et~al.}(2005){Grogin}, {Conselice}, {Chatzichristou},
  {Alexander}, {Bauer}, {Hornschemeier}, {Jogee}, {Koekemoer}, {Laidler},
  {Livio}, {Lucas}, {Paolillo}, {Ravindranath}, {Schreier}, {Simmons}, \&
  {Urry}}]{grogin05}
{Grogin}, N.~A., {Conselice}, C.~J., {Chatzichristou}, E., {et~al.} 2005,
  \apjl, 627, L97

\bibitem[{{G{\"u}ltekin} {et~al.}(2009){G{\"u}ltekin}, {Richstone}, {Gebhardt},
  {Lauer}, {Tremaine}, {Aller}, {Bender}, {Dressler}, {Faber}, {Filippenko},
  {Green}, {Ho}, {Kormendy}, {Magorrian}, {Pinkney}, \& {Siopis}}]{gueltekin09}
{G{\"u}ltekin}, K., {Richstone}, D.~O., {Gebhardt}, K., {et~al.} 2009, \apj,
  698, 198

\bibitem[{{Hasinger} {et~al.}(2005){Hasinger}, {Miyaji}, \&
  {Schmidt}}]{hasinger05}
{Hasinger}, G., {Miyaji}, T., \& {Schmidt}, M. 2005, \aap, 441, 417

\bibitem[{{Hatziminaoglou} {et~al.}(2010){Hatziminaoglou}, {Omont}, {Stevens},
  {Amblard}, {Arumugam}, {Auld}, {Aussel}, {Babbedge}, {Blain}, {Bock},
  {Boselli}, {Buat}, {Burgarella}, {Castro-Rodr{\'{\i}}guez}, {Cava},
  {Chanial}, {Clements}, {Conley}, {Conversi}, {Cooray}, {Dowell}, {Dwek},
  {Dye}, {Eales}, {Elbaz}, {Farrah}, {Fox}, {Franceschini}, {Gear}, {Glenn},
  {Gonz{\'a}lez Solares}, {Griffin}, {Halpern}, {Ibar}, {Isaak}, {Ivison},
  {Lagache}, {Levenson}, {Lu}, {Madden}, {Maffei}, {Mainetti}, {Marchetti},
  {Mortier}, {Nguyen}, {O'Halloran}, {Oliver}, {Page}, {Panuzzo},
  {Papageorgiou}, {Pearson}, {P{\'e}rez-Fournon}, {Pohlen}, {Rawlings},
  {Rigopoulou}, {Rizzo}, {Roseboom}, {Rowan-Robinson}, {Sanchez Portal},
  {Schulz}, {Scott}, {Seymour}, {Shupe}, {Smith}, {Symeonidis}, {Trichas},
  {Tugwell}, {Vaccari}, {Valtchanov}, {Vigroux}, {Wang}, {Ward}, {Wright},
  {Xu}, \& {Zemcov}}]{hatziminaoglou10}
{Hatziminaoglou}, E., {Omont}, A., {Stevens}, J.~A., {et~al.} 2010, \aap, 518,
  L33+

\bibitem[{{Hayward} {et~al.}(2011){Hayward}, {Kere{\v s}}, {Jonsson},
  {Narayanan}, {Cox}, \& {Hernquist}}]{hayward11}
{Hayward}, C.~C., {Kere{\v s}}, D., {Jonsson}, P., {et~al.} 2011, \apj, 743, 159

\bibitem[{{Heckman} {et~al.}(1984){Heckman}, {Bothun}, {Balick}, \&
  {Smith}}]{heckman84}
{Heckman}, T.~M., {Bothun}, G.~D., {Balick}, B., \& {Smith}, E.~P. 1984, \aj,
  89, 958

\bibitem[{{Hickox} {et~al.}(2009){Hickox}, {Jones}, {Forman}, {Murray},
  {Kochanek}, {Eisenstein}, {Jannuzi}, {Dey}, {Brown}, {Stern}, {Eisenhardt},
  {Gorjian}, {Brodwin}, {Narayan}, {Cool}, {Kenter}, {Caldwell}, \&
  {Anderson}}]{hickox09}
{Hickox}, R.~C., {Jones}, C., {Forman}, W.~R., {et~al.} 2009, \apj, 696, 891

\bibitem[{{Hopkins} \& {Beacom}(2006)}]{hopkins06}
{Hopkins}, A.~M. \& {Beacom}, J.~F. 2006, \apj, 651, 142

\bibitem[{{Hopkins} \& {Elvis}(2010)}]{hopkins10}
{Hopkins}, P.~F. \& {Elvis}, M. 2010, \mnras, 401, 7

\bibitem[{{Hopkins} \& {Hernquist}(2009)}]{hopkins09}
{Hopkins}, P.~F. \& {Hernquist}, L. 2009, \apj, 694, 599

\bibitem[{{Hopkins} {et~al.}(2006){Hopkins}, {Hernquist}, {Cox}, {Di Matteo},
  {Robertson}, \& {Springel}}]{hopkinsp06}
{Hopkins}, P.~F., {Hernquist}, L., {Cox}, T.~J., {et~al.} 2006, \apjs, 163, 1

\bibitem[{{Hopkins} {et~al.}(2004){Hopkins}, {Strauss}, {Hall}, {Richards},
  {Cooper}, {Schneider}, {Vanden Berk}, {Jester}, {Brinkmann}, \&
  {Szokoly}}]{hopkinsp04}
{Hopkins}, P.~F., {Strauss}, M.~A., {Hall}, P.~B., {et~al.} 2004, \aj, 128,
  1112

\bibitem[{{Hutchings} {et~al.}(2002){Hutchings}, {Frenette}, {Hanisch}, {Mo},
  {Dumont}, {Redding}, \& {Neff}}]{hutchings02}
{Hutchings}, J.~B., {Frenette}, D., {Hanisch}, R., {et~al.} 2002, \aj, 123,
  2936

\bibitem[{{Hutchings} \& {Neff}(1992)}]{hutchings92}
{Hutchings}, J.~B. \& {Neff}, S.~G. 1992, \aj, 104, 1

\bibitem[{{Hutchings} {et~al.}(2009){Hutchings}, {Scholz}, \&
  {Bianchi}}]{hutchings09}
{Hutchings}, J.~B., {Scholz}, P., \& {Bianchi}, L. 2009, \aj, 137, 3533

\bibitem[{{Ilbert} {et~al.}(2009){Ilbert}, {Capak}, {Salvato}, {Aussel},
  {McCracken}, {Sanders}, {Scoville}, {Kartaltepe}, {Arnouts}, {Le Floc'h},
  {Mobasher}, {Taniguchi}, {Lamareille}, {Leauthaud}, {Sasaki}, {Thompson},
  {Zamojski}, {Zamorani}, {Bardelli}, {Bolzonella}, {Bongiorno}, {Brusa},
  {Caputi}, {Carollo}, {Contini}, {Cook}, {Coppa}, {Cucciati}, {de la Torre},
  {de Ravel}, {Franzetti}, {Garilli}, {Hasinger}, {Iovino}, {Kampczyk},
  {Kneib}, {Knobel}, {Kovac}, {Le Borgne}, {Le Brun}, {F{\`e}vre}, {Lilly},
  {Looper}, {Maier}, {Mainieri}, {Mellier}, {Mignoli}, {Murayama}, {Pell{\`o}},
  {Peng}, {P{\'e}rez-Montero}, {Renzini}, {Ricciardelli}, {Schiminovich},
  {Scodeggio}, {Shioya}, {Silverman}, {Surace}, {Tanaka}, {Tasca}, {Tresse},
  {Vergani}, \& {Zucca}}]{ilbert09}
{Ilbert}, O., {Capak}, P., {Salvato}, M., {et~al.} 2009, \apj, 690, 1236

\bibitem[{{Ilbert} {et~al.}(2010){Ilbert}, {Salvato}, {Le Floc'h}, {Aussel},
  {Capak}, {McCracken}, {Mobasher}, {Kartaltepe}, {Scoville}, {Sanders},
  {Arnouts}, {Bundy}, {Cassata}, {Kneib}, {Koekemoer}, {Le F{\`e}vre}, {Lilly},
  {Surace}, {Taniguchi}, {Tasca}, {Thompson}, {Tresse}, {Zamojski}, {Zamorani},
  \& {Zucca}}]{ilbert10}
{Ilbert}, O., {Salvato}, M., {Le Floc'h}, E., {et~al.} 2010, \apj, 709, 644

\bibitem[{{Jahnke} {et~al.}(2004){Jahnke}, {S{\'a}nchez}, {Wisotzki}, {Barden},
  {Beckwith}, {Bell}, {Borch}, {Caldwell}, {H{\"a}ussler}, {Heymans}, {Jogee},
  {McIntosh}, {Meisenheimer}, {Peng}, {Rix}, {Somerville}, \&
  {Wolf}}]{jahnke04}
{Jahnke}, K., {S{\'a}nchez}, S.~F., {Wisotzki}, L., {et~al.} 2004, \apj, 614,
  568

\bibitem[{{Jogee}(2006)}]{jogee06}
{Jogee}, S. 2006, in Lecture Notes in Physics, Berlin Springer Verlag, Vol.
  693, Physics of Active Galactic Nuclei at all Scales, ed. {D.~Alloin}, 143--+

\bibitem[{{Johansson} {et~al.}(2009){Johansson}, {Naab}, \&
  {Burkert}}]{johansson09}
{Johansson}, P.~H., {Naab}, T., \& {Burkert}, A. 2009, \apj, 690, 802

\bibitem[{{Joseph} {et~al.}(1984){Joseph}, {Meikle}, {Robertson}, \&
  {Wright}}]{joseph84}
{Joseph}, R.~D., {Meikle}, W.~P.~S., {Robertson}, N.~A., \& {Wright}, G.~S.
  1984, \mnras, 209, 111

\bibitem[{{Karim} {et~al.}(2011){Karim}, {Schinnerer},
  {Mart{\'{\i}}nez-Sansigre}, {Sargent}, {van der Wel}, {Rix}, {Ilbert},
  {Smol{\v c}i{\'c}}, {Carilli}, {Pannella}, {Koekemoer}, {Bell}, \&
  {Salvato}}]{karim11}
{Karim}, A., {Schinnerer}, E., {Mart{\'{\i}}nez-Sansigre}, A., {et~al.} 2011,
  \apj, 730, 61

\bibitem[{{Kauffmann} {et~al.}(2003){Kauffmann}, {Heckman}, {Tremonti},
  {Brinchmann}, {Charlot}, {White}, {Ridgway}, {Brinkmann}, {Fukugita}, {Hall},
  {Ivezi{\'c}}, {Richards}, \& {Schneider}}]{kauffmann03}
{Kauffmann}, G., {Heckman}, T.~M., {Tremonti}, C., {et~al.} 2003, \mnras, 346,
  1055

\bibitem[{{Kim} {et~al.}(2006){Kim}, {Ho}, \& {Im}}]{kim06}
{Kim}, M., {Ho}, L.~C., \& {Im}, M. 2006, \apj, 642, 702

\bibitem[{{Kocevski} {et~al.}(2012){Kocevski}, {Faber}, {Mozena}, {Koekemoer},
  {Nandra}, {Rangel}, {Laird}, {Brusa}, {Wuyts}, {Trump}, {Koo}, {Somerville},
  {Bell}, {Lotz}, {Alexander}, {Bournaud}, {Conselice}, {Dahlen}, {Dekel},
  {Donley}, {Dunlop}, {Finoguenov}, {Georgakakis}, {Giavalisco}, {Guo},
  {Grogin}, {Hathi}, {Juneau}, {Kartaltepe}, {Lucas}, {McGrath}, {McIntosh},
  {Mobasher}, {Robaina}, {Rosario}, {Straughn}, {van der Wel}, \&
  {Villforth}}]{kocevski12}
{Kocevski}, D.~D., {Faber}, S.~M., {Mozena}, M., {et~al.} 2012, \apj, 744, 148

\bibitem[{{Kormendy} \& {Bender}(2009)}]{kormendy09}
{Kormendy}, J. \& {Bender}, R. 2009, \apjl, 691, L142

\bibitem[{{La Franca} {et~al.}(2005){La Franca}, {Fiore}, {Comastri}, {Perola},
  {Sacchi}, {Brusa}, {Cocchia}, {Feruglio}, {Matt}, {Vignali}, {Carangelo},
  {Ciliegi}, {Lamastra}, {Maiolino}, {Mignoli}, {Molendi}, \&
  {Puccetti}}]{lafranca05}
{La Franca}, F., {Fiore}, F., {Comastri}, A., {et~al.} 2005, \apj, 635, 864

\bibitem[{{Laird} {et~al.}(2005){Laird}, {Nandra}, {Adelberger}, {Steidel}, \&
  {Reddy}}]{laird05}
{Laird}, E.~S., {Nandra}, K., {Adelberger}, K.~L., {Steidel}, C.~C., \&
  {Reddy}, N.~A. 2005, \mnras, 359, 47

\bibitem[{{Laird} {et~al.}(2010){Laird}, {Nandra}, {Pope}, \&
  {Scott}}]{laird10}
{Laird}, E.~S., {Nandra}, K., {Pope}, A., \& {Scott}, D. 2010, \mnras, 401,
  2763

\bibitem[{{Le Floc'h} {et~al.}(2009){Le Floc'h}, {Aussel}, {Ilbert},
  {Riguccini}, {Frayer}, {Salvato}, {Arnouts}, {Surace}, {Feruglio},
  {Rodighiero}, {Capak}, {Kartaltepe}, {Heinis}, {Sheth}, {Yan}, {McCracken},
  {Thompson}, {Sanders}, {Scoville}, \& {Koekemoer}}]{lefloch09}
{Le Floc'h}, E., {Aussel}, H., {Ilbert}, O., {et~al.} 2009, \apj, 703, 222

\bibitem[{{Lee} {et~al.}(2010){Lee}, {Ferguson}, {Somerville}, {Wiklind}, \&
  {Giavalisco}}]{lee10}
{Lee}, S.-K., {Ferguson}, H.~C., {Somerville}, R.~S., {Wiklind}, T., \&
  {Giavalisco}, M. 2010, \apj, 725, 1644

\bibitem[{{Luo} {et~al.}(2008){Luo}, {Bauer}, {Brandt}, {Alexander}, {Lehmer},
  {Schneider}, {Brusa}, {Comastri}, {Fabian}, {Finoguenov}, {Gilli},
  {Hasinger}, {Hornschemeier}, {Koekemoer}, {Mainieri}, {Paolillo}, {Rosati},
  {Shemmer}, {Silverman}, {Smail}, {Steffen}, \& {Vignali}}]{luo08}
{Luo}, B., {Bauer}, F.~E., {Brandt}, W.~N., {et~al.} 2008, \apjs, 179, 19

\bibitem[{{Luo} {et~al.}(2010){Luo}, {Brandt}, {Xue}, {Brusa}, {Alexander},
  {Bauer}, {Comastri}, {Koekemoer}, {Lehmer}, {Mainieri}, {Rafferty},
  {Schneider}, {Silverman}, \& {Vignali}}]{luo10}
{Luo}, B., {Brandt}, W.~N., {Xue}, Y.~Q., {et~al.} 2010, \apjs, 187, 560

\bibitem[{{Lusso} {et~al.}(2010){Lusso}, {Comastri}, {Vignali}, {Zamorani},
  {Brusa}, {Gilli}, {Iwasawa}, {Salvato}, {Civano}, {Elvis}, {Merloni},
  {Bongiorno}, {Trump}, {Koekemoer}, {Schinnerer}, {Le Floc'h}, {Cappelluti},
  {Jahnke}, {Sargent}, {Silverman}, {Mainieri}, {Fiore}, {Bolzonella}, {Le
  F{\`e}vre}, {Garilli}, {Iovino}, {Kneib}, {Lamareille}, {Lilly}, {Mignoli},
  {Scodeggio}, \& {Vergani}}]{lusso10}
{Lusso}, E., {Comastri}, A., {Vignali}, C., {et~al.} 2010, \aap, 512, A34+

\bibitem[{{Lutz} {et~al.}(2010){Lutz}, {Mainieri}, {Rafferty}, {Shao},
  {Hasinger}, {Wei{\ss}}, {Walter}, {Smail}, {Alexander}, {Brandt}, {Chapman},
  {Coppin}, {F{\"o}rster Schreiber}, {Gawiser}, {Genzel}, {Greve}, {Ivison},
  {Koekemoer}, {Kurczynski}, {Menten}, {Nordon}, {Popesso}, {Schinnerer},
  {Silverman}, {Wardlow}, \& {Xue}}]{lutz10}
{Lutz}, D., {Mainieri}, V., {Rafferty}, D., {et~al.} 2010, \apj, 712, 1287

\bibitem[{{Lutz} {et~al.}(2011){Lutz}, {Poglitsch}, {Altieri}, {Andreani},
  {Aussel}, {Berta}, {Bongiovanni}, {Brisbin}, {Cava}, {Cepa}, {Cimatti},
  {Daddi}, {Dominguez-Sanchez}, {Elbaz}, {F{\"o}rster Schreiber}, {Genzel},
  {Grazian}, {Gruppioni}, {Harwit}, {Le Floc'h}, {Magdis}, {Magnelli},
  {Maiolino}, {Nordon}, {P{\'e}rez Garc{\'{\i}}a}, {Popesso}, {Pozzi},
  {Riguccini}, {Rodighiero}, {Saintonge}, {Sanchez Portal}, {Santini}, {Shao},
  {Sturm}, {Tacconi}, {Valtchanov}, {Wetzstein}, \& {Wieprecht}}]{lutz11}
{Lutz}, D., {Poglitsch}, A., {Altieri}, B., {et~al.} 2011, \aap, 532, A90+

\bibitem[{{Lutz} {et~al.}(2008){Lutz}, {Sturm}, {Tacconi}, {Valiante},
  {Schweitzer}, {Netzer}, {Maiolino}, {Andreani}, {Shemmer}, \&
  {Veilleux}}]{lutz08}
{Lutz}, D., {Sturm}, E., {Tacconi}, L.~J., {et~al.} 2008, \apj, 684, 853

\bibitem[{{Lutz} {et~al.}(2005){Lutz}, {Yan}, {Armus}, {Helou}, {Tacconi},
  {Genzel}, \& {Baker}}]{lutz05}
{Lutz}, D., {Yan}, L., {Armus}, L., {et~al.} 2005, \apjl, 632, L13

\bibitem[{{Magdis} {et~al.}(2011){Magdis}, {Elbaz}, {Dickinson}, {Hwang},
  {Charmandaris}, {Armus}, {Daddi}, {Le Floc'h}, {Aussel}, {Dannerbauer},
  {Rigopoulou}, {Buat}, {Morrison}, {Mullaney}, {Lutz}, {Scott}, {Coia},
  {Pope}, {Pannella}, {Altieri}, {Burgarella}, {Bethermin}, {Dasyra},
  {Kartaltepe}, {Leiton}, {Magnelli}, {Popesso}, \& {Valtchanov}}]{magdis11}
{Magdis}, G.~E., {Elbaz}, D., {Dickinson}, M., {et~al.} 2011, \aap, 534, A15+

\bibitem[{{Magnelli} {et~al.}(2009){Magnelli}, {Elbaz}, {Chary}, {Dickinson},
  {Le Borgne}, {Frayer}, \& {Willmer}}]{magnelli09}
{Magnelli}, B., {Elbaz}, D., {Chary}, R.~R., {et~al.} 2009, \aap, 496, 57

\bibitem[{{Mainieri} {et~al.}(2011){Mainieri}, {Bongiorno}, {Merloni}, {Aller},
  {Carollo}, {Iwasawa}, {Koekemoer}, {Mignoli}, {Silverman}, {Bolzonella},
  {Brusa}, {Comastri}, {Gilli}, {Halliday}, {Ilbert}, {Lusso}, {Salvato},
  {Vignali}, {Zamorani}, {Contini}, {Kneib}, {Le Fevre}, {Lilly}, {Renzini},
  {Scodeggio}, {Balestra}, {Bardelli}, {Caputi}, {Coppa}, {Cucciati}, {de la
  Torre}, {de Ravel}, {Franzetti}, {Garilli}, {Iovino}, {Kampczyk}, {Knobel},
  {Kovac}, {Lamareille}, {Le Borgne}, {.~Le Brun}, {Maier}, {Nair}, {Pello},
  {Peng}, {Perez Montero}, {Pozzetti}, {Ricciardelli}, {Tanaka}, {Tasca},
  {Tresse}, {Vergani}, {Zucca}, {Aussel}, {Capak}, {Cappelluti}, {Elvis},
  {Fiore}, {Hasinger}, {Impey}, {Le Floc'h}, {Scoville}, {Taniguchi}, \&
  {Trump}}]{mainieri11}
{Mainieri}, V., {Bongiorno}, A., {Merloni}, A., {et~al.} 2011, \aap, 535, A80+

\bibitem[{{Mainieri} {et~al.}(2007){Mainieri}, {Hasinger}, {Cappelluti},
  {Brusa}, {Brunner}, {Civano}, {Comastri}, {Elvis}, {Finoguenov}, {Fiore},
  {Gilli}, {Lehmann}, {Silverman}, {Tasca}, {Vignali}, {Zamorani},
  {Schinnerer}, {Impey}, {Trump}, {Lilly}, {Maier}, {Griffiths}, {Miyaji},
  {Capak}, {Koekemoer}, {Scoville}, {Shopbell}, \& {Taniguchi}}]{mainieri07}
{Mainieri}, V., {Hasinger}, G., {Cappelluti}, N., {et~al.} 2007, \apjs, 172,
  368

\bibitem[{{Maiolino} {et~al.}(2007){Maiolino}, {Shemmer}, {Imanishi}, {Netzer},
  {Oliva}, {Lutz}, \& {Sturm}}]{maiolino07}
{Maiolino}, R., {Shemmer}, O., {Imanishi}, M., {et~al.} 2007, \aap, 468, 979

\bibitem[{{Maraston}(2005)}]{maraston05}
{Maraston}, C. 2005, \mnras, 362, 799

\bibitem[{{Maraston} {et~al.}(2010){Maraston}, {Pforr}, {Renzini}, {Daddi},
  {Dickinson}, {Cimatti}, \& {Tonini}}]{maraston10}
{Maraston}, C., {Pforr}, J., {Renzini}, A., {et~al.} 2010, \mnras, 407, 830

\bibitem[{{Marconi} \& {Hunt}(2003)}]{marconi03}
{Marconi}, A. \& {Hunt}, L.~K. 2003, \apjl, 589, L21

\bibitem[{{Marconi} {et~al.}(2004){Marconi}, {Risaliti}, {Gilli}, {Hunt},
  {Maiolino}, \& {Salvati}}]{marconi04}
{Marconi}, A., {Risaliti}, G., {Gilli}, R., {et~al.} 2004, \mnras, 351, 169

\bibitem[{{Martini}(2004)}]{martini04}
{Martini}, P. 2004, in IAU Symposium, Vol. 222, The Interplay Among Black
  Holes, Stars and ISM in Galactic Nuclei, ed. {T.~Storchi-Bergmann, L.~C.~Ho,
  \& H.~R.~Schmitt}, 235--241

\bibitem[{{McCracken} {et~al.}(2010){McCracken}, {Capak}, {Salvato}, {Aussel},
  {Thompson}, {Daddi}, {Sanders}, {Kneib}, {Willott}, {Mancini}, {Renzini},
  {Cook}, {Le F{\`e}vre}, {Ilbert}, {Kartaltepe}, {Koekemoer}, {Mellier},
  {Murayama}, {Scoville}, {Shioya}, \& {Tanaguchi}}]{mccracken10}
{McCracken}, H.~J., {Capak}, P., {Salvato}, M., {et~al.} 2010, \apj, 708, 202

\bibitem[{{Menci} {et~al.}(2008){Menci}, {Fiore}, {Puccetti}, \&
  {Cavaliere}}]{menci08}
{Menci}, N., {Fiore}, F., {Puccetti}, S., \& {Cavaliere}, A. 2008, \apj, 686,
  219

\bibitem[{{Merloni} {et~al.}(2010){Merloni}, {Bongiorno}, {Bolzonella},
  {Brusa}, {Civano}, {Comastri}, {Elvis}, {Fiore}, {Gilli}, {Hao}, {Jahnke},
  {Koekemoer}, {Lusso}, {Mainieri}, {Mignoli}, {Miyaji}, {Renzini}, {Salvato},
  {Silverman}, {Trump}, {Vignali}, {Zamorani}, {Capak}, {Lilly}, {Sanders},
  {Taniguchi}, {Bardelli}, {Carollo}, {Caputi}, {Contini}, {Coppa}, {Cucciati},
  {de la Torre}, {de Ravel}, {Franzetti}, {Garilli}, {Hasinger}, {Impey},
  {Iovino}, {Iwasawa}, {Kampczyk}, {Kneib}, {Knobel}, {Kova{\v c}},
  {Lamareille}, {Le Borgne}, {Le Brun}, {Le F{\`e}vre}, {Maier}, {Pello},
  {Peng}, {Perez Montero}, {Ricciardelli}, {Scodeggio}, {Tanaka}, {Tasca},
  {Tresse}, {Vergani}, \& {Zucca}}]{merloni10}
{Merloni}, A., {Bongiorno}, A., {Bolzonella}, M., {et~al.} 2010, \apj, 708, 137

\bibitem[{{Mullaney} {et~al.}(2011){Mullaney}, {Alexander},
  {Goulding}, \& {Hickox}}]{mullaney11}
{Mullaney}, J.~R., {Alexander}, D.~M., {Goulding}, A.~D., \& {Hickox}, R.~C.
  2011, \mnras, 414, 1082

\bibitem[{{Mullaney} {et~al.}(2010){Mullaney}, {Alexander}, {Huynh},
  {Goulding}, \& {Frayer}}]{mullaney10}
{Mullaney}, J.~R., {Alexander}, D.~M., {Huynh}, M., {Goulding}, A.~D., \&
  {Frayer}, D. 2010, \mnras, 401, 995

\bibitem[{{Mullaney} {et~al.}(2012){Mullaney}, {Pannella},
  {Daddi}, {Alexander}, {Elbaz}, {Hickox}, {Bournaud}, {Altieri}, {Aussel},
  {Coia}, {Dannerbauer}, {Dasyra}, {Dickinson}, {Hwang}, {Kartaltepe},
  {Leiton}, {Magdis}, {Magnelli}, {Popesso}, {Valtchanov}, {Bauer}, {Del Moro},
  {Hanish}, {Ivison}, {Juneau}, {Lutz}, \& {Sargent}}]{mullaney12}
{Mullaney}, J.~R., {Pannella}, M., {Daddi}, E., {et~al.} 2012,
  \mnras, 419, 95

\bibitem[{{Nandra} {et~al.}(2007){Nandra}, {Georgakakis}, {Willmer}, {Cooper},
  {Croton}, {Davis}, {Faber}, {Koo}, {Laird}, \& {Newman}}]{nandra07}
{Nandra}, K., {Georgakakis}, A., {Willmer}, C.~N.~A., {et~al.} 2007, \apjl,
  660, L11

\bibitem[{{Netzer}(2009)}]{netzer09}
{Netzer}, H. 2009, \mnras, 399, 1907

\bibitem[{{Netzer} {et~al.}(2007){Netzer}, {Lutz}, {Schweitzer}, {Contursi},
  {Sturm}, {Tacconi}, {Veilleux}, {Kim}, {Rupke}, {Baker}, {Dasyra},
  {Mazzarella}, \& {Lord}}]{netzer07}
{Netzer}, H., {Lutz}, D., {Schweitzer}, M., {et~al.} 2007, \apj, 666, 806

\bibitem[{{Noeske} {et~al.}(2007){Noeske}, {Weiner}, {Faber}, {Papovich},
  {Koo}, {Somerville}, {Bundy}, {Conselice}, {Newman}, {Schiminovich}, {Le
  Floc'h}, {Coil}, {Rieke}, {Lotz}, {Primack}, {Barmby}, {Cooper}, {Davis},
  {Ellis}, {Fazio}, {Guhathakurta}, {Huang}, {Kassin}, {Martin}, {Phillips},
  {Rich}, {Small}, {Willmer}, \& {Wilson}}]{noeske07}
{Noeske}, K.~G., {Weiner}, B.~J., {Faber}, S.~M., {et~al.} 2007, \apjl, 660,
  L43

\bibitem[{{Omont} {et~al.}(2003){Omont}, {Beelen}, {Bertoldi}, {Cox},
  {Carilli}, {Priddey}, {McMahon}, \& {Isaak}}]{omont03}
{Omont}, A., {Beelen}, A., {Bertoldi}, F., {et~al.} 2003, \aap, 398, 857

\bibitem[{{Page} {et~al.}(2004){Page}, {Stevens}, {Ivison}, \&
  {Carrera}}]{page04}
{Page}, M.~J., {Stevens}, J.~A., {Ivison}, R.~J., \& {Carrera}, F.~J. 2004,
  \apjl, 611, L85

\bibitem[{{Pannella} {et~al.}(2009){Pannella}, {Carilli}, {Daddi}, {McCracken},
  {Owen}, {Renzini}, {Strazzullo}, {Civano}, {Koekemoer}, {Schinnerer},
  {Scoville}, {Smol{\v c}i{\'c}}, {Taniguchi}, {Aussel}, {Kneib}, {Ilbert},
  {Mellier}, {Salvato}, {Thompson}, \& {Willott}}]{pannella09}
{Pannella}, M., {Carilli}, C.~L., {Daddi}, E., {et~al.} 2009, \apjl, 698, L116

\bibitem[{{Pilbratt et al.}(2010)}]{pilbratt10}
{Pilbratt et al.} 2010, \aap, Special Issue

\bibitem[{{Poglitsch et al.}(2010)}]{poglitsch10}
{Poglitsch et al.} 2010, \aap, Special Issue

\bibitem[{{Pope} {et~al.}(2008){Pope}, {Chary}, {Alexander}, {Armus},
  {Dickinson}, {Elbaz}, {Frayer}, {Scott}, \& {Teplitz}}]{pope08}
{Pope}, A., {Chary}, R.-R., {Alexander}, D.~M., {et~al.} 2008, \apj, 675, 1171

\bibitem[{{Priddey} {et~al.}(2003){Priddey}, {Isaak}, {McMahon}, \&
  {Omont}}]{priddey03}
{Priddey}, R.~S., {Isaak}, K.~G., {McMahon}, R.~G., \& {Omont}, A. 2003,
  \mnras, 339, 1183

\bibitem[{{Ranalli} {et~al.}(2003){Ranalli}, {Comastri}, \&
  {Setti}}]{ranalli03}
{Ranalli}, P., {Comastri}, A., \& {Setti}, G. 2003, \aap, 399, 39

\bibitem[{{Richstone} {et~al.}(1998){Richstone}, {Ajhar}, {Bender}, {Bower},
  {Dressler}, {Faber}, {Filippenko}, {Gebhardt}, {Green}, {Ho}, {Kormendy},
  {Lauer}, {Magorrian}, \& {Tremaine}}]{richstone98}
{Richstone}, D., {Ajhar}, E.~A., {Bender}, R., {et~al.} 1998, \nat, 395, A14+

\bibitem[{{Rodighiero} {et~al.}(2010){Rodighiero}, {Cimatti}, {Gruppioni},
  {Popesso}, {Andreani}, {Altieri}, {Aussel}, {Berta}, {Bongiovanni},
  {Brisbin}, {Cava}, {Cepa}, {Daddi}, {Dominguez-Sanchez}, {Elbaz}, {Fontana},
  {F{\"o}rster Schreiber}, {Franceschini}, {Genzel}, {Grazian}, {Lutz},
  {Magdis}, {Magliocchetti}, {Magnelli}, {Maiolino}, {Mancini}, {Nordon},
  {Perez Garcia}, {Poglitsch}, {Santini}, {Sanchez-Portal}, {Pozzi},
  {Riguccini}, {Saintonge}, {Shao}, {Sturm}, {Tacconi}, {Valtchanov},
  {Wetzstein}, \& {Wieprecht}}]{rodighiero10}
{Rodighiero}, G., {Cimatti}, A., {Gruppioni}, C., {et~al.} 2010, \aap, 518,
  L25+

\bibitem[{{Rosario} {et~al.}(2011){Rosario}, {Mozena}, {Wuyts}, {Nandra},
  {Koekemoer}, {McGrath}, {Hathi}, {Dekel}, {Donley}, {Dunlop}, {Faber},
  {Ferguson}, {Giavalisco}, {Grogin}, {Guo}, {Newman}, {Kocevski}, {Koo}, \&
  {Somerville}}]{rosario11}
{Rosario}, D.~J., {Mozena}, M., {Wuyts}, S., {et~al.} 2011, ArXiv: 1110.3816

\bibitem[{{Rowan-Robinson}(1995)}]{rowanrobinson95}
{Rowan-Robinson}, M. 1995, \mnras, 272, 737

\bibitem[{{Sajina} {et~al.}(2008){Sajina}, {Yan}, {Lutz}, {Steffen}, {Helou},
  {Huynh}, {Frayer}, {Choi}, {Tacconi}, \& {Dasyra}}]{sajina08}
{Sajina}, A., {Yan}, L., {Lutz}, D., {et~al.} 2008, \apj, 683, 659

\bibitem[{{Salimbeni} {et~al.}(2008){Salimbeni}, {Giallongo}, {Menci},
  {Castellano}, {Fontana}, {Grazian}, {Pentericci}, {Trevese}, {Cristiani},
  {Nonino}, \& {Vanzella}}]{salimbeni08}
{Salimbeni}, S., {Giallongo}, E., {Menci}, N., {et~al.} 2008, \aap, 477, 763

\bibitem[{{Salvato} {et~al.}(2009){Salvato}, {Hasinger}, {Ilbert}, {Zamorani},
  {Brusa}, {Scoville}, {Rau}, {Capak}, {Arnouts}, {Aussel}, {Bolzonella},
  {Buongiorno}, {Cappelluti}, {Caputi}, {Civano}, {Cook}, {Elvis}, {Gilli},
  {Jahnke}, {Kartaltepe}, {Impey}, {Lamareille}, {Le Floc'h}, {Lilly},
  {Mainieri}, {McCarthy}, {McCracken}, {Mignoli}, {Mobasher}, {Murayama},
  {Sasaki}, {Sanders}, {Schiminovich}, {Shioya}, {Shopbell}, {Silverman},
  {Smol{\v c}i{\'c}}, {Surace}, {Taniguchi}, {Thompson}, {Trump}, {Urry}, \&
  {Zamojski}}]{salvato09}
{Salvato}, M., {Hasinger}, G., {Ilbert}, O., {et~al.} 2009, \apj, 690, 1250

\bibitem[{{Sanders} \& {Mirabel}(1996)}]{sanders96}
{Sanders}, D.~B. \& {Mirabel}, I.~F. 1996, \araa, 34, 749

\bibitem[{{Sanders} {et~al.}(2007){Sanders}, {Salvato}, {Aussel}, {Ilbert},
  {Scoville}, {Surace}, {Frayer}, {Sheth}, {Helou}, {Brooke}, {Bhattacharya},
  {Yan}, {Kartaltepe}, {Barnes}, {Blain}, {Calzetti}, {Capak}, {Carilli},
  {Carollo}, {Comastri}, {Daddi}, {Ellis}, {Elvis}, {Fall}, {Franceschini},
  {Giavalisco}, {Hasinger}, {Impey}, {Koekemoer}, {Le F{\`e}vre}, {Lilly},
  {Liu}, {McCracken}, {Mobasher}, {Renzini}, {Rich}, {Schinnerer}, {Shopbell},
  {Taniguchi}, {Thompson}, {Urry}, \& {Williams}}]{sanders07}
{Sanders}, D.~B., {Salvato}, M., {Aussel}, H., {et~al.} 2007, \apjs, 172, 86

\bibitem[{{Sanders} {et~al.}(1989){Sanders}, {Scoville}, {Zensus}, {Soifer},
  {Wilson}, {Zylka}, \& {Steppe}}]{sanders89}
{Sanders}, D.~B., {Scoville}, N.~Z., {Zensus}, A., {et~al.} 1989, \aap, 213, L5

\bibitem[{{Sanders} {et~al.}(1988{\natexlab{a}}){Sanders}, {Soifer}, {Elias},
  {Madore}, {Matthews}, {Neugebauer}, \& {Scoville}}]{sanders88a}
{Sanders}, D.~B., {Soifer}, B.~T., {Elias}, J.~H., {et~al.} 1988{\natexlab{a}},
  \apj, 325, 74

\bibitem[{{Sanders} {et~al.}(1988{\natexlab{b}}){Sanders}, {Soifer}, {Elias},
  {Neugebauer}, \& {Matthews}}]{sanders88b}
{Sanders}, D.~B., {Soifer}, B.~T., {Elias}, J.~H., {Neugebauer}, G., \&
  {Matthews}, K. 1988{\natexlab{b}}, \apjl, 328, L35

\bibitem[{{Santini} {et~al.}(2009){Santini}, {Fontana}, {Grazian}, {Salimbeni},
  {Fiore}, {Fontanot}, {Boutsia}, {Castellano}, {Cristiani}, {de Santis},
  {Gallozzi}, {Giallongo}, {Menci}, {Nonino}, {Paris}, {Pentericci}, \&
  {Vanzella}}]{santini09}
{Santini}, P., {Fontana}, A., {Grazian}, A., {et~al.} 2009, \aap, 504, 751

\bibitem[{{Santini} {et~al.}(2012){Santini}, {Fontana}, {Grazian}, {Salimbeni},
  {Fontanot}, {Paris}, {Boutsia}, {Castellano}, {Fiore}, {Gallozzi},
  {Giallongo}, {Koekemoer}, {Menci}, {Pentericci}, \& {Somerville}}]{santini12}
{Santini}, P., {Fontana}, A., {Grazian}, A., {et~al.} 2012, \aap, 538, A33+

\bibitem[{{Schawinski} {et~al.}(2011){Schawinski}, {Treister}, {Urry},
  {Cardamone}, {Simmons}, \& {Yi}}]{schawinski11}
{Schawinski}, K., {Treister}, E., {Urry}, C.~M., {et~al.} 2011, \apjl, 727,
  L31+

\bibitem[{{Schweitzer} {et~al.}(2006){Schweitzer}, {Lutz}, {Sturm}, {Contursi},
  {Tacconi}, {Lehnert}, {Dasyra}, {Genzel}, {Veilleux}, {Rupke}, {Kim},
  {Baker}, {Netzer}, {Sternberg}, {Mazzarella}, \& {Lord}}]{schweitzer06}
{Schweitzer}, M., {Lutz}, D., {Sturm}, E., {et~al.} 2006, \apj, 649, 79

\bibitem[{{Serjeant} \& {Hatziminaoglou}(2009)}]{serjeant09}
{Serjeant}, S. \& {Hatziminaoglou}, E. 2009, \mnras, 397, 265

\bibitem[{{Shao} {et~al.}(2010){Shao}, {Lutz}, {Nordon}, {Maiolino},
  {Alexander}, {Altieri}, {Andreani}, {Aussel}, {Bauer}, {Berta},
  {Bongiovanni}, {Brandt}, {Brusa}, {Cava}, {Cepa}, {Cimatti}, {Daddi},
  {Dominguez-Sanchez}, {Elbaz}, {F{\"o}rster Schreiber}, {Geis}, {Genzel},
  {Grazian}, {Gruppioni}, {Magdis}, {Magnelli}, {Mainieri}, {P{\'e}rez
  Garc{\'{\i}}a}, {Poglitsch}, {Popesso}, {Pozzi}, {Riguccini}, {Rodighiero},
  {Rovilos}, {Saintonge}, {Salvato}, {Sanchez Portal}, {Santini}, {Sturm},
  {Tacconi}, {Valtchanov}, {Wetzstein}, \& {Wieprecht}}]{shao10}
{Shao}, L., {Lutz}, D., {Nordon}, R., {et~al.} 2010, \aap, 518, L26+

\bibitem[{{Silk}(2005)}]{silk05}
{Silk}, J. 2005, \mnras, 364, 1337

\bibitem[{{Silk} \& {Norman}(2009)}]{silk09}
{Silk}, J. \& {Norman}, C. 2009, \apj, 700, 262

\bibitem[{{Silva} {et~al.}(2004){Silva}, {Maiolino}, \& {Granato}}]{silva04}
{Silva}, L., {Maiolino}, R., \& {Granato}, G.~L. 2004, \mnras, 355, 973

\bibitem[{{Silverman} {et~al.}(2009{\natexlab{a}}){Silverman}, {Kova{\v c}},
  {Knobel}, {Lilly}, {Bolzonella}, {Lamareille}, {Mainieri}, {Brusa},
  {Cappelluti}, {Peng}, {Hasinger}, {Zamorani}, {Scodeggio}, {Contini},
  {Carollo}, {Jahnke}, {Kneib}, {Le Fevre}, {Bardelli}, {Bongiorno}, {Brunner},
  {Caputi}, {Civano}, {Comastri}, {Coppa}, {Cucciati}, {de la Torre}, {de
  Ravel}, {Elvis}, {Finoguenov}, {Fiore}, {Franzetti}, {Garilli}, {Gilli},
  {Griffiths}, {Iovino}, {Kampczyk}, {Koekemoer}, {Le Borgne}, {Le Brun},
  {Maier}, {Mignoli}, {Pello}, {Perez Montero}, {Ricciardelli}, {Tanaka},
  {Tasca}, {Tresse}, {Vergani}, {Vignali}, {Zucca}, {Bottini}, {Cappi},
  {Cassata}, {Marinoni}, {McCracken}, {Memeo}, {Meneux}, {Oesch}, {Porciani},
  \& {Salvato}}]{silverman09a}
{Silverman}, J.~D., {Kova{\v c}}, K., {Knobel}, C., {et~al.}
  2009{\natexlab{a}}, \apj, 695, 171

\bibitem[{{Silverman} {et~al.}(2009{\natexlab{b}}){Silverman}, {Lamareille},
  {Maier}, {Lilly}, {Mainieri}, {Brusa}, {Cappelluti}, {Hasinger}, {Zamorani},
  {Scodeggio}, {Bolzonella}, {Contini}, {Carollo}, {Jahnke}, {Kneib}, {Le
  F{\`e}vre}, {Merloni}, {Bardelli}, {Bongiorno}, {Brunner}, {Caputi},
  {Civano}, {Comastri}, {Coppa}, {Cucciati}, {de la Torre}, {de Ravel},
  {Elvis}, {Finoguenov}, {Fiore}, {Franzetti}, {Garilli}, {Gilli}, {Iovino},
  {Kampczyk}, {Knobel}, {Kova{\v c}}, {Le Borgne}, {Le Brun}, {Mignoli},
  {Pello}, {Peng}, {Perez Montero}, {Ricciardelli}, {Tanaka}, {Tasca},
  {Tresse}, {Vergani}, {Vignali}, {Zucca}, {Bottini}, {Cappi}, {Cassata},
  {Fumana}, {Griffiths}, {Kartaltepe}, {Koekemoer}, {Marinoni}, {McCracken},
  {Memeo}, {Meneux}, {Oesch}, {Porciani}, \& {Salvato}}]{silverman09b}
{Silverman}, J.~D., {Lamareille}, F., {Maier}, C., {et~al.} 2009{\natexlab{b}},
  \apj, 696, 396

\bibitem[{{Silverman} {et~al.}(2008){Silverman}, {Mainieri}, {Lehmer},
  {Alexander}, {Bauer}, {Bergeron}, {Brandt}, {Gilli}, {Hasinger}, {Schneider},
  {Tozzi}, {Vignali}, {Koekemoer}, {Miyaji}, {Popesso}, {Rosati}, \&
  {Szokoly}}]{silverman08}
{Silverman}, J.~D., {Mainieri}, V., {Lehmer}, B.~D., {et~al.} 2008, \apj, 675,
  1025

\bibitem[{{Springel} {et~al.}(2005){Springel}, {Di Matteo}, \&
  {Hernquist}}]{springel05b}
{Springel}, V., {Di Matteo}, T., \& {Hernquist}, L. 2005, \mnras, 361, 776

\bibitem[{{Stark} {et~al.}(2009){Stark}, {Ellis}, {Bunker}, {Bundy}, {Targett},
  {Benson}, \& {Lacy}}]{stark09}
{Stark}, D.~P., {Ellis}, R.~S., {Bunker}, A., {et~al.} 2009, \apj, 697, 1493

\bibitem[{{Stevens} {et~al.}(2005){Stevens}, {Page}, {Ivison}, {Carrera},
  {Mittaz}, {Smail}, \& {McHardy}}]{stevens05}
{Stevens}, J.~A., {Page}, M.~J., {Ivison}, R.~J., {et~al.} 2005, \mnras, 360,
  610

\bibitem[{{Stockton}(1982)}]{stockton82}
{Stockton}, A. 1982, \apj, 257, 33

\bibitem[{{Stockton} \& {Ridgway}(1991)}]{stockton91}
{Stockton}, A. \& {Ridgway}, S.~E. 1991, \aj, 102, 488

\bibitem[{{Tacconi} {et~al.}(2010){Tacconi}, {Genzel}, {Neri}, {Cox}, {Cooper},
  {Shapiro}, {Bolatto}, {Bouch{\'e}}, {Bournaud}, {Burkert}, {Combes},
  {Comerford}, {Davis}, {Schreiber}, {Garcia-Burillo}, {Gracia-Carpio}, {Lutz},
  {Naab}, {Omont}, {Shapley}, {Sternberg}, \& {Weiner}}]{tacconi10}
{Tacconi}, L.~J., {Genzel}, R., {Neri}, R., {et~al.} 2010, \nat, 463, 781

\bibitem[{{Tacconi} {et~al.}(2008){Tacconi}, {Genzel}, {Smail}, {Neri},
  {Chapman}, {Ivison}, {Blain}, {Cox}, {Omont}, {Bertoldi}, {Greve},
  {F{\"o}rster Schreiber}, {Genel}, {Lutz}, {Swinbank}, {Shapley}, {Erb},
  {Cimatti}, {Daddi}, \& {Baker}}]{tacconi08}
{Tacconi}, L.~J., {Genzel}, R., {Smail}, I., {et~al.} 2008, \apj, 680, 246

\bibitem[{{Veilleux} {et~al.}(2009{\natexlab{a}}){Veilleux}, {Kim}, {Rupke},
  {Peng}, {Tacconi}, {Genzel}, {Lutz}, {Sturm}, {Contursi}, {Schweitzer},
  {Dasyra}, {Ho}, {Sanders}, \& {Burkert}}]{veilleux09b}
{Veilleux}, S., {Kim}, D.-C., {Rupke}, D.~S.~N., {et~al.} 2009{\natexlab{a}},
  \apj, 701, 587

\bibitem[{{Veilleux} {et~al.}(2009{\natexlab{b}}){Veilleux}, {Rupke}, {Kim},
  {Genzel}, {Sturm}, {Lutz}, {Contursi}, {Schweitzer}, {Tacconi}, {Netzer},
  {Sternberg}, {Mihos}, {Baker}, {Mazzarella}, {Lord}, {Sanders}, {Stockton},
  {Joseph}, \& {Barnes}}]{veilleux09a}
{Veilleux}, S., {Rupke}, D.~S.~N., {Kim}, D., {et~al.} 2009{\natexlab{b}},
  \apjs, 182, 628

\bibitem[{{Wada}(2004)}]{wada04}
{Wada}, K. 2004, Coevolution of Black Holes and Galaxies, 186

\bibitem[{{Wuyts} {et~al.}(2011){Wuyts}, {Forster Schreiber}, {van der Wel},
  {Magnelli}, {Guo}, {Genzel}, {Lutz}, {Aussel}, {Berta}, {Cava},
  {Gracia-Carpio}, {Kocevski}, {Koekemoer}, {Lee}, {Le Floc'h}, {McGrath},
  {Nordon}, {Popesso}, {Pozzi}, {Riguccini}, {Rodighiero}, {Saintonge}, \&
  {Tacconi}}]{wuyts11}
{Wuyts}, S., {Forster Schreiber}, N.~M., {van der Wel}, A., {et~al.} 2011, \apj, 742, 96

\bibitem[{{Xue} {et~al.}(2010){Xue}, {Brandt}, {Luo}, {Rafferty}, {Alexander},
  {Bauer}, {Lehmer}, {Schneider}, \& {Silverman}}]{xue10}
{Xue}, Y.~Q., {Brandt}, W.~N., {Luo}, B., {et~al.} 2010, \apj, 720, 368

\end{thebibliography}

\begin{appendix}

\section{The main contributor to the SF enhancement}

As we demonstrated in Sect. \ref{sec:distrib}, the SF enhancement in AGN hosts compared to inactive galaxies of similar stellar mass is due to a combination of two effects: modestly brighter FIR stacked emission in faint FIR sources (i.e. undetected by PACS) and higher PACS detection rate among AGN hosts. We run a simple calculation to understand what is the dominant effect to produce the  offset observed between the average \lums of the two populations. 

Throughout the paper, the average \lums in each redshift and mass bin is defined as 
\begin{equation}
\langle \nu L _\nu (60\mu m)\rangle = \frac{\sum_{i=1}^{N_{DET}} \nu L^i_\nu (60\mu m) + N_{STACK}\times \nu L^{STACK}_\nu (60\mu m)}{N_{tot}}
\end{equation}
To simplify the calculation, we assumed constant \lums distributions for the detected sources, such as the value of each $\nu L^i_\nu (60\mu $m$)$ is  equal to the average of their distribution. The offset between AGN hosts and non-AGNs can therefore be approximated by
\begin{eqnarray}
 \Delta (\nu L _\nu (60\mu m))  = & \\
  = & \log \langle \nu L _\nu (60\mu m)\rangle_{AGN} - \log \langle \nu L _\nu (60\mu m)\rangle_{non-AGN} \nonumber 
 \end{eqnarray}
 where 
\begin{eqnarray}
 \langle \nu L _\nu (60\mu m)\rangle_{Y}  =  &    \\
   = &\frac{N_{DET,Y} \times \langle \nu L^{DET}_\nu (60\mu m)\rangle_{Y}  + N_{STACK,Y} \times \nu L^{STACK}_\nu (60\mu m)_Y }{N_{tot,Y} }  \nonumber 
\end{eqnarray}

We adopted the following assumptions which can be considered as fairly representative of the real distributions observed with our dataset (Sect. \ref{sec:distrib}):
\begin{itemize}
\item $\langle \nu L^{DET}_\nu (60\mu m)\rangle_{AGN} = 10^{45}$\ergs;
\item $\langle \nu L^{DET}_\nu (60\mu m)\rangle_{non-AGN} =\langle \nu L^{DET}_\nu (60\mu m)\rangle_{AGN}$: FIR bright AGN hosts are not enhanced in \lums with respect to inactive galaxies (as suggested by the Kolmogorov Smirnov test); 
\item $\nu L^{STACK}_\nu (60\mu m)_{AGN} = 10^{44}$\ergs;
\item $\nu L^{STACK}_\nu (60\mu m)_{non-AGN} = (0.4 \div 1) \times \nu L^{STACK}_\nu (60\mu m)_{AGN}$: faint (i.e. undetected by PACS) AGN hosts have a modestly larger stacked FIR emission than inactive galaxies (see Fig. \ref{fig:indiv});
\item the average detection rate ($N_{DET,AGN}/N_{tot,AGN}$) for the AGN population is 20\% (10\%), indicative of the GOODS (COSMOS) fields; 
\item $N_{DET,non-AGN} = (0.1 \div 1) \times N_{DET,AGN}$: 
AGN hosts have larger PACS detection rates compared to the control galaxies (see Fig. \ref{fig:detfrac}).
\end{itemize}
Finally, $N_{STACK,Y}=N_{tot,Y}-N_{DET,Y}$. 

We computed the values of $\Delta (\nu L _\nu (60\mu m))$ as a function of $\nu L^{STACK}_\nu (60\mu m)_{non-AGN}$ and $N_{DET,non-AGN}$. 
We first fixed the  detection rates for the two populations to the same value and set $\nu L^{STACK}_\nu (60\mu m)_{non-AGN}$ as a free parameter: we obtained $\Delta (\nu L _\nu (60\mu m)) \lesssim 0.08$  (0.14) dex. 
Afterwards, we fixed $\nu L^{STACK}_\nu (60\mu m)_{non-AGN}$ to the value assumed for AGN hosts and allowed the detection rate for non-AGNs to vary in the allowed range: we found a maximum offset of 0.25 (0.17) dex. 

Although very simple, this simulation tells that the main driver of the measured SF enhancement in GOODS is the higher PACS detection rate. Indeed, for a given detection rate, the linear average is highly dominated by the brightest sources, with PACS undetected sources only contributing in a minor fashion to the average offset. In the shallower COSMOS field, given the larger fraction of PACS undetected sources, the brighter FIR stacked emission and the higher PACS detection rate among AGNs can give comparable contributions to the average \lums  offset between the two populations.

\end{appendix}

\end{document}